# On the binding and stability of $CeO_2$ molecules on low-index Au surfaces

Ying-Cheng Chen[a*], Ahmad Ahmad[a], Anter El-Azab[a*]

[a] School of Materials Engineering, Purdue University, West Lafayette, IN 47907, USA

*Corresponding author: Ying-Cheng Chen, chen5305@purdue.edu; Anter El-Azab, aelazab@purdue.edu

**Abstract**

Many multi-phase thin films grown by self-assembly involve co-deposition of both metallic and oxide phases that grow into the same film. Under such conditions, the oxide species may be present in stoichiometric molecular form on the surface of the metallic phase. Quantifying the stability, binding, and diffusion of such molecular oxide states on metallic surfaces is critical for understanding self-assembly. In this work, we investigate the binding and stability of $CeO_2$ molecules on Au surfaces using spin-polarized density functional theory calculations. Specifically, we examine how the configuration of the $CeO_2$ molecule, its binding energy and binding free energy, structural stability, effective charge, and projected electronic density of states vary among various binding sites on Au(001), Au(110), and Au(111) surfaces. Molecular $CeO_2$ was found to be more favorable on all Au surfaces than its atomic constituents. It was also found that the bent $CeO_2$ molecule configuration is more stable than the linear counterpart, with the Ce atom on the hollow site and the O atoms on the top sites. In addition, the configurations of the $CeO_2$ molecule on Au surfaces with a binding energy around or weaker than -1 eV were found to have a low probability of existence above room temperature. Furthermore, the charge transfer was found to occur in such a way that the $CeO_2$ molecule becomes polarized on all surfaces, but the polarization is stronger only for the strongly bonded states on Au(001) and Au(110) surfaces. Finally, three different cases of how the $CeO_2$ molecule binds to Au(001), Au(110), and Au(111) were identified: case 1: chemisorption with symmetric bonds, case 2: chemisorption with asymmetric bonds, and case 3: physisorption with symmetric bonds. These three cases of $CeO_2$ binding are present on Au(001) and Au(110), but only cases 2 and 3 occur on Au(111).

## 1. Introduction

Many thin film systems are produced to have multiple phases on top of substrates.[1,2] Examples of such systems include films consisting of metal and oxide phases produced by pulsed laser deposition.[3] In these systems, film synthesis relies on selection of materials that phase-separate upon deposition, forming combinations of metal and oxide regions, which show as distinct parts of the film surface during the growth process, as in Au-$CeO_2$,[4] Au-$BaTiO_3$,[5] and Ag-$Al_2O_3$[6] nano-composite films. To understand the kinetics and thermodynamics of synthesis of these films, it is important to understand the energetics of binding and diffusion of the deposited species, which themselves are the atomic or molecular constituents of the topmost layers of the films. In systems involving the deposition of oxide phases, it is important to understand the oxide state on surfaces, specifically, whether the oxide binds and diffuses to the substrate and the top film surface in atomic or molecular forms. For example, in the Au-$CeO_2$ film system,[4] it is expected that both Au and $CeO_2$ species will be diffusing and segregating on the film surface, which has Au and $CeO_2$ regions. In this case, the bonding of Au adatoms and $CeO_2$ admolecules to Au and $CeO_2$ surfaces and the hopping of the two species on these surfaces must be carefully analyzed at the atomic scale to explain the self-assembly and growth of the multi-phase film itself. While single metallic atoms pose no difficulty in this regard, the state of the oxide species is not as clear. For example, it is important to figure out whether the oxide species is molecular or dissociated, and in the former case, which configuration it may take on the surface, what is the stability of this(these) configuration(s), and how hopping takes place. In the current work, we focus on the Au-$CeO_2$ system and investigate the binding and stability of $CeO_2$ molecules on various Au surfaces.

The case of oxide molecules on metallic surfaces is interesting not just in the context of multi-phase thin film growth as indicated above. For instance, inverse catalysts with the oxide-metal interaction make it possible not only to increase catalytic activities but also to investigate reactions microscopically.[7] $Cu_2O$ nanostructures on Pt, Ag, and Au surfaces,[8] FeO islands on Pt(111) surface,[9] and $CeO_x$ adlayers on Ag surfaces[10] provide insights into CO oxidation reactions. $CeO_x$ and $TiO_x$ nanoparticles on the Au(111) surface catalyze water-gas shift reactions with reaction pathways discovered theoretically.[11] Both experimental and theoretical results of $CeO_x$/Cu(111) surface catalyzing the methanol synthesis from carbon dioxide are also found.[12] For the Au-$CeO_2$ system, usually Au works as a catalyst, and $CeO_2$ works as a support for the Au catalyst.[13–15] Although Au is nonreactive, it has been extensively studied for its catalytic properties as a single atom and nanoparticles.[16] On the other hand, $CeO_2$ is utilized as a catalyst or a non-inert support for other catalysts because of its reversible redox capability from $Ce^{4+}$ to $Ce^{3+}$ and oxygen storage capacity due to oxygen vacancy formation.[17]

To the best of our knowledge, there is no literature on the energetic and electronic interactions of a $CeO_2$ molecule on Au surfaces. Moreover, there is limited literature on free-standing $CeO_2$ molecules, and both experimental and theoretical studies show that the stable form of a $CeO_2$ molecule is bent rather than linear.[18–21] The binding and migration of a $CeO_2$ molecule on the Au(001) surface were studied in a previous work that assumed there was only one rigid and bent $CeO_2$ configuration, as in bulk $CeO_2$ crystals, with no rotation or dissociation allowed during the migration between different hollow sites where the molecule binds to as its low-energy site.[22,23] Nonetheless, a $CeO_2$ molecule may have a variety of configurations when binding to different sites of surfaces such as Au, which can be inferred from a similar but well-known case such as the binding and dissociation of a $CO_2$ molecule on metal surfaces.[24,25]

This work is motivated by our previous study of growth of Au pillars in a $CeO_2$ matrix on a $SrTiO_3$ substrate, [22] as well as inspired by studies of binding of a $CO_2$ molecule on metal surfaces.[24,25] By utilizing density functional theory (DFT) calculations and our intuition of chemistry and symmetry, a $CeO_2$ molecule in vacuum is studied first, then placed onto Au(001), Au(110), and Au(111) to search for possible configurations and corresponding binding (free) energy, stability, and electronic properties of the bonded configurations. It was reported that a $CeO_2$ molecule in vacuum had a dissociation energy of 350±15 kcal mol$^{-1}$ (15.177±0.650 eV) experimentally,[26] and its bent geometry is more stable than the linear geometry by 1-3 kcal mol$^{-1}$ (0.043-0.130 eV) with an O1-Ce-O2 angle between 120-132 degrees in a Hartree-Fock-Slater study,[20] which seems to support our previous assumption. As DFT calculations typically suggest that $CeO_2$ exhibits metallic characteristics with delocalized Ce $f$ orbitals, the inclusion of a Hubbard term denoted as "$U$" is necessary to account for the electron-electron repulsion within the localized Ce $f$ orbitals.[27] By incorporating this into DFT+$U$ calculations, it allows for the prediction of $CeO_2$ as an insulator with localized Ce $f$ orbitals.[28] However, as presented later, it was found here that the $CeO_2$ molecule is more stable in a linear geometry in vacuum when Hubbard $U$ corrections $U_{\text{eff}}$ of 3 eV and 5 eV are added to include the repulsion among Ce $f$ electrons, while it prefers to be bent in three cases: when there is no Hubbard $U$ correction; when there are Hubbard $U$ plus thermal corrections; and when it is absorbed on the Au surfaces with Hubbard $U$ correction. Moreover, we observed that DFT calculations of a $CeO_2$ molecule on Au surfaces without $U$ could not converge, while the same calculations with $U$ could converge. These findings motivated the use of the Hubbard $U$ correction in all our DFT calculations of a $CeO_2$ molecule on Au surfaces. To place the Ce, O1, and O2 atoms of a $CeO_2$ molecule on Au(001), Au(110), and Au(111), four kinds of binding sites—bridge, hollow, top, and vacuum—were defined.

To theoretically verify the results of the literature for a $CeO_2$ molecule and contribute to the understanding of the interaction between the $CeO_2$ molecule and Au surfaces, the configuration, thermodynamics, and electronic properties of the $CeO_2$ molecule in vacuum and on low-index Au surfaces such as Au(001), Au(110), and Au(111) are investigated. This paper is organized as follows: The computational details of the DFT calculations and the equations used in the analyses are presented in Section 2. Results and discussion regarding geometrical, thermodynamic, and electronic investigations of a single $CeO_2$ molecule in vacuum and on Au(001), Au(110), and Au(111) are given in Section 3, followed by concluding remarks in Section 4.

## 2. Computational methods

DFT calculations were performed using the Vienna *Ab initio* Simulation Package (VASP)[29,30] with the projector augmented-wave (PAW)[31,32] pseudopotential and Perdew-Burke-Ernzerhof (PBE)[33] exchange-correlation functional. The atomic species, valence electron count, and configurations used in the PAW pseudopotential are as follows: Ce includes 12 electrons in $4f^1$ $5s^2$ $5p^6$ $5d^1$ $6s^2$, O includes 6 electrons in $2s^2$ $2p^4$, and Au includes 11 electrons in $5d^{10}$ $6s^1$. The converged energy cutoff for the plane-wave basis set in Ce-included and pure Au structures is 500 and 400 eV, respectively. To account for the localized electrons in the Ce $4f$ orbital, the simplified Hubbard $U$ correction DFT+$U$ introduced by Dudarev *et al.*[34] is used with a $U_{\mathrm{eff}}$ of 3 eV and 5 eV in all $CeO_2$/Au calculations.[27,35] However, when calculating a $CeO_2$ molecule in vacuum, introducing a Hubbard $U$ correction results in a linear molecular configuration of $CeO_2$ being more stable than a bent $CeO_2$ structure, which contradicts the experimental and theoretical literature that the bent $CeO_2$ molecule is more stable than the linear $CeO_2$ molecule.[18–21] Only when Hubbard $U$ and thermal corrections are considered simultaneously do the results align with the literature. Nevertheless, we find that without the Hubbard $U$ correction, DFT calculations for Ce atoms or $CeO_2$ molecules on Au surfaces struggle to converge. Since the focus of this study is to understand the binding and stability of the $CeO_2$ molecule on Au surfaces, rather than the geometry of the $CeO_2$ molecule in vacuum, DFT calculations with a $U_{\mathrm{eff}}$ of 3 eV and 5 eV are conducted in this work, except for verifying that the bent $CeO_2$ molecule is more stable in vacuum than the linear one according to the literature. A Gaussian smearing with a width of 0.05 eV was used across all calculations. All the structures were optimized using a conjugate gradient algorithm[36,37] until the deviations of electronic energies and ionic forces were smaller than $10^{-5}$ eV and $10^{-2}$ eV Å$^{-1}$, respectively. Spin polarization was considered only in $CeO_2$ and $CeO_2$/Au structures.[38] The supercell cell dimensions are as follows: For $CeO_2$ molecule in vacuum, a cubic supercell of dimensions a = b = c = 20.000 Å was used; for Au(001) slab (Figure 1(a)), the supercell dimensions are a = b = 8.312 Å and c = 23.312 Å along [100], [010], and [001] where the solid part is 6.240 Å along the [001] direction; for Au(110) slab (Figure 1(b)), the dimensions are a = 8.312 Å, b = 8.816 Å, and c = 20.877 Å along [001], [$1\bar{1}0$], and [110] with the solid part being 4.408 Å along the [110] direction; and for Au(111) slab (Figure 1(c)), the dimensions are a = 8.816, b = 10.180, and c = 24.598 Å along [$1\bar{1}0$], [$11\bar{2}$], and [111], with solid thickness of 7.198 Å along the [111] direction.

All slab models had a vacuum layer in the range of 16-18 Å, four layers of atoms, and only the upper two layers of atoms, including the $CeO_2$ admolecule, were relaxed. The hollow (H), bridge (B), top (T), and vacuum (V) sites on the Au(001), Au(110), and Au(111) slabs on which the constituent atoms of the $CeO_2$ molecule are placed upon are indicated in Figure 1 with blue and capital letters in circles. The vacuum (V) site refers to where the O atom is located when the Ce-O bond extends from the Au surface toward vacuum. The converged Γ-centered k-points meshes of 1×1×1, 4×4×1, 4×4×1, and 4×3×1 are used for the $CeO_2$ molecule, Au(001), Au(110), and Au(111) slabs calculations, respectively. The finite and central differences without symmetry in 3D were applied in the phonon calculations.[39–41] Thermal Corrections for Gas and Thermal Corrections for Adsorbates in the Catalysis-ElectroChem Kit within the Structural Utilities of VASPKIT were utilized to analyze phonon calculations and calculate thermal correction to $G(T)$ of $CeO_2$ molecules in vacuum and on Au surfaces, respectively.[42] Equations equivalent to those in the commercial program Gaussian are used in these two kinds of thermal corrections in VASPKIT.[43,44] A spin multiplicity of 1 was used since there is no spin and no unpaired electrons for $CeO_2$ molecules in vacuum. A standard pressure of 1 atm was used to calculate thermal corrections from translation of $CeO_2$ molecules in vacuum. IdealGasThermo and HarmonicThermo classes in the thermochemistry module of the Atomic Simulation Environment (ASE) were used to verify thermal corrections to $G(T)$ of $CeO_2$ molecules in vacuum and on Au surfaces from VASPKIT,[45] imaginary frequencies were ignored, and discrepancies between VASPKIT and ASE were smaller than 1 meV. The 3D visualization of structures was done using VESTA.[46] Matplotlib was used to plot figures.[47]

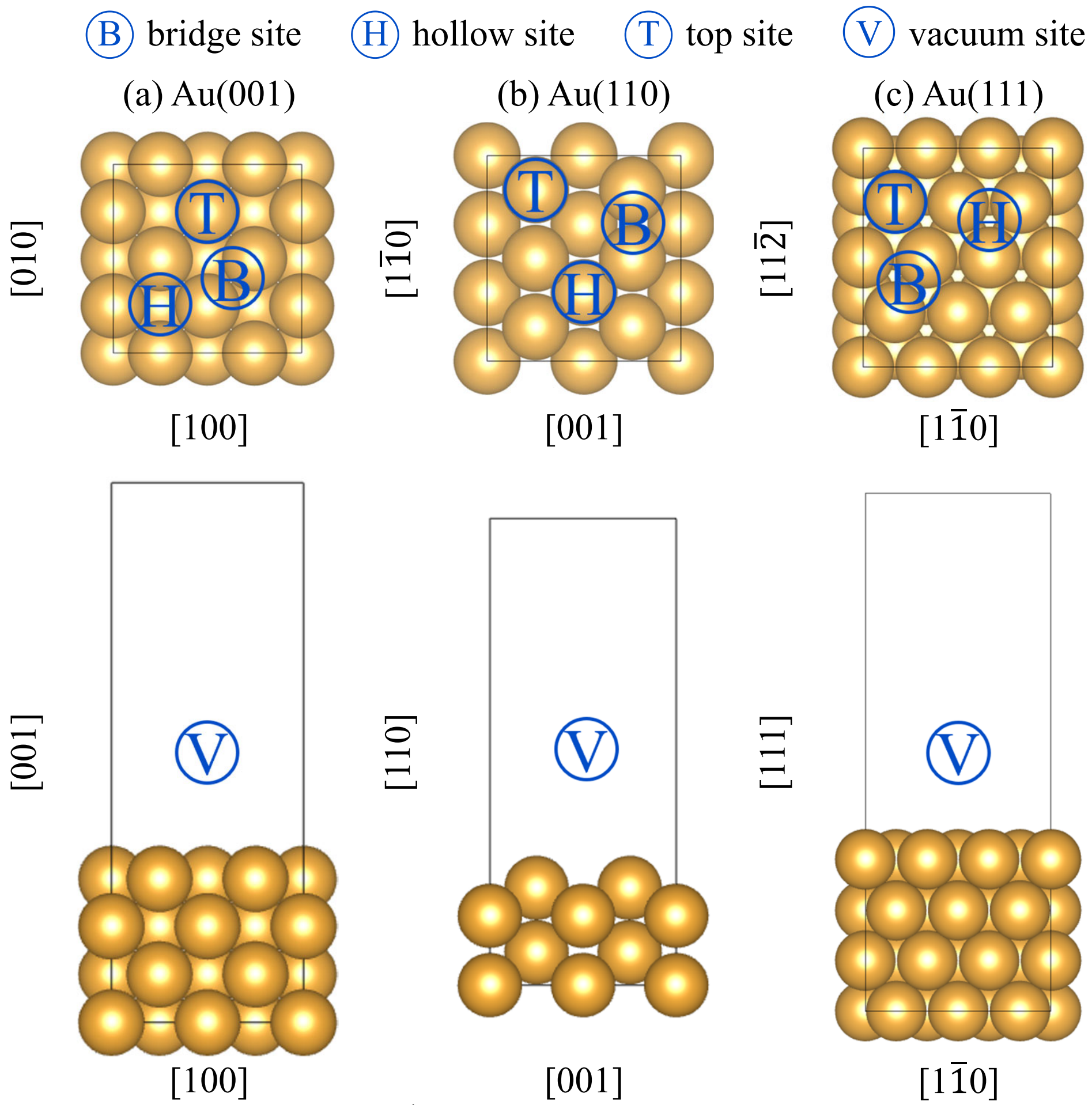

Figure 1. The top (1st row) and side (2nd row) views of unit cells of (a) Au(001), (b) Au(110), and (c) Au(111) slabs. The gold and space-filling balls represent Au atoms. The blue B, H, T, and V in circles represent the bridge, hollow, top, and vacuum sites on Au surfaces.

The strength of interaction of the $CeO_2$ molecule with the Au surfaces is characterized by the binding energy, $E_\mathrm{b}$, which is defined by:

$$E_\mathrm{b} = E_{\mathrm{CeO_2/Au}} - E_{\mathrm{CeO_2}} - E_{\mathrm{Au}}. \quad (1)$$

In the above, $E_{\mathrm{CeO_2/Au}}$, $E_{\mathrm{CeO_2}}$, and $E_{\mathrm{Au}}$ are the DFT total energies of the $CeO_2$/Au slab with $U_\mathrm{eff}$ of 3 eV and 5 eV on Ce *f* orbital, the stable linear $CeO_2$ molecule in vacuum with $U_\mathrm{eff}$ of 3 eV and 5 eV on Ce *f* orbital, and the free slab, respectively. To consider the temperature effect on the binding of a $CeO_2$ molecule on Au surfaces, it is required to add thermal corrections. In this case, the internal energy of a given state is given by:

$$E(T) = E_{\mathrm{DFT}} + E_{\mathrm{thermal}}(T). \tag{2}$$

in which $E_{\mathrm{DFT}}$ is the DFT total energy computed using DFT and $E_{\mathrm{thermal}}$ is the thermal correction to the internal energy. Gibbs free energy, $G$, is also expressed in the form:

$$G(T) = H - TS_{\mathrm{thermal}} = (E + PV) - TS_{\mathrm{thermal}}, \tag{3}$$

where $S_{\mathrm{thermal}}$ is the entropy. As the pressure is taken to be zero, Gibbs energy reduces to:

$$G = E_{\mathrm{DFT}} + E_{\mathrm{thermal}} - TS_{\mathrm{thermal}} = E_{\mathrm{DFT}} + G_{\mathrm{thermal}}(T), \tag{4}$$

which is the same as the Helmholtz free energy, $F$. In the above, $G_{\mathrm{thermal}}(T) = E_{\mathrm{thermal}} - TS_{\mathrm{thermal}}$ is the thermal contribution to free energy. The binding free energy $G_{\mathrm{b}}$ of a $CeO_2$ molecule on Au is now given by:

$$G_{\mathrm{b}}(T) = G_{\mathrm{CeO_2/Au}} - G_{\mathrm{CeO_2}} - G_{\mathrm{Au}} = E_{\mathrm{b}} + \Delta G_{\mathrm{thermal}}(T), \tag{5}$$

where $\Delta G_{\mathrm{thermal}}(T)$ is the thermal correction to the binding free energy; it is given by:

$$\Delta G_{\mathrm{thermal}}(T) = G_{\mathrm{thermal,CeO_2/Au}} - G_{\mathrm{thermal,CeO_2}} - G_{\mathrm{thermal,Au}}. \tag{6}$$

According to the Phonopy developer,[48] since selective dynamics is used when modeling the surface using the slab model in DFT calculations, the assumption of the Taylor expansion of crystal potential energy with zero first-order coefficients is not satisfied.[49] Therefore, in this work, it was assumed that $G_{\mathrm{thermal,CeO_2/Au}} = G_{\mathrm{thermal,CeO_2@Au}} + G_{\mathrm{thermal,Au@CeO_2}}$ and any vibrational degrees of freedom of the Au surface caused by the presence of a $CeO_2$ molecule, $G_{\mathrm{thermal,Au@CeO_2}}$, were minimal.[50,51] In this case,

$$\Delta G_{\mathrm{thermal}}(T) = G_{\mathrm{thermal,CeO_2@Au}} - G_{\mathrm{thermal,CeO_2}} - G_{\mathrm{thermal,Au}}, \tag{7}$$

where $G_{\mathrm{thermal,CeO_2@Au}}$ is the free energy of the $CeO_2$ molecule itself when placed on the Au surface. If the free energy of the phonon in an Au surface, $G_{\mathrm{thermal,Au}}$, in Equation (7) is also neglected, the binding free energy can be formulated as Equation (8), which is the final equation being used to calculate the binding free energy $G_{\mathrm{b}}$ in this work:

$$G_{\mathrm{b}}(T) = E_{\mathrm{b}} + G_{\mathrm{thermal,CeO_2@Au}} - G_{\mathrm{thermal,CeO_2}} = E_{\mathrm{b}} + \Delta G_{\mathrm{thermal,CeO_2}}(T), \tag{8}$$

where $G_{\mathrm{thermal,CeO_2@Au}}$ and $G_{\mathrm{thermal,CeO_2}}$ are thermal corrections to $G(T)$ of $CeO_2$ molecules in vacuum and on Au surfaces, respectively.

The probability $P_i(T)$ of a certain $CeO_2$ molecular configuration $i$ on the Au surface as a function of temperature $T$ is calculated using the Boltzmann factor[52] in terms of the binding free energy $G_{\mathrm{b},i}$ of the $CeO_2$ molecule in configuration $i$ calculated from Equation (8),

$$P_i = \frac{e^{-\frac{G_{\mathrm{b},i}}{k_{\mathrm{B}}T}}}{\sum_j e^{-\frac{G_{\mathrm{b},j}}{k_{\mathrm{B}}T}}}, \tag{9}$$

with $k_{\mathrm{B}}$ being the Boltzmann constant with a value of $8.617333262\times10^{-5}$ eV $K^{-1}$.

The stability of various $CeO_2$ molecular configurations was assessed by calculating the energy difference $\Delta E$ between a non-dissociated $CeO_2$ molecule and its constituents on the Au slab. This energy is expressed in the form:

$$\Delta E = E_{\mathrm{CeO_2/Au}} - E_{\mathrm{Ce/Au}} - E_{\mathrm{2O/Au}} + E_{\mathrm{Au}}, \tag{10}$$

where $E_{\mathrm{CeO_2/Au}}$, $E_{\mathrm{Ce/Au}}$, and $E_{\mathrm{2O/Au}}$ are the DFT total energy of a $CeO_2$ on Au slab with a $U_{\mathrm{eff}}$ of 3 eV and 5 eV on Ce *f* orbital, a Ce atom on Au slab with a $U_{\mathrm{eff}}$ of 3 eV and 5 eV on Ce *f* orbital, and two O atoms on the Au slab. Our calculations assume that the atomic form of O is more stable than the molecular form. According to the literature, 2O chemisorbed on Au surfaces are the strongest by binding on hollow sites and forming a linear O-Au-O structure.[53] In this work, the most stable configurations of 2O on Au(001) and Au(111) match the literature,[53] as shown in Figure S1(a) and Figure S3(a), respectively. However, a -0.080 eV more stable configuration of 2O on Au(110) is found in this work (Figure S2(a)), compared to that matches the literature (Figure S2(b)).

Effective charge, charge density difference, and projected density of states (PDOS) are analyzed to present the electronic perspective on the varying binding energies among different configurations of a $CeO_2$ molecule on Au surfaces. To quantify electron transfer, the effective charges of Ce, O1, and O2 atoms in a $CeO_2$/Au configuration, $Z_{\mathrm{eff,Ce}}$, $Z_{\mathrm{eff,O1}}$, and $Z_{\mathrm{eff,O2}}$, are calculated using:

$$Z_{\mathrm{eff}} = Z_{\mathrm{val}} - Q_{\mathrm{Bader}}, \tag{11}$$

where the number of valence electrons in Ce, O1, and O2 atoms considered in the pseudopotential of the DFT calculations, $Z_{\mathrm{val,Ce1}}$, $Z_{\mathrm{val,O1}}$, and $Z_{\mathrm{val,O2}}$, are 12, 6, and 6, respectively.[54] The Bader charges of Ce, O1, and O2 atoms, $Q_{\mathrm{Bader,Ce}}$, $Q_{\mathrm{Bader,O1}}$, and $Q_{\mathrm{Bader,O2}}$, are within the Bader region surrounded by the zero-flux surfaces of charges.[55–58] When $Z_{\mathrm{eff}}$ is negative, the atom gains electrons, and vice versa. To visualize the transfer of the electrons, the charge density difference $\Delta\rho$ is computed according to:

$$\Delta\rho = \rho_{\mathrm{CeO_2/Au}} - \rho_{\mathrm{CeO_2}} - \rho_{\mathrm{Au}}, \tag{12}$$

where $\rho_{\mathrm{CeO_2/Au}}$, $\rho_{\mathrm{CeO_2}}$, and $\rho_{\mathrm{Au}}$ are the charge densities of the $CeO_2$/Au surfaces, the $CeO_2$ molecule, and the corresponding Au surfaces.[42] In the figures of the charge density difference, yellow and blue isosurfaces mean 0.005 e Bohr$^{-3}$ and -0.005 e Bohr$^{-3}$ of $\Delta\rho$, respectively. When $\Delta\rho$ is positive in the $CeO_2$/Au supercell, the charges accumulate in that region, and vice versa. Electronic PDOS $D_{\vec{R}L}(E)$ of the atom at site $\vec{R}$ with the local quantum number $L$ across $n$ bands is calculated using:

$$D_{\vec{R}L}(E) = \sum_{\vec{k}} \sum_n \delta\left(E - E_n(\vec{k})\right) \left|P(\vec{k})_{\vec{R}Ln}\right|^2, \tag{13}$$

where the Dirac delta distribution $\delta\left(E - E_n(\vec{k})\right)$ is the energy difference between arbitrary energy value $E$ and the eigenvalue $E_n(\vec{k})$.[59] $P(\vec{k})_{\vec{R}Ln} = \langle\tilde{\pi}_{\vec{R}L}|\tilde{\Psi}_{n\vec{k}}\rangle$ is the orthogonal projector function projecting the pseudo Kohn-Sham states $|\tilde{\Psi}_{n\vec{k}}\rangle$ onto a local basis function $|\pi_L\rangle$, where $\langle\tilde{\pi}_{\vec{R}L}|$ are the complex conjugate of $|\tilde{\pi}_{\vec{R}L}\rangle$ and $\vec{k}$ represents k-points.[60] The Dirac delta is approximated by the Gaussian form:

$$\delta\left(E - E_n(\vec{k})\right) = \frac{1}{\sigma\sqrt{2\pi}} \exp\left[-\frac{\left(E - E_n(\vec{k})\right)^2}{2\sigma^2}\right], \tag{14}$$

with a smearing width of $\sigma$ to broaden the Dirac delta function in Equation (13).[61]

To find out how Au surface features impact on target properties $E_{\mathrm{b}}$, $\Delta E$, and $Z_{\mathrm{eff,Ce}}$, surface Au atoms $Au_s$ are defined as Au atoms satisfying $z \geq z_{\mathrm{Au,max}} - 0.5$ Å where $z_{\mathrm{Au,max}}$ is the maximum Cartesian $z$-coordinate among all Au atoms, and four Au surface features $N_{\mathrm{Ce-Au_s}}$, $N_{\mathrm{O-Au_s}}$, $N_{\mathrm{Au_{Ce}-Au}}$, and $N_{\mathrm{Au_O-Au}}$ are defined as well. $N_{\mathrm{Ce-Au_s}}$ is the number of $Au_s$ within a cutoff radius from Ce atom in $CeO_2$ molecule, which quantifies the extent of direct cation-metal interaction and reflects how strongly Ce atom interacts with $Au_s$:

$$N_{\mathrm{Ce-Au_s}} = \sum_{i \in \mathrm{Au_s}} H(r_{\mathrm{Ce-Au}} - d_{\mathrm{Ce-Au_{s,i}}}), \tag{15}$$

where $r_{\mathrm{Ce-Au}}$ = 3.4 Å is the cutoff radius between Ce atom and Au atom, $d_{\mathrm{Ce-Au}_{\mathrm{s},i}}$ is the distance between Ce atom and $i^{\mathrm{th}}$ $Au_s$ atom. $N_{\mathrm{O-Au_s}}$ denotes the number of $Au_s$ within a cutoff radius from the O atom binding to Au atom in $CeO_2$ molecule, which characterizes the strength of O binding on the Au surface:

$$N_{\mathrm{O-Au_s}} = \sum_{i \in \mathrm{Au_s}} H(r_{\mathrm{O-Au}} - d_{\mathrm{O-Au}_{\mathrm{s},i}}), \tag{16}$$

where $r_{\mathrm{O-Au}}$ = 2.34646 Å is the cutoff radius to visualize the O-Au bonds in VESTA, $d_{\mathrm{O-Au}_{\mathrm{s},i}}$ is the distance between the O atom bound to the Au atom in the $CeO_2$ molecule and $i^{\mathrm{th}}$ $Au_s$ atom. $N_{\mathrm{Au_{Ce}-Au}}$ represents the number of neighboring Au atoms within a cutoff radius from Au atom closest to the Ce atom $Au_{Ce}$, which characterizes the local Au coordination environment surrounding the Ce-proximal $Au_s$ site:

$$N_{\mathrm{Au_{Ce}-Au}} = \sum_{i \in \mathrm{Au}} H(r_{\mathrm{Au-Au}} - d_{\mathrm{Au_{Ce}-Au}_i}), \tag{17}$$

where $r_{\mathrm{Au-Au}}$ = 6.5 Å including $1^{\mathrm{st}}$ to $4^{\mathrm{th}}$ and partial $5^{\mathrm{th}}$ Au nearest neighbors, $d_{\mathrm{Au_{Ce}-Au}_i}$ being the distance between $Au_{Ce}$ atom and $i^{\mathrm{th}}$ Au atom. $N_{\mathrm{Au_O-Au}}$ is defined as the number of neighboring Au atoms within a cutoff radius from $Au_O$ that binds to the anchoring oxygen O, which encodes the local surface topology at the oxide anchoring $Au_O$ site:

$$N_{\mathrm{Au_O-Au}} = \sum_{i \in \mathrm{Au}} H(r_{\mathrm{Au-Au}} - d_{\mathrm{Au_O-Au}_i}). \tag{18}$$

Random forest[62] machine learning models are trained using RandomForestRegressor in scikit-learn[63] to interpret how these four Au surface features $N_{\mathrm{Ce-Au_s}}$, $N_{\mathrm{O-Au_s}}$, $N_{\mathrm{Au_{Ce}-Au}}$, and $N_{\mathrm{Au_O-Au}}$ affect target properties $E_b$, $\Delta E$, and $Z_{\mathrm{eff,Ce}}$. TreeExplainer[64] of SHapley Additive exPlanations (SHAP)[65] is then used to quantitatively interpret the random forest model results. For each target property, a Random Forest regression model was trained using the same set of Au surface features.

## 3. Results and discussion

### 3.1. $CeO_2$ molecules in vacuum

To study the binding and stability of the $CeO_2$ molecule on Au surfaces, the configuration of the $CeO_2$ molecule in vacuum was investigated first using DFT calculations by fixing the Ce atom, relaxing the O1 atom along $x$-axis only and the O2 atom on $xy$ plane. According to a previous Hartree-Fock-Slater study, a $CeO_2$ molecule is more stable in a bent rather than linear geometry by 1-3 kcal $mol^{-1}$ (0.043-0.130 eV), with an O1-Ce-O2 angle between 120-132 degrees.[20] Another study using the local-density approximation (LDA) functional with the gradient corrections to the exchange-correlation from Becke and Perdew shows that a $CeO_2$ molecule is also more stable in bent form in quasirelativistic calculations, while it is more stable in linear form in nonrelativistic calculations.[18] Yet, another theoretical study also showed that a bent $CeO_2$ molecule is more stable than a linear form using calculations of the multiconfigurational second-order perturbation theory (CASPT2) and DFT/B3LYP functional.[19] Infrared spectra also demonstrated that the $CeO_2$ molecule is bent.[21] The results of the literature are organized in Table 1 to compare with the results of this paper.

Table 1. The Ce-O bond length and O-Ce-O bond angle of the $CeO_2$ molecule in this work and from the literature using different methods. The first six rows with no citations are our work.

| Method | Ce-O (Å) | O-Ce-O (°) | $E_{CeO_2}$ (eV) |
|---|---|---|---|
| DFT/PBE | 1.838 | 129.5 | -21.508 |
| DFT/PBE | 1.838 | 180.0 | -21.363 |
| DFT/PBE $U_{eff}$ = 3 eV | 1.849 | 162.5 | -19.884 |
| DFT/PBE $U_{eff}$ = 3 eV | 1.843 | 180.0 | -19.907 |
| DFT/PBE $U_{eff}$ = 5 eV | 1.881 | 130.2 | -18.850 |
| DFT/PBE $U_{eff}$ = 5 eV | 1.847 | 180.0 | -19.030 |
| Nonrelativistic DFT/LDA+BP[18] | 1.800 | 180.0 | |
| Quasirelativistic DFT/LDA+BP[18] | 1.843 | 131.5 | |
| DFT/B3LYP[19] | 1.822 | 140 | |
| Hartree-Fock-Slater[20] | 1.82 | 132 | |
| Experiment[21] | 1.88 | 146±2 | |

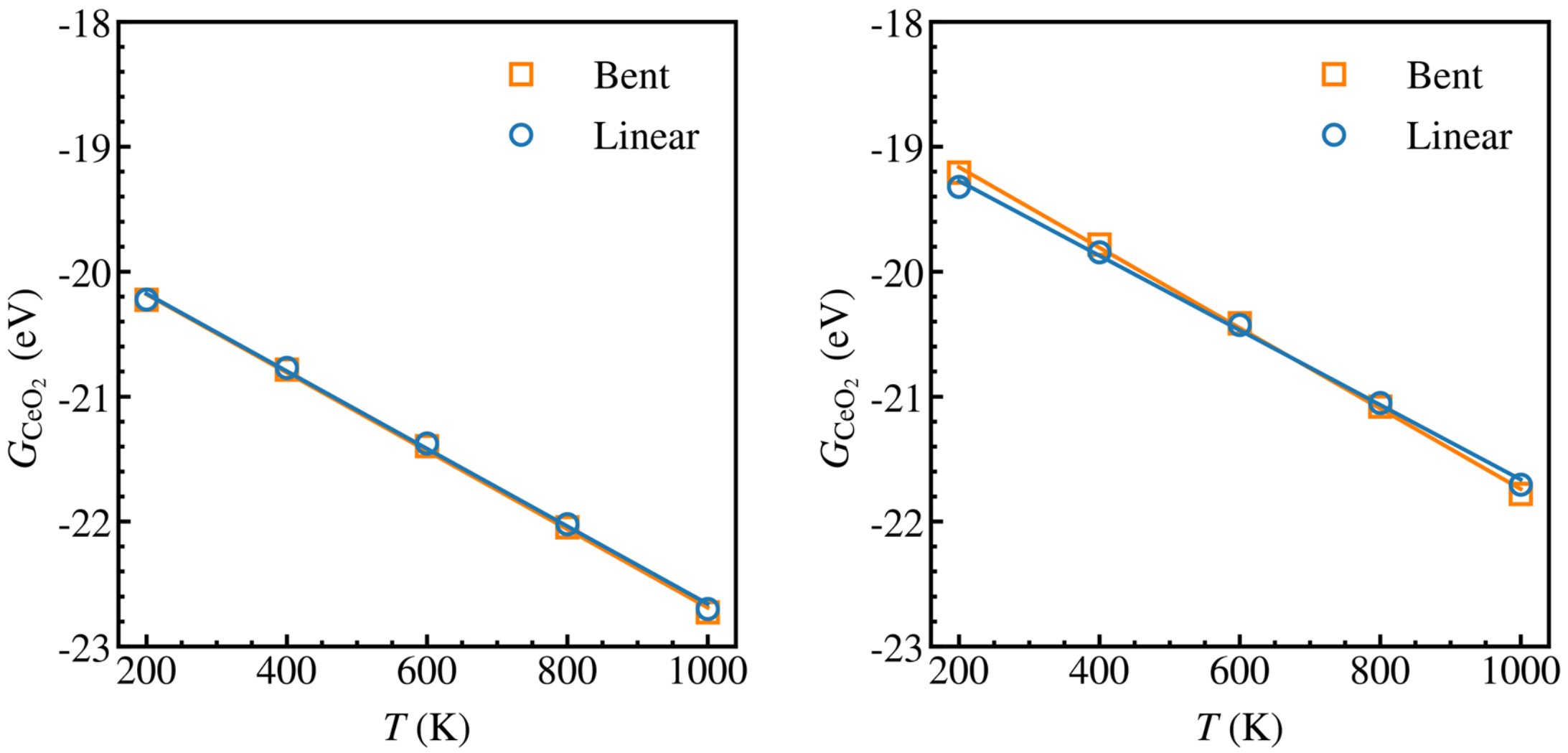

Figure 2. $G_{CeO_2}$ as a function of $T$ with a $U_{eff}$ of 3 eV (left) and 5 eV (right), respectively.

Table 2. Energy difference between bent and linear $CeO_2$ molecules at 0, 200, 400, 600, 800, and 1000 K with $U_{eff}$ of 3 eV and 5 eV.

| | DFT/PBE $U_{eff}$ = 3 eV | DFT/PBE $U_{eff}$ = 5 eV |
|---|---|---|
| $E_{CeO_2,bent} - E_{CeO_2,linear}$ (eV) | 0.022 | 0.179 |
| $G(200)_{CeO_2,bent} - G(200)_{CeO_2,linear}$ (eV) | -0.002 | 0.114 |
| $G(400)_{CeO_2,bent} - G(400)_{CeO_2,linear}$ (eV) | -0.014 | 0.062 |
| $G(600)_{CeO_2,bent} - G(600)_{CeO_2,linear}$ (eV) | -0.021 | 0.014 |
| $G(800)_{CeO_2,bent} - G(800)_{CeO_2,linear}$ (eV) | -0.025 | -0.030 |
| $G(1000)_{CeO_2,bent} - G(1000)_{CeO_2,linear}$ (eV) | -0.027 | -0.073 |

From Table 1, including $U_{eff}$ = 3 eV and 5 eV in DFT/PBE calculations can affect the relative stability of the bent and linear $CeO_2$ molecules. Without $U_{eff}$ = 3 eV and 5 eV, the relative stability of the $CeO_2$ molecule matches the results of the literature, but DFT calculations of Ce/Au or $CeO_2$/Au cannot converge. However, with $U_{eff}$ = 3 eV and 5 eV, the results contradict the literature, while DFT calculations of Ce/Au or $CeO_2$/Au can converge. Notably, with Hubbard $U$ and thermal corrections, a bent $CeO_2$ becomes more stable than a linear one when temperature rises, as shown in Figure 2 and

Table 2. Since the focus of this study is the interaction of the $CeO_2$ molecule and Au surfaces instead of the bare $CeO_2$ molecule, the lowest $E_{CeO_2}$ and $G_{CeO_2}$ in Table 1, Figure 2, and

Table 2 are respectively used to calculate $E_b$ and $G_b$ under certain $U_{eff}$ and $T$. The electronic PDOS of the $CeO_2$ molecule without and with a $U_{eff}$ of 3 eV and 5 eV are plotted in Figure S4 to find out the changes in the electronic structures of the $CeO_2$ molecule. As $U_{eff}$ increases from 0 eV to 3 eV and 5 eV, the number of excited PDOS peaks above $E - E_\mathrm{F}$ of the linear $CeO_2$ molecule decreases compared to that of the bent one.

### 3.2. $CeO_2$ molecules on Au surfaces

To find the possible configuration of the $CeO_2$ molecule on Au surfaces, the atoms of $CeO_2$ were put on hollow (H), bridge (B), top (T), and vacuum (V) sites of the Au(001), Au(110), and Au(111) surfaces and relaxed together with the upper two layers of Au surfaces using a $U_\mathrm{eff}$ of 3 eV first. Optimization is then continued from the converged structures with a $U_\mathrm{eff}$ of 3 eV using a $U_\mathrm{eff}$ of 5 eV to verify and compare the results. Here, the vacuum (V) site only applies to the position of the O atom when the Ce-O bond is pointing from the Au surface to the vacuum.

To facilitate the distinction between various configurations of the $CeO_2$ molecules on Au surfaces, we introduce a naming rule for the configuration of the $CeO_2$ molecule on Au surfaces, indicating the position of atoms after the structure is optimized. The name of the configuration consists of four letters and one number, e.g., HTT-B1. The first letter specifies the position of the Ce atom (H, T, and B). The second and third letters denote the positions of O1 and O2 atoms (H, T, B, and V). The fourth letter indicates the geometry of the $CeO_2$ molecule, specifying whether it exhibits bent (B) or linear (L) geometry, with the bond angle being less than or greater than 170 degrees, respectively. The number at the end is used for ordering the configurations from strong to weak $E_\mathrm{b}$ when the first through fourth letters are identical (1, 2, 3, …). The positions of constituent atoms of the $CeO_2$ molecule on the Au(001), Au(110), and Au(111) surfaces before and after the optimization with a $U_\mathrm{eff}$ of 3 eV and 5 eV are respectively listed in Table S1 and S2, S5 and S6, S9 and S10. By putting the constituent atoms of the $CeO_2$ molecule on the hollow (H), bridge (B), top (T), and vacuum (V) sites of the Au slab models, the study determined that the possible configurations of $CeO_2$ on Au(001), Au(110), and Au(111) are eleven (Figure 7 and Figure 8;

Section 3.2.1), eight (Figure 11 and Figure 12; Section 3.2.2), and nine (Figure 15 and Figure 16; Section 3.2.3) with a $U_{\mathrm{eff}}$ of 3 eV and 5 eV, respectively. Since the planar packing fraction and the symmetry of Au(001), Au(110), and Au(111) are different, according to Table 3, the binding of $CeO_2$ will not be the same amongst these surfaces.

Table 3. The planar packing fraction and the symmetry of Au(001), Au(110), and Au(111).

| Surface | Planar packing fraction | Symmetry |
|---|---|---|
| Au(001) | 0.785 | 4-fold rotational |
| Au(110) | 0.555 | 2-fold rotational |
| Au(111) | 0.907 | 6-fold rotational |

Observing the behavior of $E_{\mathrm{b}}$ and $G_{\mathrm{b}}$ of $CeO_2$ molecules on Au(110), Au(001), and Au(111) from low to high planar packing fraction and symmetry, it is noticed that both the sample means of $E_{\mathrm{b}}$ and $G_{\mathrm{b}}$ become stronger, while the sample standard deviation becomes smaller, as shown in Figure 3. The reason behind this behavior is that the diversity of configurations of a $CeO_2$ molecule on Au surfaces decreases when the planar packing fraction and symmetry of Au surfaces increase, as shown in the following subsections. With a $U_{\mathrm{eff}}$ = 3 eV, the energy difference $\Delta E$ (see Equation (10)) becomes more negative and more stable when a $CeO_2$ molecule binds on Au(110), Au(001), and Au(111) from low to high planar packing fraction and symmetry, but there is no unified trend of how increasing $U_{\mathrm{eff}}$ to 5 eV affects $\Delta E$, as shown in Figure 4. When $U_{\mathrm{eff}}$ is increased from 3 eV to 5 eV, $Z_{\mathrm{eff,Ce}}$, as defined in Equation (11), becomes more positive (Figure 5) indicating that the Ce atom becomes more oxidized and less reduced in terms of losing electrons, which makes $CeO_2$ have stronger $E_{\mathrm{b}}$ and $G_{\mathrm{b}}$ (Figure 3).

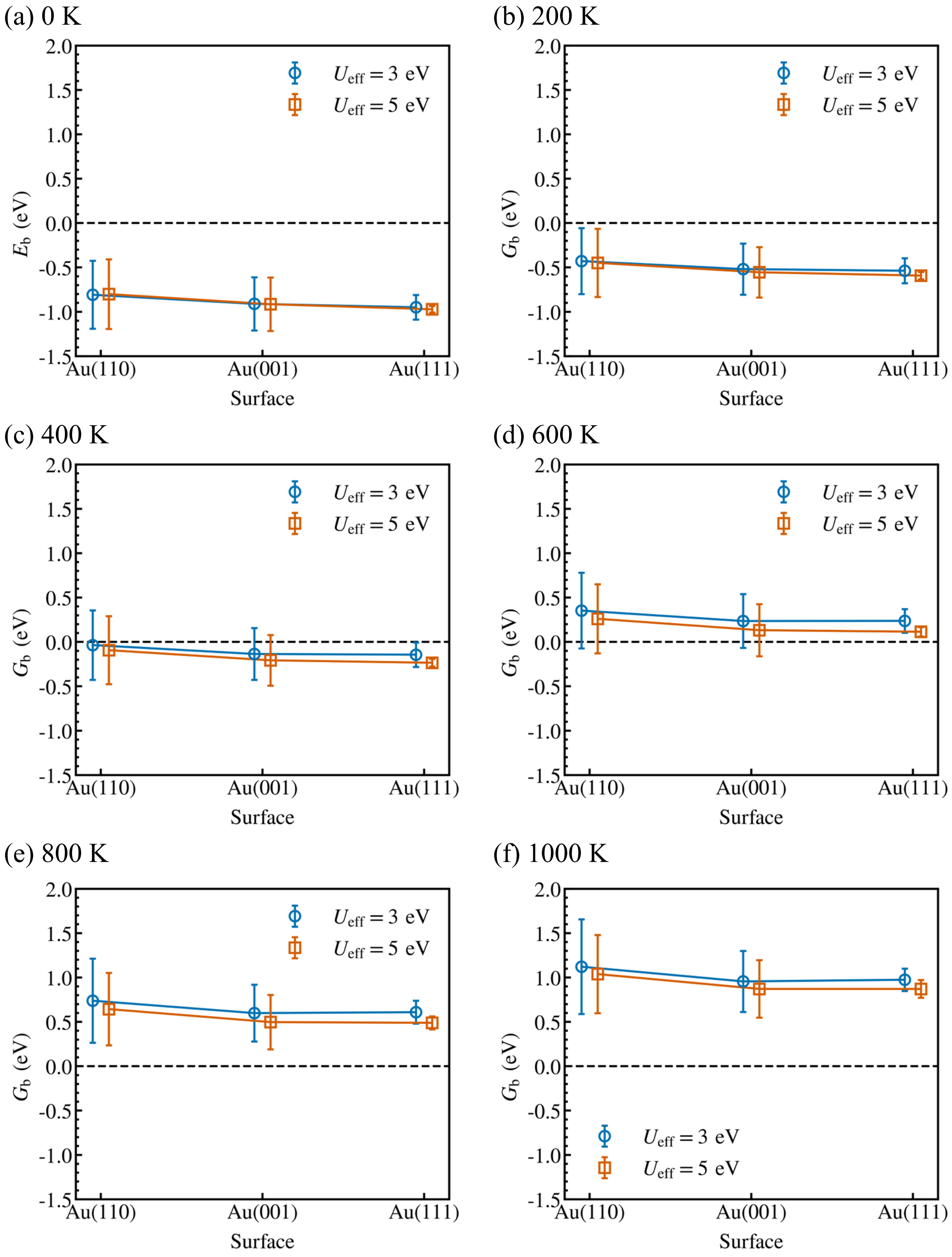

Figure 3. (a) $E_b$ at 0 K and $G_b$ at (b) 200 K, (c) 400 K, (d) 600 K, (e) 800 K, and (f) 1000 K of a $CeO_2$ molecule on Au(110), Au(001), and Au(111) from low to high planar packing fraction and symmetry with a $U_{eff}$ of 3 eV (blue hollow circle) and 5 eV (orange hollow square). Data points and error bars represent the sample means and sample standard deviations, respectively.

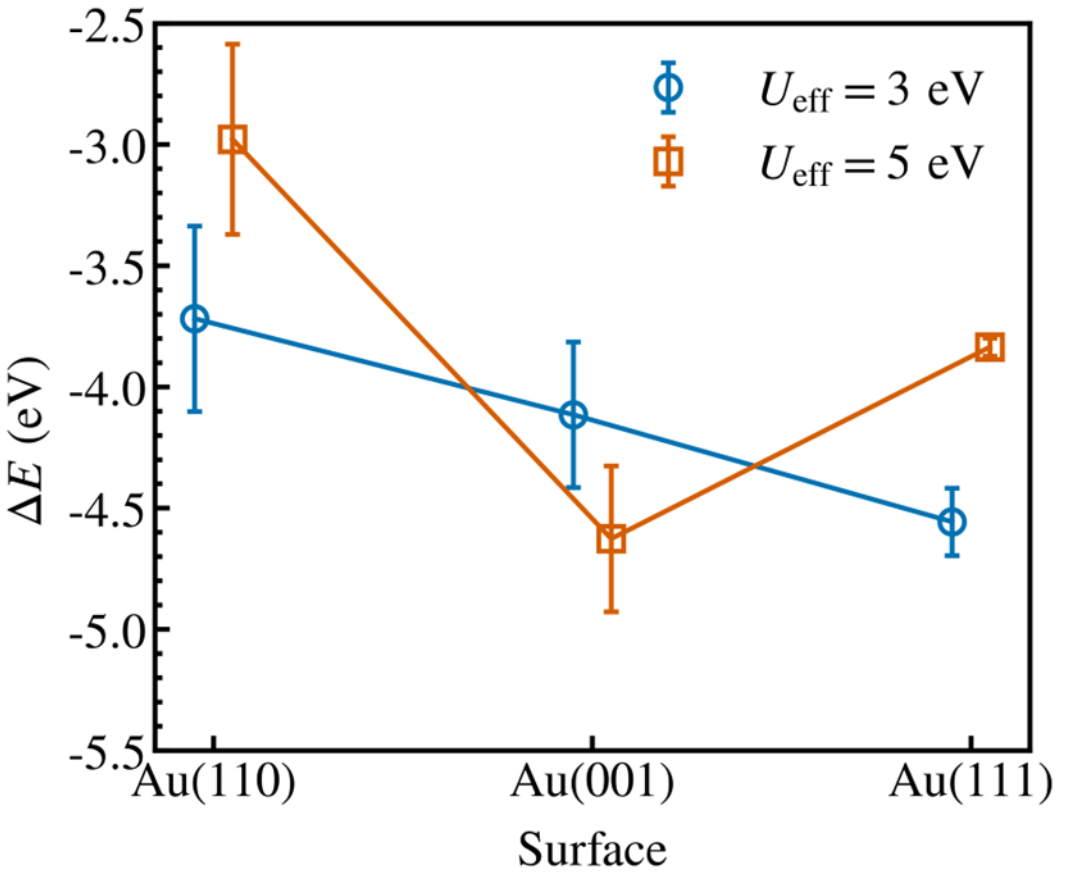

Figure 4. $\Delta E$ of a $CeO_2$ molecule on Au(110), Au(001), and Au(111) with a $U_{\mathrm{eff}}$ of 3 eV (blue hollow circle) and 5 eV (orange hollow square). Data points and error bars represent the sample means and sample standard deviations, respectively.

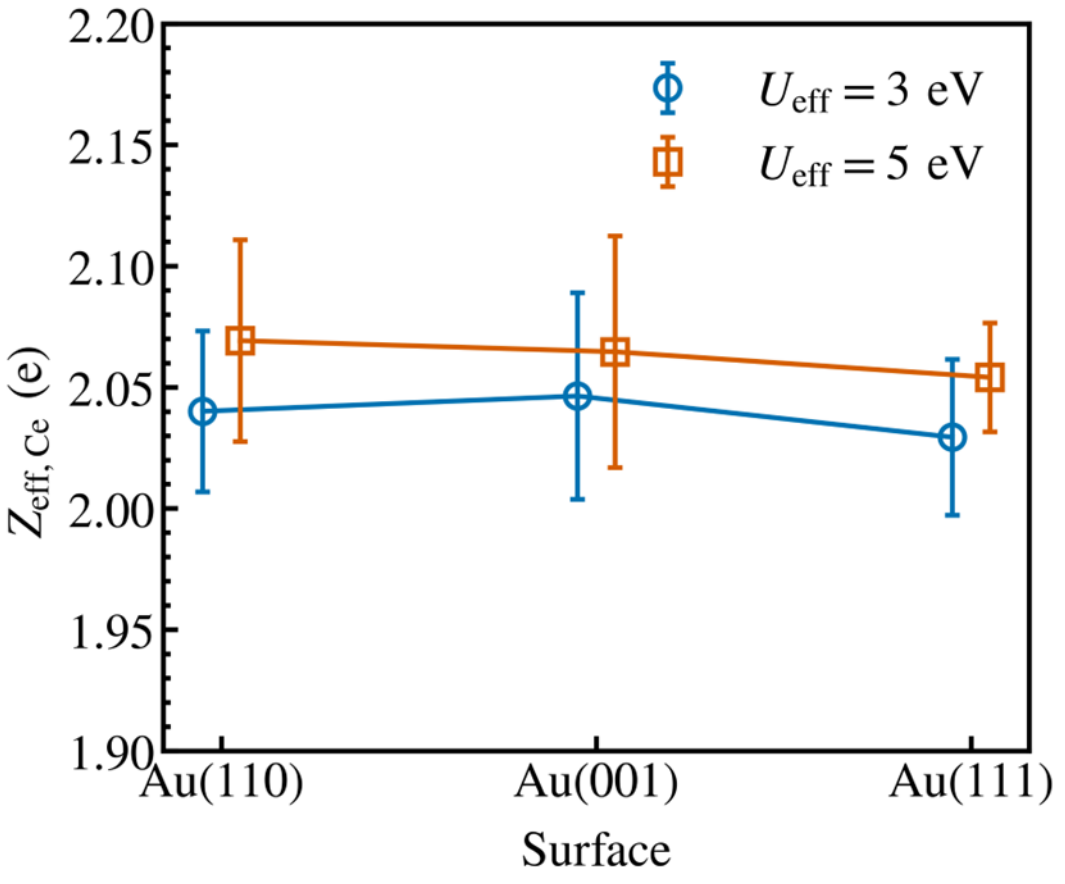

Figure 5. $Z_{\mathrm{eff,Ce}}$ of a $CeO_2$ molecule on Au(110), Au(001), and Au(111) with a $U_{\mathrm{eff}}$ of 3 eV (blue hollow circle) and 5 eV (orange hollow square). Data points and error bars represent the sample means and sample standard deviations, respectively.

Random forest models were trained to regress all four Au surface features $N_{\mathrm{Ce-Au_s}}$, $N_{\mathrm{O-Au_s}}$, $N_{\mathrm{Au_{Ce}-Au}}$, and $N_{\mathrm{Au_O-Au}}$ against each three target properties $E_{\mathrm{b}}$, $\Delta E$, and $Z_{\mathrm{eff,Ce}}$ with data of three Au surfaces mixed but with different $U_{\mathrm{eff}}$ equaling to 3 eV and 5 eV, respectively. In Figure 6(a) and (b), it can be observed that higher $N_{\mathrm{O-Au_s}}$ and $N_{\mathrm{Ce-Au_s}}$ contribute to more negative $E_{\mathrm{b}}$ values, while $N_{\mathrm{Au_{Ce}-Au}}$ and $N_{\mathrm{Au_{O}-Au}}$ have smaller impact on $E_{\mathrm{b}}$. As for $\Delta E$, when $U_{\mathrm{eff}}$ = 3 eV, higher $N_{\mathrm{O-Au_{s}}}$ following by $N_{\mathrm{Au_{O}-Au}}$, $N_{\mathrm{Au_{Ce}-Au}}$, and $N_{\mathrm{Ce-Au_{s}}}$ make $\Delta E$ more negative, as shown in Figure 6(c). However, when $U_{\mathrm{eff}}$ = 5 eV, only higher $N_{\mathrm{O-Au_{s}}}$ makes $\Delta E$ more negative and impacts of $N_{\mathrm{Au_{O}-Au}}$, $N_{\mathrm{Au_{Ce}-Au}}$, and $N_{\mathrm{Ce-Au_{s}}}$ on $\Delta E$ lack trends, according to Figure 6(d). Figure 6(e) and (f) show that higher $N_{\mathrm{Ce-Au_{s}}}$ cause negative $Z_{\mathrm{eff,Ce}}$ values, followed by $N_{\mathrm{O-Au_{s}}}$, while $N_{\mathrm{Au_{O}-Au}}$ and $N_{\mathrm{Au_{Ce}-Au}}$ have smaller and mixed impacts, respectively.

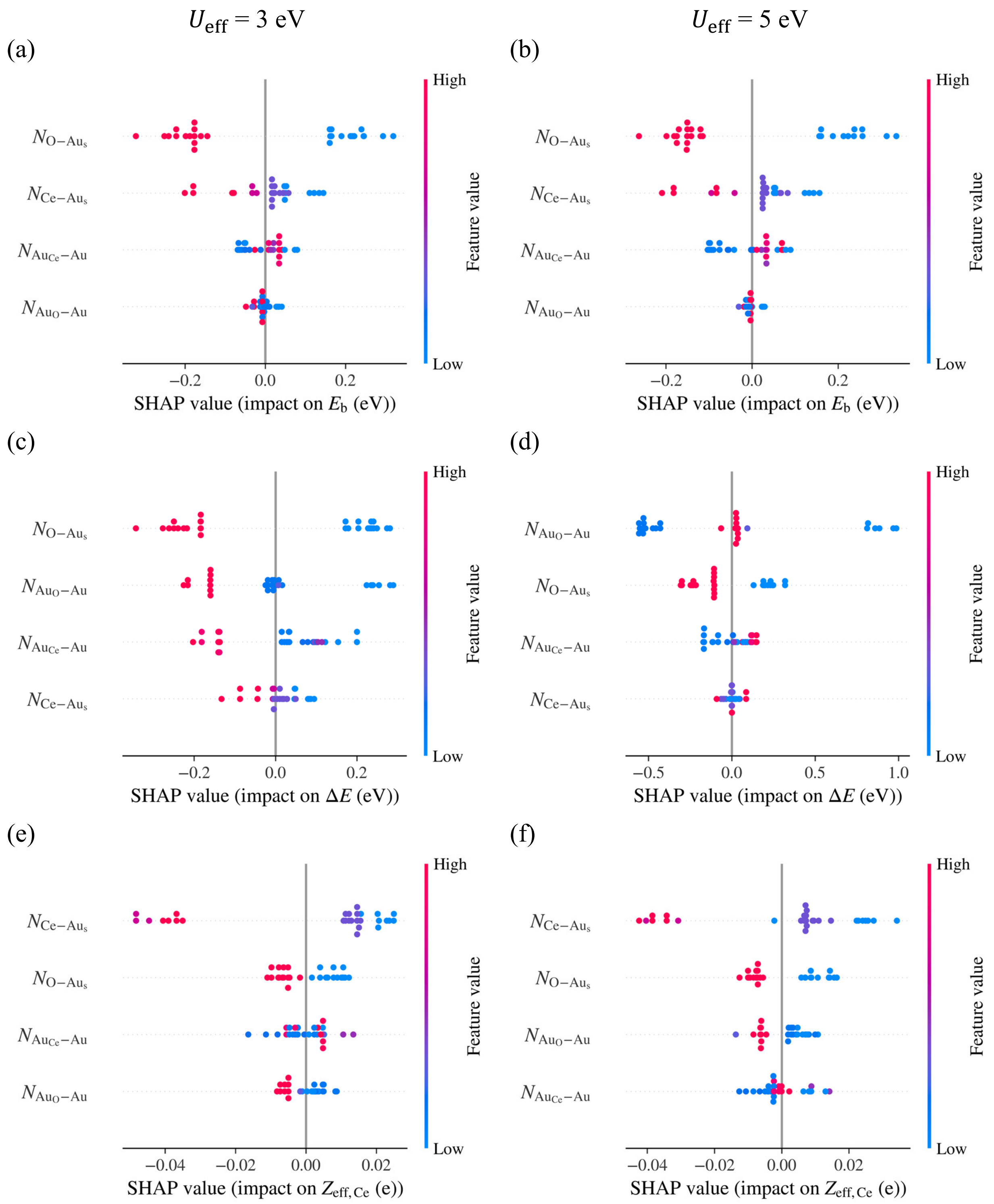

Figure 6. Beeswarm plots of SHAP values of the impact of four Au surface features $N_{Ce-Au_s}$, $N_{O-Au_s}$, $N_{Au_{Ce}-Au}$, and $N_{Au_O-Au}$ on target properties: (a, b) $E_b$, (c, d) $\Delta E$, and (e, f) $Z_{eff,Ce}$ for $CeO_2$ on Au(001), Au(110), and Au(111), where (a, c, e) correspond to a $U_{eff}$ of 3 eV and (b, d, f) correspond to a $U_{eff}$ of 5 eV.

The binding of the $CeO_2$ molecule on Au surfaces can be categorized into three cases: case 1 is the chemisorption with symmetric bonds, case 2 is the chemisorption with asymmetric bonds, and case 3 is the physisorption with symmetric bonds. The configurations before and after optimization, as well as $E_b$, $\Delta E$, bond length, bond angle, and case of binding of a $CeO_2$ molecule on Au(001), Au(110), and Au(111) are collected in Table S1 ($U_{eff}$ = 3 eV) and S2 ($U_{eff}$ = 5 eV), S5 ($U_{eff}$ = 3 eV) and S6 ($U_{eff}$ = 5 eV), S9 ($U_{eff}$ = 3 eV) and S10 ($U_{eff}$ = 5 eV). The correlation between bond lengths and $E_b$ of $CeO_2$ on Au(001) (Figure S5(a) for $U_{eff}$ = 3 eV and (b) for $U_{eff}$ = 5 eV), Au(110) (Figure S10(a) for $U_{eff}$ = 3 eV and (b) for $U_{eff}$ = 5 eV), and Au(111) (Figure S15(a) for $U_{eff}$ = 3 eV and (b) for $U_{eff}$ = 5 eV) helps to categorize all cases of binding. For case 1, the chemisorption with symmetric bonds, O atoms tend to be on the top site and bind with the Au atom to form two O-Au bonds with a length less than 2.864 Å, and the binding energy $E_b$ is more negative than -1 eV. Moreover, two Ce-O bonds of the $CeO_2$ molecule are symmetric and vary between 1.933 Å and 1.944 Å. The bond angle of the $CeO_2$ molecule tends to be bent between 106 and 176 degrees. For case 2, the chemisorption with asymmetric bonds, the preferred binding sites of the $CeO_2$ molecule and the binding energy $E_b$ are like case 1. However, the two Ce-O bonds of the $CeO_2$ molecule are asymmetric and deviate between ~1.85 Å and ~2.00 Å. The bond angle of the $CeO_2$ molecule tends to be bent between 119 and 145 degrees. In case 3, the physisorption with symmetric bonds, there is no covalent bonding between the O atoms and the Au atom, $E_b$ becomes more positive than -1 eV. The Ce-O bonds are symmetric and close to ~1.85 Å, which is also the bond length of the $CeO_2$ molecule in vacuum (Table 1). The geometry of the $CeO_2$ molecule tends to be linear, as its bond angles are larger than 170 degrees. $CeO_2$ molecule tends to be bent rather than linear when binding on Au(001), Au(110), and Au(111), as shown in the correlation between bond angle and $E_b$ of $CeO_2$ on Au(001) (Figure S5(c) for $U_{eff}$ = 3 eV and (d) for $U_{eff}$ = 5 eV), Au(110) (Figure S10(c) for $U_{eff}$ = 3 eV and (d) for $U_{eff}$ = 5 eV) and Au(111) (Figure S15(c) for $U_{eff}$ = 3 eV and (d) for $U_{eff}$ = 5 eV). Their corresponding binding sites for constituent atoms, the binding energy $E_b$, the Ce-O bond length, the bond angle O1-Ce-O2, and the bond symmetry are summarized in Table 4.

Table 4. Three cases of the binding of the $CeO_2$ molecule on Au(001), Au(110), and Au(111) surfaces, together with preferred sites of their constituent atoms, binding energy $E_b$, the Ce-O bond lengths, O1-Ce-O2 bond angles, and bond symmetry. In columns of preferred Ce, O1, and O2 sites, B, H, T, V, and N/A stand for bridge, hollow, top, vacuum, and not applicable sites on Au surfaces, as indicated in Figure 1. Preferred Ce, O1, and O2 sites on Au surfaces are listed in descending order from left to right.

| Case | Ce site | O1 site | O2 site | $E_b$ (eV) | Ce-O (Å) | O1-Ce-O2 (°) | Bond symmetry |
|---|---|---|---|---|---|---|---|
| 1 | (001) H<br>(110) H<br>(111) N/A | (001) T<br>(110) B<br>(111) N/A | (001) T<br>(110) B<br>(111) N/A | < -1 | ~1.95 | 106-176 | Symmetric |
| 2 | (001) H<br>(110) H, T<br>(111) B, T | (001) T<br>(110) T<br>(111) T | (001) V<br>(110) V<br>(111) V | < -1 | ~1.85-2.00 | 119-145 | Asymmetric |
| 3 | (001) H, B, T<br>(110) H, B, T<br>(111) H, B | (001) H, T<br>(110) H, T<br>(111) T, H | (001) H, T<br>(110) H, T<br>(111) T, H | > -1 | ~1.85 | > 170 | Symmetric |

The charge density difference of the $CeO_2$ molecule on the Au(001), Au(110), and Au(111) surfaces were calculated using Equation (12) and the visualized results are shown in Figure S6 ($U_{eff}$ = 3 eV) and S7 ($U_{eff}$ = 5 eV) for Au(001), Figure S11 ($U_{eff}$ = 3 eV) and S12 ($U_{eff}$ = 5 eV) for Au(110), and Figure S16 ($U_{eff}$ = 3 eV) and S17 ($U_{eff}$ = 5 eV) for Au(111). Note that the increase of $U_{\mathrm{eff}}$ makes the region of charge density difference $\Delta\rho = \pm 0.005$ e Bohr$^{-3}$ more localized in space, where positive and negative $\Delta\rho$ imply charge accumulation and depletion. In cases 1 and 2 bindings, there is charge accumulation near the Au atom and depletion near the Ce atom between the Ce and Au atoms as well as depletion between the O and Au atoms. In case 3 binding, there is charge accumulation near the Au atom and depletion near the Ce atom between the Ce and Au atoms, but slight depletion to none between the O and Au atoms, and accumulation occurs on the O atom.

To see the orbital changes of the $CeO_2$ molecule after different configurations chemisorbing and physisorbing on the Au(001), (110), and (111) surfaces, the electronic projected density of states for Ce *d*, Ce *f*, O1 *p*, O2 *p*, and Au *d* orbitals of different $CeO_2$ configurations on Au surfaces are plotted in Figure S8 ($U_{\mathrm{eff}}$ = 3 eV) and S9 ($U_{\mathrm{eff}}$ = 5 eV) for Au(001), Figure S13 ($U_{\mathrm{eff}}$ = 3 eV) and S14 ($U_{\mathrm{eff}}$ = 5 eV) for Au(110), and Figure S18 ($U_{\mathrm{eff}}$ = 3 eV) and S19 ($U_{\mathrm{eff}}$ = 5 eV) for Au(111). The electronic PDOS of Au *d* spans from -7 eV to 0 eV. Chemisorption, with the binding energy $E_{\mathrm{b}}$ stronger than -1 eV, has delocalized Ce *d*, Ce *f*, O1 *p*, and O2 *p* spanning from -7 eV to 0 eV as well. Physisorption, with $E_{\mathrm{b}}$ weaker than -1 eV, has localized Ce *d*, Ce *f*, O1 *p*, and O2 *p* between -5 eV and 0 eV. When the $CeO_2$ molecule is chemisorbed on Au surfaces with only one of the O atoms, only one of the O *p* orbitals is delocalized between -7 eV and 0 eV, and another O *p* orbital is localized between -5 eV and 0 eV, which also correlates with the asymmetric Ce-O bond length. On the other hand, when the $CeO_2$ molecule is chemisorbed or physisorbed on Au surfaces with two O atoms, both O1 and O2 *p* orbitals are delocalized within the same energy range, which also implies symmetric Ce-O bond lengths.

### 3.2.1. $CeO_2$ molecules on Au(001)

The Ce atom and the O atoms of the $CeO_2$ molecule have different preferences for binding sites when the $CeO_2$ molecule adsorbs on the Au(001) surface, as shown in Figure 7, with a $U_{\text{eff}}$ of 3 eV, and Figure 8, with a $U_{\text{eff}}$ of 5 eV, ranking from strong to weak $E_{\text{b}}$. $CeO_2$ configurations on Au(001) showed no significant changes after further optimization with $U_{\text{eff}}$ = 5 eV following initial optimization at $U_{\text{eff}}$ = 3 eV. However, the order of $CeO_2$ configurations from strong to weak $E_{\text{b}}$ was updated after this subsequent optimization. The Ce atom of the $CeO_2$ molecule tends to be on the hollow site first, then on the bridge site, then on the top site of the Au(001) surface. This observation may have to do with the crystal structure of the Ce atom, since the Ce atom is more stable in the close-packed crystal structures.[66] The configurations after and before optimization, as well as $E_{\text{b}}$, $\Delta E$, bond length, bond angle, and cases of binding of a $CeO_2$ molecule on Au(001) are tabulated in Table S1 ($U_{\text{eff}}$ = 3 eV) and S2 ($U_{\text{eff}}$ = 5 eV).

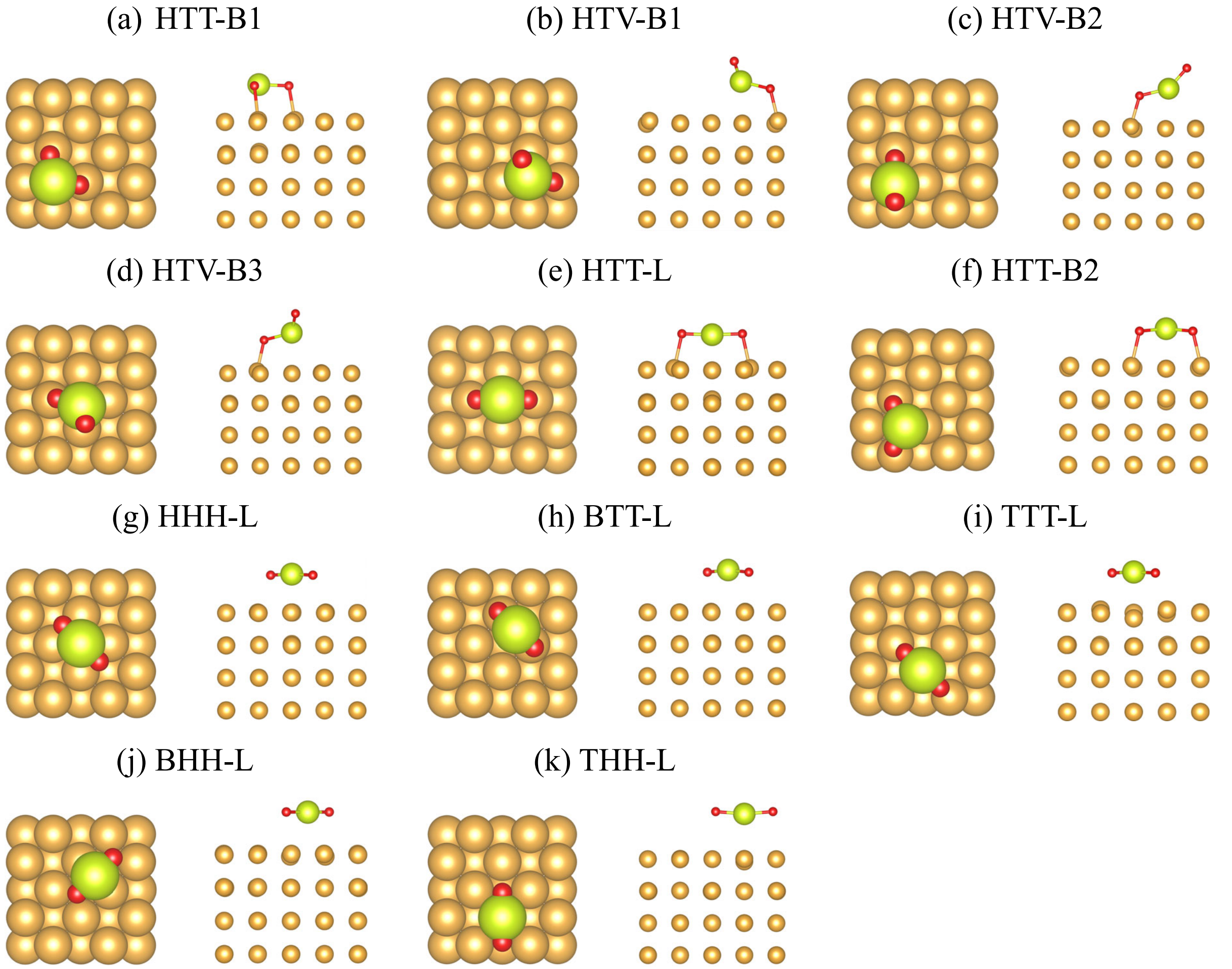

Figure 7. Different configurations of the $CeO_2$ molecule on the Au(001) slab model with a $U_{\mathrm{eff}}$ of 3 eV. The green, red, and orange balls represent Ce, O, and Au atoms, respectively. The maximum plotted length of the Ce-O bond is 2.864 Å.

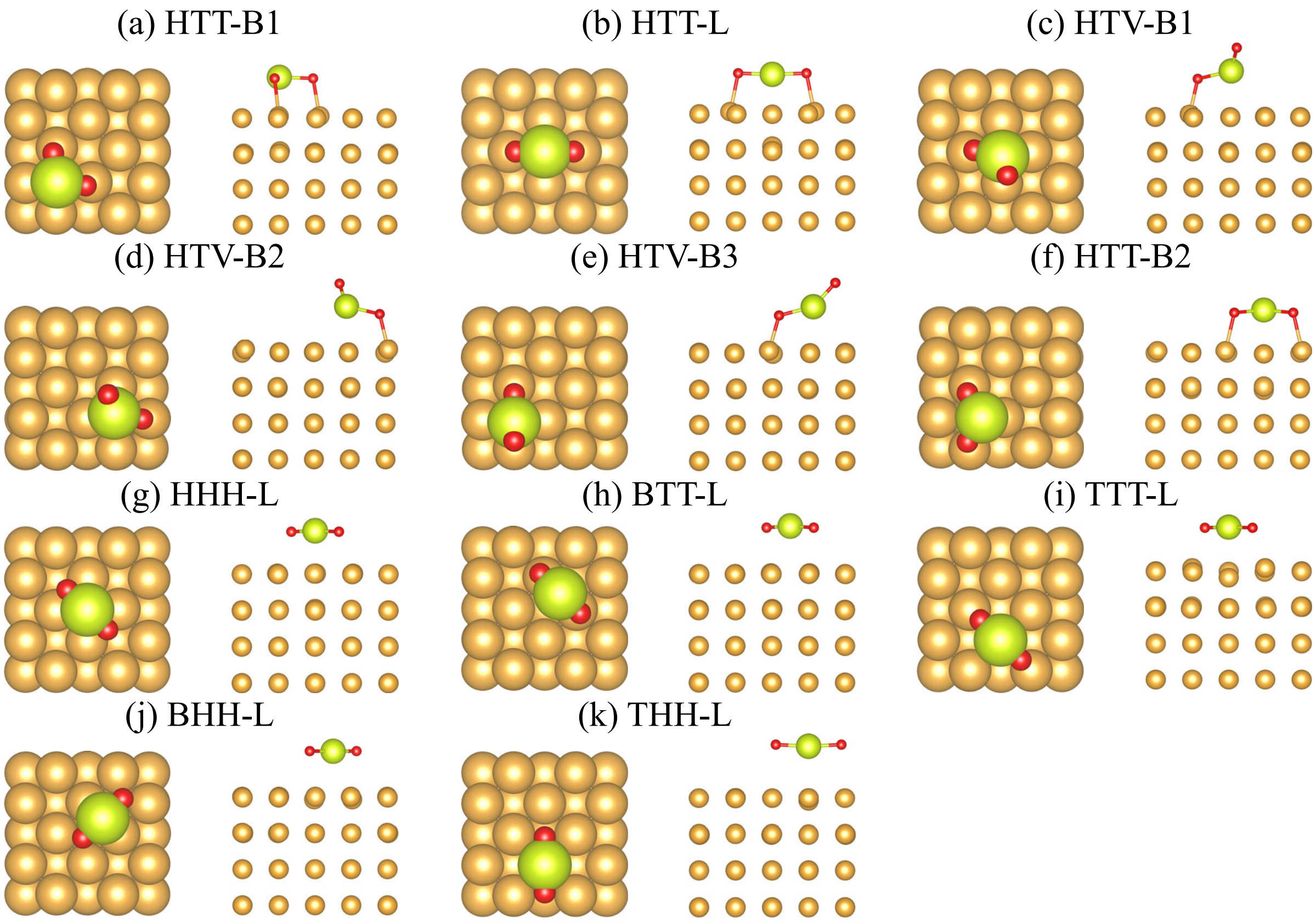

Figure 8. Different configurations of the $CeO_2$ molecule on the Au(001) slab model with a $U_{\text{eff}}$ of 5 eV. The green, red, and orange balls represent Ce, O, and Au atoms. The maximum plotted length of the Ce-O bond is 2.864 Å. Notice that configurations in Figure 8 were optimized from configurations in Figure 7, the order of binding energy strength changes due to different $U_{\text{eff}}$ used.

The binding free energy $G_{\text{b}}$ of $CeO_2$ molecules on the Au(001) surface was calculated using Equation (8) and shown in Figure 9(a) and (b), with a $U_{\text{eff}}$ of 3 eV and 5 eV, respectively. When $T$ increases, $G_{\text{b}}$ becomes weaker; all configurations have positive $G_{\text{b}}$ after $T$ = 800 K, even though increasing $U_{\text{eff}}$ makes $G_{\text{b}}$ stronger. It should be noted here that positive values of $G_{\text{b}}$ refer to states in which the $CeO_2$ molecules are not bonded to the surface. The probability $P_i$ and binding energy $E_{\text{b}}$ of each configuration of the $CeO_2$ molecule on Au(001) were calculated using Equation (9) and (1) and plotted in Figure 9 (c) and (d), with a $U_{\text{eff}}$ of 3 eV and 5 eV, respectively. Overall, configurations with values of $E_{\text{b}}$ stronger than -1 eV have probabilities that are not negligible. This implies that values of $E_b$ stronger than -1 eV can be a criterion to find significant configurations when the temperature is not zero.

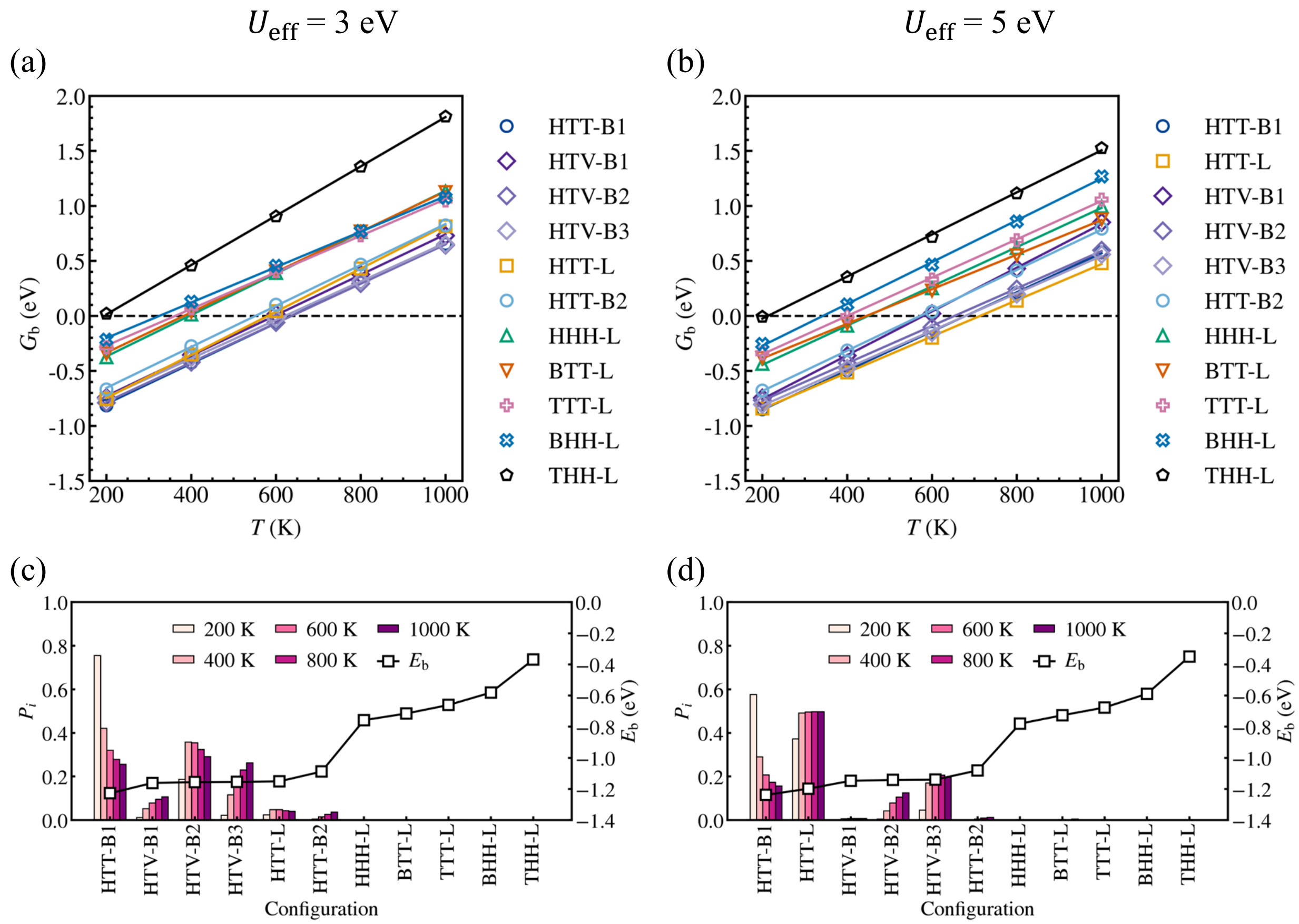

Figure 9. (a, b) Binding free energy $G_b$ and (c, d) probability of molecular configurations $P_i$ versus binding energy $E_b$ of $CeO_2$ molecules on the Au(001) surface at temperatures of 200, 400, 600, 800, and 1000 K. (a, c) correspond to a $U_{eff}$ of 3 eV and (b, d) correspond to a $U_{eff}$ of 5 eV.

Random forest models are trained to regress all four Au surface features $N_{Ce-Au_s}$, $N_{O-Au_s}$, $N_{Au_{Ce}-Au}$, and $N_{Au_O-Au}$ against the three target properties $E_b$, $\Delta E$, and $Z_{eff,Ce}$, with data of only Au(001) but with different $U_{eff}$ equaling to 3 eV and 5 eV. The effective charge $Z_{eff,Ce}$, $Z_{eff,O1}$, and $Z_{eff,O2}$ in the $CeO_2$ molecule on the Au(001) surface were calculated using Equation (11) and shown in Table S3 ($U_{eff}$ = 3 eV) and S4 ($U_{eff}$ = 5 eV). It can be seen in Figure 10(a) and (b) that higher $N_{\mathrm{O-Au_s}}$ and lower $N_{\mathrm{Au_{Ce}-Au}}$ contribute toward a more negative value of $E_{\mathrm{b}}$, while $N_{\mathrm{Ce-Au_s}}$ has relatively little impact. As for $\Delta E$ in Figure 10(c) and (d), higher $N_{\mathrm{O-Au_s}}$ makes $\Delta E$ most negative, which can be explained by more binding between O and $Au_s$ atoms, while lower $N_{\mathrm{Au_{Ce}-Au}}$ and higher $N_{\mathrm{Ce-Au_s}}$ also make $\Delta E$ slightly negative and stable, indicating that O following by Ce forming bonds with $Au_s$ stabilize molecular $CeO_2$ on Au(001), lower coordination of $Au_{Ce}$ also contributes to higher stability. Figure 10(c) and (d) share the high similarity pattern compared to Figure 10(a) and (b), indicating that these four features affect $\Delta E$ in the same way as to $E_{\mathrm{b}}$. Interestingly, from Figure 10(e) and (f), it can be seen that $U_{\mathrm{eff}}$ values of 3 eV and 5 eV make different features significant to $Z_{\mathrm{eff,Ce}}$. With a $U_{\mathrm{eff}}$ of 3 eV, higher $N_{\mathrm{O-Au_s}}$ and $N_{\mathrm{Ce-Au_s}}$ make $Z_{\mathrm{eff,Ce}}$ negative, meaning that Ce gains electrons from $Au_s$ and becomes less oxidized and more reduced. However, higher $N_{\mathrm{Au_{Ce}-Au}}$ makes $Z_{\mathrm{eff,Ce}}$ positive, implying that Ce is more oxidized and less reduced, which may be due to more Au around $Au_{Ce}$ share electrons from Ce. As for a $U_{\mathrm{eff}}$ of 5 eV, Ce-related features such as $N_{\mathrm{Au_{Ce}-Au}}$ and $N_{\mathrm{Ce-Au_s}}$ become more dominant. $N_{\mathrm{Au_O-Au}}$ has no effect on $E_{\mathrm{b}}$, $\Delta E$, and $Z_{\mathrm{eff,Ce}}$ under a $U_{\mathrm{eff}}$ of 3 eV and 5 eV.

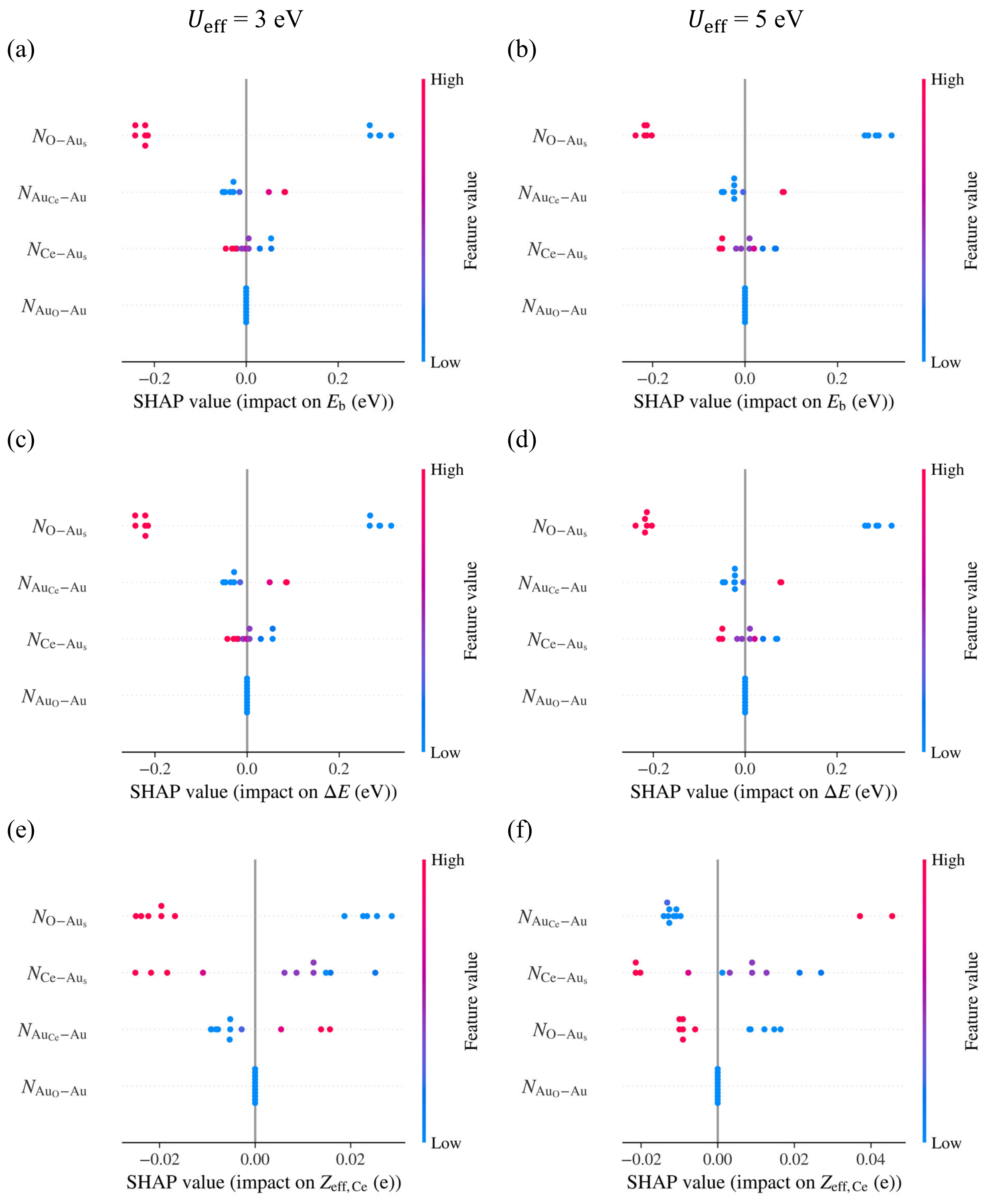

Figure 10. Beeswarm plots of SHAP values of the impact of four Au surface features $N_{\mathrm{Ce-Au_s}}$, $N_{\mathrm{O-Au_s}}$, $N_{\mathrm{Au_{Ce}-Au}}$, and $N_{\mathrm{Au_O-Au}}$ on target properties: (a, b) $E_b$, (c, d) $\Delta E$, and (e, f) $Z_{\mathrm{eff,Ce}}$ for $CeO_2$ on Au(001) surface, where (a, c, e) correspond to $U_{\mathrm{eff}}$ of 3 eV and (b, d, f) correspond to $U_{\mathrm{eff}}$ of 5 eV.

### 3.2.2. $CeO_2$ molecules on Au(110)

Ce and O atoms of $CeO_2$ molecules exhibit similar site preferences on Au(110) surfaces as observed on Au(001) surfaces, as shown in Figure 11, with a $U_{\mathrm{eff}}$ of 3 eV, and Figure 12, with a $U_{\mathrm{eff}}$ of 5 eV, which rank $E_{\mathrm{b}}$ from strong to weak. $CeO_2$ configurations, as well as $E_{\mathrm{b}}$ on Au(110) showed no noticeable changes after further optimization with $U_{\mathrm{eff}}$ = 5 eV following initial optimization at $U_{\mathrm{eff}}$ = 3 eV. Therefore, the order of $CeO_2$ configurations from strong to weak $E_{\mathrm{b}}$ was not changed after this subsequent optimization. However, the number of configurations of the $CeO_2$ molecule on Au(110) is fewer than on Au(001), which may be due to the lower planar packing fraction and reduced symmetry of Au(110) compared to Au(001) (Table 3). The configurations before and after optimization, $E_{\mathrm{b}}$, $\Delta E$, bond length, bond angle, and binding cases of $CeO_2$ molecules on Au(110) surfaces are organized in Table S5 ($U_{\mathrm{eff}}$ = 3 eV) and S6 ($U_{\mathrm{eff}}$ = 5 eV), respectively.

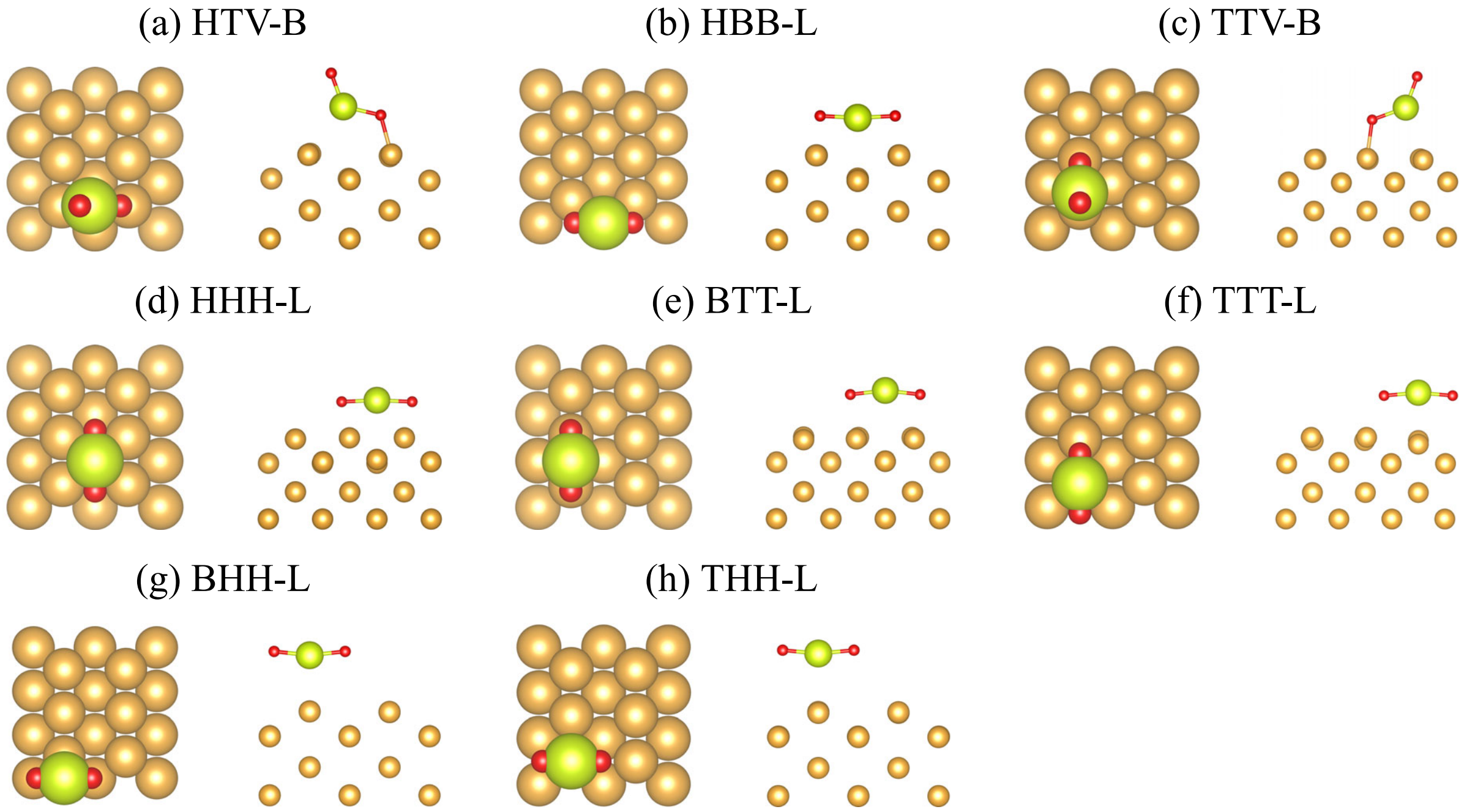

Figure 11. Different configurations of the $CeO_2$ molecule on the Au(110) slab model with a $U_{\text{eff}}$ of 3 eV. The green, red, and orange balls represent Ce, O, and Au atoms, respectively. The maximum plotted length of the Ce-O bond is 2.864 Å.

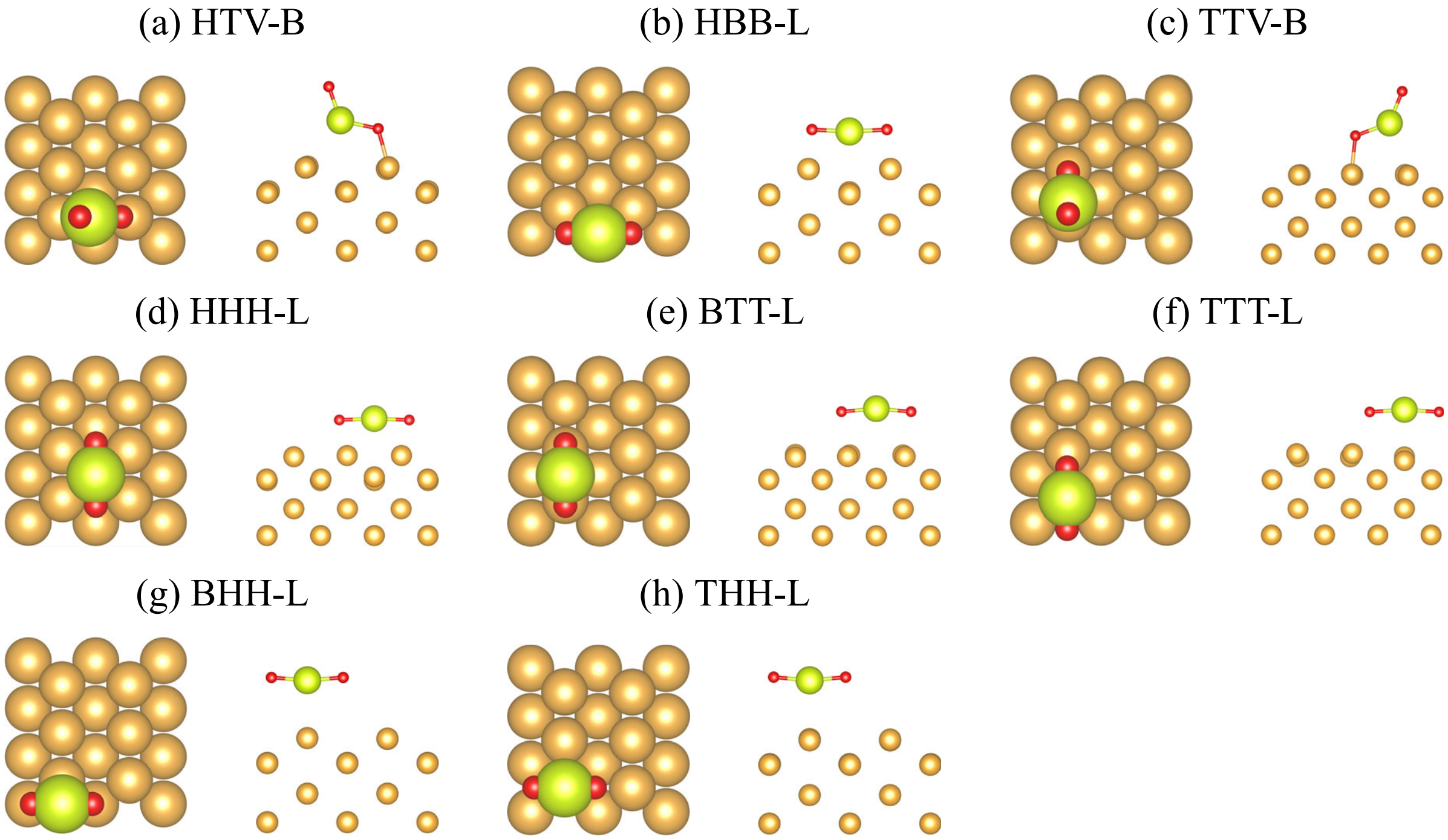

Figure 12. Different configurations of the $CeO_2$ molecule on the Au(110) slab model with a $U_{\text{eff}}$ of 5 eV. The green, red, and orange balls represent Ce, O, and Au atoms, respectively. The maximum plotted length of the Ce-O bond is 2.864 Å.

Equation (8) was used to calculate the binding free energy $G_{\text{b}}$ of $CeO_2$ molecules on Au(110) surfaces and results are shown in Figure 13(a) and (b) with a $U_{\text{eff}}$ of 3 eV and 5 eV, respectively. All the configurations have weaker $G_{\text{b}}$ when $T$ increases or $U_{\text{eff}}$ decreases, which are the same as those on Au(001). $E_{\text{b}}$ and $P_i$ of each $CeO_2$ molecule configuration on Au(110) is calculated using Equation (1) and (9) and displayed in Figure 13(c) ($U_{\text{eff}}$ = 3 eV) and (d) ($U_{\text{eff}}$ = 5 eV). Only two configurations, HTV-B and HBB-L, with $E_{\text{b}}$ stronger than -1 eV have probabilities that are not negligible. This again implies that $E_{\text{b}}$ values stronger than -1 eV can be a criterion to find significant configurations when $T$ is not zero.

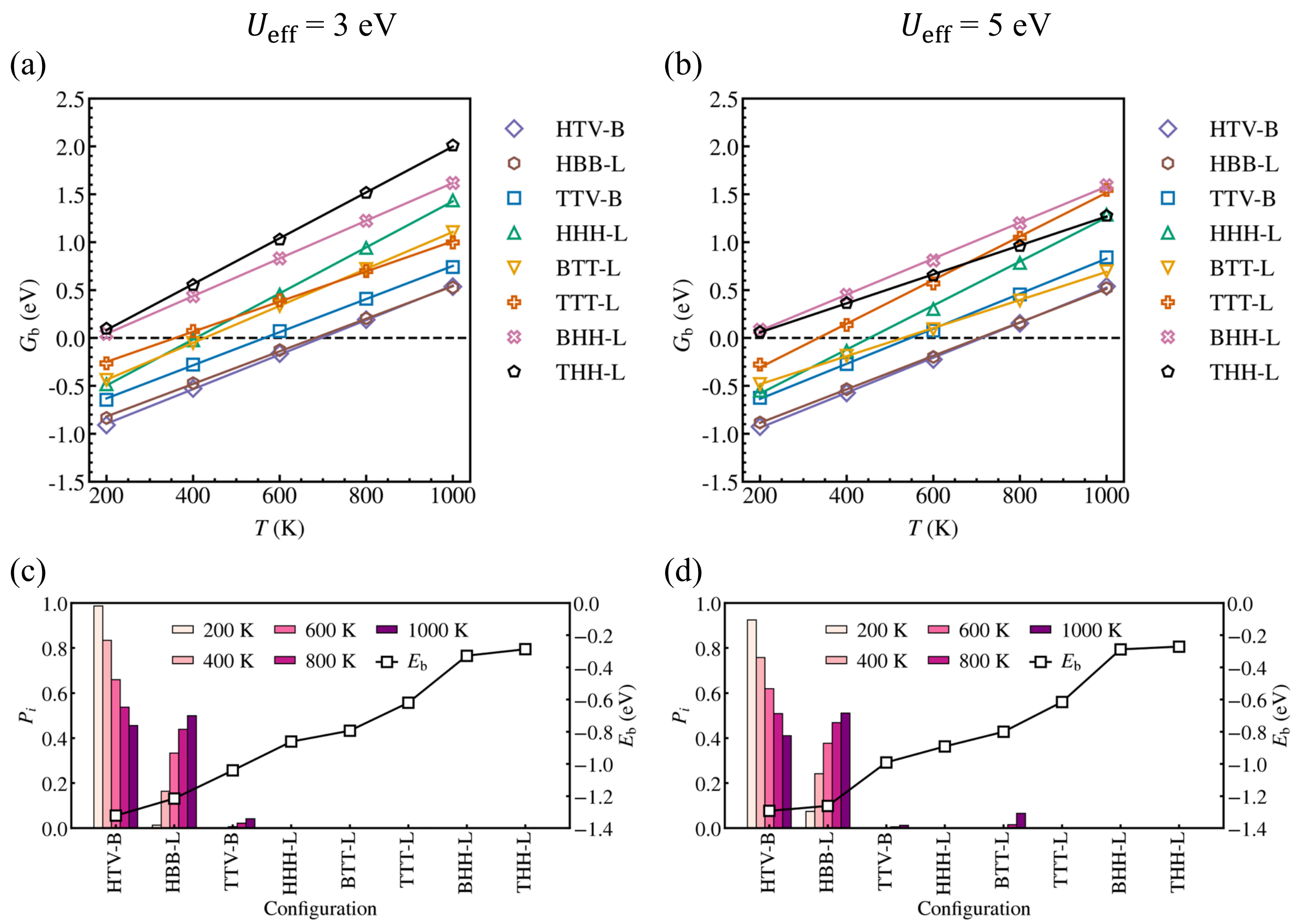

Figure 13. (a, b) Binding free energy $G_\mathrm{b}$ and (c, d) probability of molecular configurations $P_i$ versus binding energy $E_\mathrm{b}$ of the $CeO_2$ molecule on the Au(110) surface at temperatures of 200, 400, 600, 800, and 1000 K. (a) and (c) correspond to $U_\mathrm{eff}$ of 3 eV and (b) and (d) correspond to $U_\mathrm{eff}$ of 5 eV.

To investigate how four Au(110) surface features $N_{\mathrm{Ce-Au_s}}$, $N_{\mathrm{O-Au_s}}$, $N_{\mathrm{Au_{Ce}-Au}}$, and $N_{\mathrm{Au_O-Au}}$ affect three target properties $E_\mathrm{b}$, $\Delta E$, and $Z_{\mathrm{eff,Ce}}$, data of $CeO_2$ on Au(110) with $U_\mathrm{eff}$ values of 3 eV and 5 eV were used to train six random forest models interpreted by SHAP values in Figure 14. The effective charges $Z_{\mathrm{eff,Ce}}$, $Z_{\mathrm{eff,O1}}$, and $Z_{\mathrm{eff,O2}}$ were calculated using Equation (11) and shown in Table S7 ($U_\mathrm{eff}$ = 3 eV) and S8 ($U_\mathrm{eff}$ = 5 eV), respectively. Observing Figure 14(a) and (b), higher $N_{\mathrm{O-Au_s}}$ and $N_{\mathrm{Ce-Au_s}}$ contribute to more negative $E_\mathrm{b}$ values, while $N_{\mathrm{Au_{Ce}-Au}}$ and $N_{\mathrm{Au_O-Au}}$ have relative lower impact, implying that the number of $Au_s$ on Au(110) within cutoff from $CeO_2$ plays a more important role than the number of Au nearest neighbors from $Au_{Ce}$ and $Au_O$. Similar to $CeO_2$ on Au(001), Figure 14(c) and (d) share a high similarity pattern compared to Figure 14(a) and (b), indicating that impacts of these four Au surface features on $E_b$ and $\Delta E$ are the same for both $E_b$ and $\Delta E$ are energy differences except with different energy references. Nevertheless, from Figure 14(e) and (f), it can be seen that $U_{eff}$ values of 3 eV and 5 eV make different features significant to $Z_{eff,Ce}$. With $U_{eff}$ equaling to 3 eV, higher $N_{Ce-Au_s}$ and $N_{Au_{Ce}-Au}$ make $Z_{eff,Ce}$ negative, meaning that Ce gains electrons from $Au_s$ and becomes less oxidized and more reduced, while $N_{Au_O-Au}$ and $N_{O-Au_s}$ have no clear effect on $Z_{eff,Ce}$. As for $U_{eff}$ = 5 eV, higher $N_{Ce-Au_s}$ followed by $N_{Au_O-Au}$, $N_{Au_{Ce}-Au}$, and $N_{O-Au_s}$ contribute to more negative $Z_{eff,Ce}$.

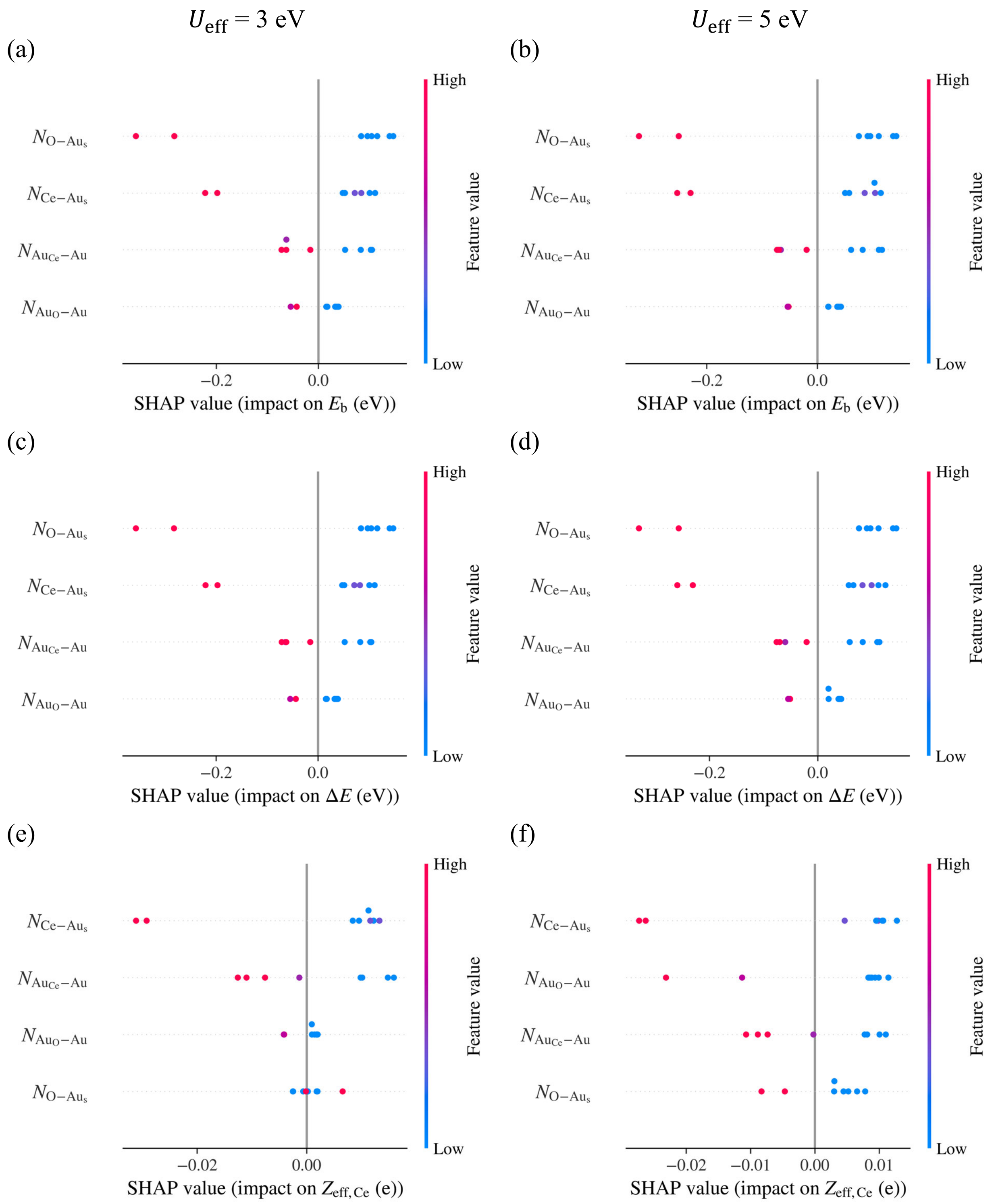

Figure 14. Beeswarm plots of SHAP values of the impact of four Au surface features $N_{\mathrm{Ce-Au_s}}$, $N_{\mathrm{O-Au_s}}$, $N_{\mathrm{Au_{Ce}-Au}}$, and $N_{\mathrm{Au_O-Au}}$ on target properties: (a, b) $E_{\mathrm{b}}$, (c, d) $\Delta E$, and (e, f) $Z_{\mathrm{eff,Ce}}$ for $CeO_2$ on Au(110), where (a, c, e) correspond to a $U_{\mathrm{eff}}$ of 3 eV and (b, d, f) correspond to a $U_{\mathrm{eff}}$ of 5 eV.

### 3.2.3. $CeO_2$ molecules on Au(111)

Nine $CeO_2$ configurations on Au(111) in Figure 15 ($U_{\mathrm{eff}}$ = 3 eV) and Figure 16 ($U_{\mathrm{eff}}$ = 5 eV) are ranked from stronger to weaker $E_{\mathrm{b}}$, and constituent atoms of $CeO_2$ have the same preference of binding sites as on Au(001) and Au(111) surfaces. Configurations of $CeO_2$ molecules on Au(111) surfaces showed no significant changes after further optimization with a $U_{\mathrm{eff}}$ = 5 eV following initial optimization at a $U_{\mathrm{eff}}$ = 3 eV. However, the order of configurations from strong to weak $E_{\mathrm{b}}$ was updated after this subsequent optimization. The planar packing fraction of 0.907 and the six-fold rotational symmetry of Au(111) (Table 3) are believed to result in five ($U_{\mathrm{eff}}$ = 3 eV) to six ($U_{\mathrm{eff}}$ = 5 eV) similar BTV-B configurations out of the nine. Table S9 ($U_{\mathrm{eff}}$ = 3 eV) and S10 ($U_{\mathrm{eff}}$ = 5 eV) include configurations before and after optimization, $E_{\mathrm{b}}$, $\Delta E$, bond length, bond angle, and binding cases of $CeO_2$ molecules on Au(111) surfaces.

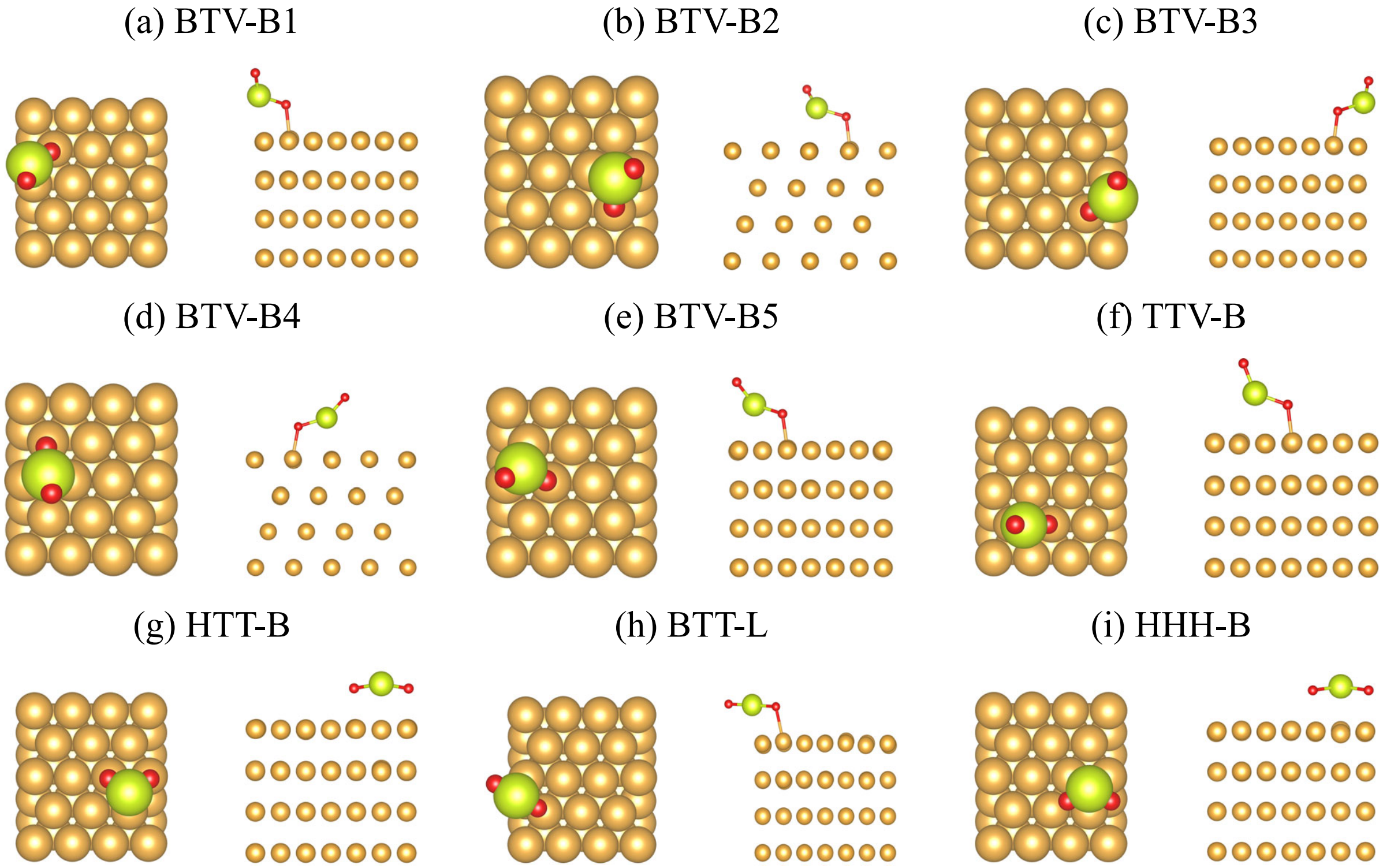

Figure 15. Different configurations of the $CeO_2$ molecule on the Au(111) surface with a $U_{\mathrm{eff}}$ of 3 eV. The green, red, and orange balls represent Ce, O, and Au atoms, respectively. The maximum plotted length of the Ce-O bond is 2.864 Å.

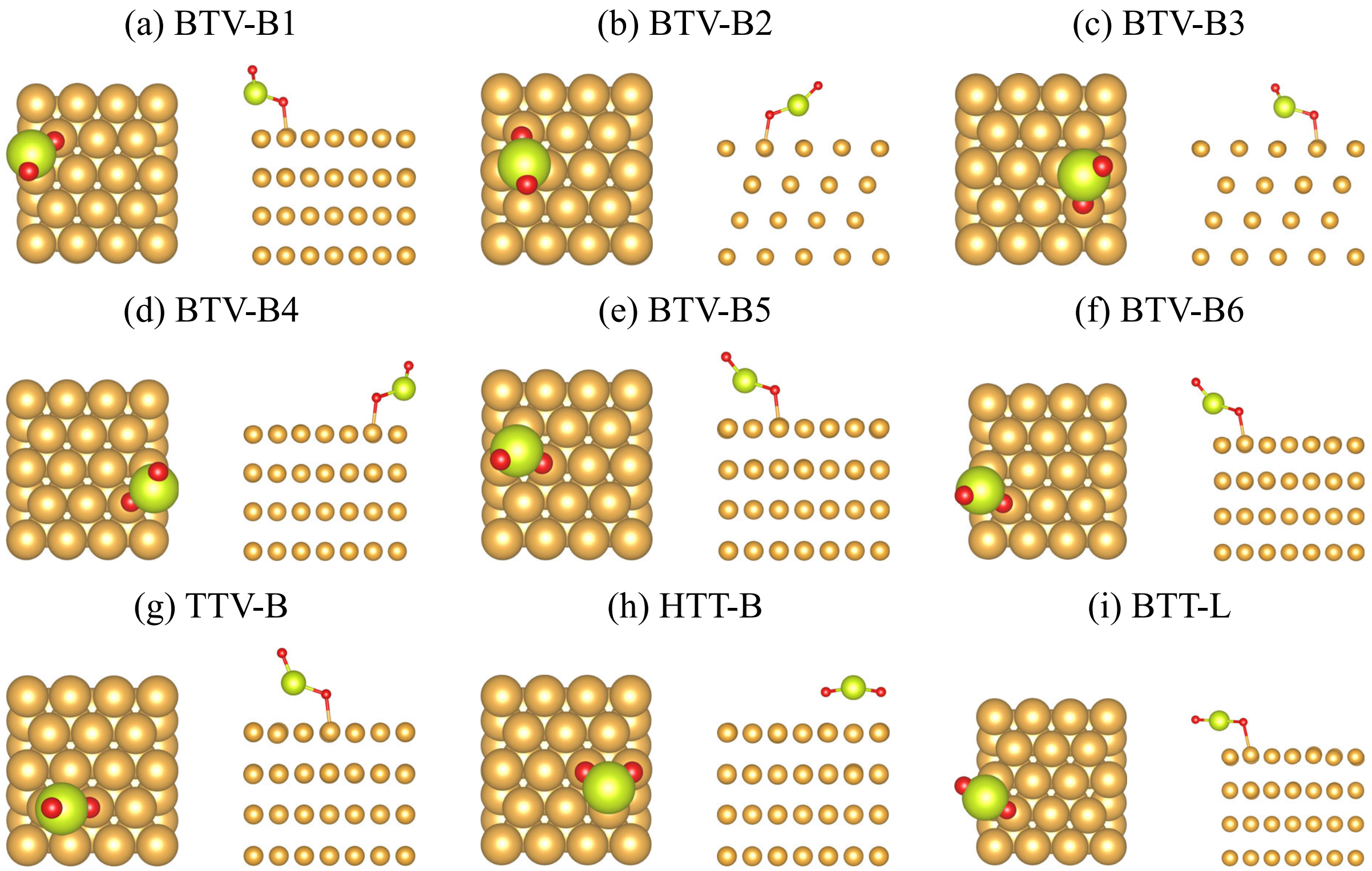

Figure 16. Different configurations of the $CeO_2$ molecule on the Au(111) surface with a $U_{\mathrm{eff}}$ of 5 eV. The green, red, and orange balls represent Ce, O, and Au atoms, respectively. The maximum plotted length of the Ce-O bond is 2.864 Å. Notice that configurations in Figure 16 were optimized from configurations in Figure 15, the order of binding energy strength changes due to different $U_{\mathrm{eff}}$ used.

Figure 17(a) ($U_{\mathrm{eff}}$ = 3 eV) and (b) ($U_{\mathrm{eff}}$ = 5 eV) respectively show binding free energies $G_{\mathrm{b}}$ of $CeO_2$ molecules on Au(111) surfaces calculated using Equation (8). Identical to configurations on Au(001) and Au(110), $G_{\mathrm{b}}$ of configurations on Au(111) becomes weaker when $T$ increases or $U_{\mathrm{eff}}$ decreases. $E_{\mathrm{b}}$ and $P_i$ of $CeO_2$ molecules on Au(111) surfaces are shown in Figure 17(c) ($U_{\mathrm{eff}}$ = 3 eV) and (d) ($U_{\mathrm{eff}}$ = 5 eV), respectively. Since most configurations have $E_{\mathrm{b}}$ more negative than -1 eV and their $G_{\mathrm{b}}$ are close, their $P_i$ are also comparable.

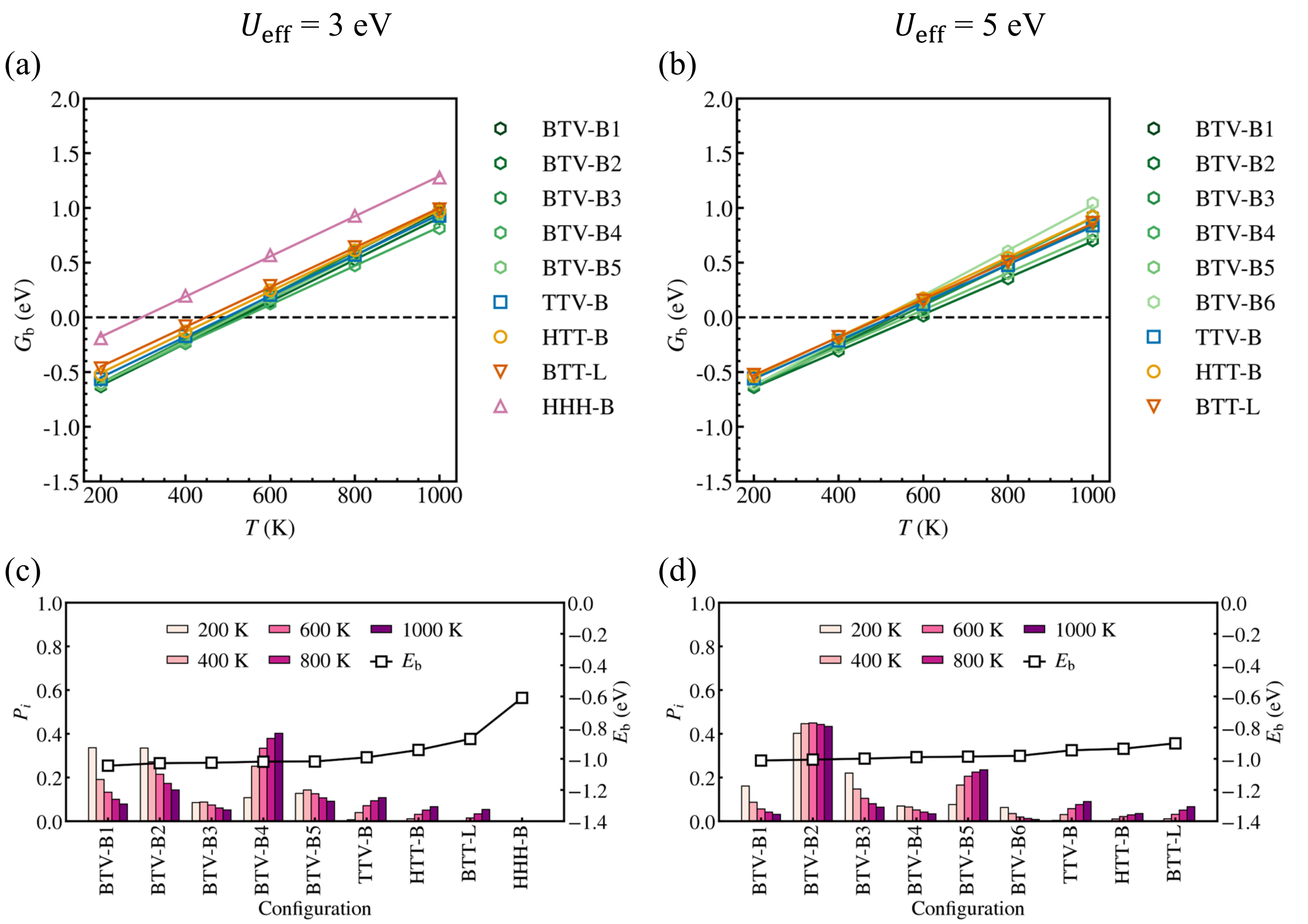

Figure 17. (a, b) Binding free energy $G_{\mathrm{b}}$ and (c, d) probability of molecular configurations $P_i$ versus binding energy $E_{\mathrm{b}}$ of the $CeO_2$ molecule on the Au(111) surface at temperatures of 200, 400, 600, 800, and 1000 K. (a) and (c) correspond to a $U_{\mathrm{eff}}$ of 3 eV and (b) and (d) correspond to a $U_{\mathrm{eff}}$ of 5 eV.

Impacts of four Au(111) surface features $N_{\mathrm{Ce-Au_s}}$, $N_{\mathrm{O-Au_s}}$, $N_{\mathrm{Au_{Ce}-Au}}$, and $N_{\mathrm{Au_O-Au}}$ on each three target properties $E_{\mathrm{b}}$, $\Delta E$, and $Z_{\mathrm{eff,Ce}}$ were investigated by training six random forest models using data of $CeO_2$ on Au(111) only but with two kinds of $U_{\mathrm{eff}}$ values (3 eV and 5 eV) respectively, as shown in Figure 18. Effective charges $Z_{\mathrm{eff,Ce}}$, $Z_{\mathrm{eff,O1}}$, and $Z_{\mathrm{eff,O2}}$ calculated using Equation (11) were organized in Table S11 ($U_{\mathrm{eff}}$ = 3 eV) and S12 ($U_{\mathrm{eff}}$ = 5 eV). With a $U_{\mathrm{eff}}$ of 3 eV (Figure 18(a)), only higher $N_{\mathrm{O-Au_s}}$ contributes to more negative $E_{\mathrm{b}}$ values, $N_{\mathrm{Ce-Au_s}}$ has no proportional impact clearly, while $N_{\mathrm{Au_{Ce}-Au}}$ and $N_{\mathrm{Au_O-Au}}$ have mixed and no impact on $E_{\mathrm{b}}$, which tells that only the coordination number between O of $CeO_2$ and surface Au atoms $Au_s$ impacts $E_{\mathrm{b}}$ proportionally. However, different from Au(001) and Au(110), impacts of four Au surface features on $E_{\mathrm{b}}$ change significantly between a $U_{\mathrm{eff}}$ of 3 eV and 5 eV, as shown in Figure 18(a) and (b), respectively. Similar to $CeO_2$ on Au(001) and Au(110), Figure 18(c) and (d) share the high similarity pattern compared to Figure 18(a) and (b), because $E_{\mathrm{b}}$ and $\Delta E$ are both energy differences but evaluated with different energy references. With a $U_{\mathrm{eff}}$ equaling to 3 eV (Figure 18(e)), higher $N_{\mathrm{Ce-Au_s}}$, lower $N_{\mathrm{Au_{Ce}-Au}}$ and lower $N_{\mathrm{O-Au_s}}$ make $Z_{\mathrm{eff,Ce}}$ negative. As for a $U_{\mathrm{eff}}$ equaling to 5 eV (Figure 18(f)), only lower $N_{\mathrm{O-Au_s}}$ contributes to more negative $Z_{\mathrm{eff,Ce}}$, the rest of Au(111) surface features do not clearly have proportional impacts on $Z_{\mathrm{eff,Ce}}$.

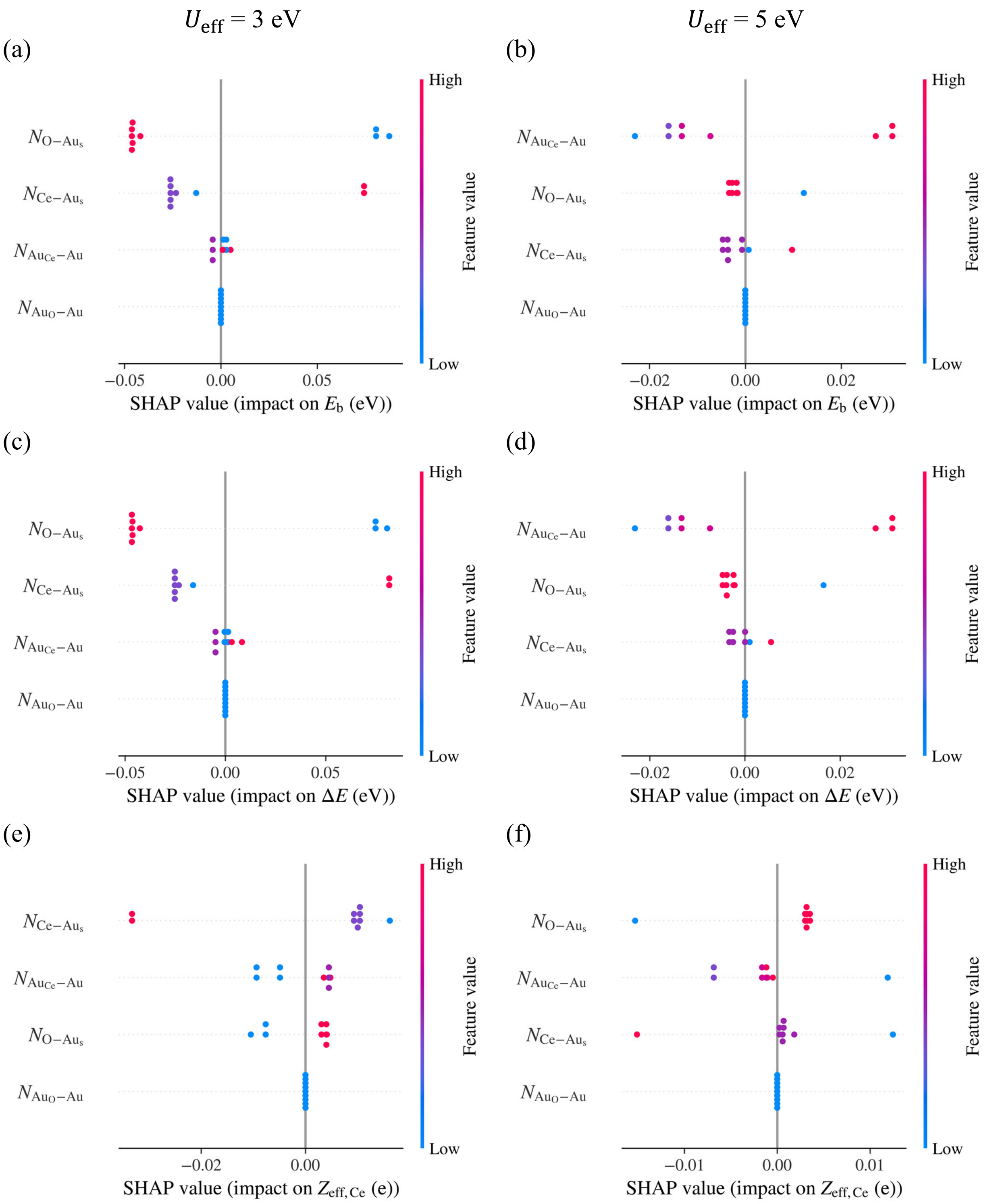

Figure 18. Beeswarm plots of SHAP values of the impact of four Au surface features $N_{Ce-Au_s}$, $N_{O-Au_s}$, $N_{Au_{Ce}-Au}$, and $N_{Au_O-Au}$ on target properties: (a, b) $E_b$, (c, d) $\Delta E$, and (e, f) $Z_{eff,Ce}$ for $CeO_2$ on Au(111), where (a, c, e) correspond to a $U_{eff}$ of 3 eV and (b, d, f) correspond to a $U_{eff}$ of 5 eV.

## 3.3. A short discussion

Several remarks regarding the results presented above are in order. First, to investigate configurations of $CeO_2$ molecules on Au surfaces, we started with the study of $CeO_2$ molecules in vacuum. Among academic communities, it is common to use Hubbard $U$ corrections on Ce $f$ orbital when calculating $CeO_2$ bulk using DFT calculations. We demonstrated that the commonly used PBE exchange-correlation functional can only predict a bent $CeO_2$ molecule to be more stable than its linear counterpart when Hubbard $U$ and thermal corrections are both considered, which reveals that the bent $CeO_2$ molecule becomes less stable than the linear one when the temperature is low.

Second, we defined binding sites and constructed configurations of $CeO_2$ molecules on Au(001), Au(110), and Au(111) surfaces. Based on optimized configurations of $CeO_2$ molecules on Au surfaces, it was found that $CeO_2$ molecules tend to maintain the bent state on Au surfaces as they are in vacuum, which matches the assumption in our previous works.[22,23] Moreover, Ce prefers binding to hollow sites over bridge sites and top sites of Au surfaces, while O prefers to bind on top sites over vacuum sites over hollow sites. We also found that $CeO_2$ molecules have less distinct configurations and no case 1 binding on the Au(111) surface, which may be due to the close-packed nature of Au(111) surfaces. A binding energy $E_{\mathrm{b}}$ of approximately -1 eV was found to be a dividing line between chemisorption (case 1 and case 2) and physisorption (case 3). Configurations with binding energy $E_{\mathrm{b}}$ more negative than -1 eV have stronger binding free energies $G_{\mathrm{b}}$ and higher probability $P_i$ to exist under a certain temperature, which helps us to sample crucial configurations during investigation of high-temperature diffusion or deposition of thin films. Above 800 K, all configurations of a $CeO_2$ molecule on Au surfaces become unbonded, which implies an upper limit of the surface temperature for $CeO_2$ molecules to deposit on Au surfaces. The finding of $\Delta E<0$ indicates that all $CeO_2$ configurations are more stable than the atomic constituents on Au surfaces, implying that the dissociated $CeO_2$ will be hard to discover on Au surfaces. We also discovered a linear O-Au-O configuration on Au(110) that -0.080 eV more stable than that reported in the literature[53] (Figure S2 (b)). Overall, $E_b$, $G_b$, and $\Delta E$ of $CeO_2$ molecules have more negative mean values and lower standard deviation going from Au(001), Au(110), to Au(111), meaning that the higher the planar symmetry, the less diverse and the more stable the $CeO_2$ configurations on Au surfaces.

Third, as explained earlier, we trained random forest models to quantify how four Au surface features, $N_{\mathrm{Ce-Au_s}}$, $N_{\mathrm{O-Au_s}}$, $N_{\mathrm{Au_{Ce}-Au}}$, and $N_{\mathrm{Au_O-Au}}$, impact the target properties $E_b$, $\Delta E$, and $Z_{\mathrm{eff,Ce}}$, then utilized SHAP values to explain models and quantify impacts of features on one target at a time. The local environment between $CeO_2$ and surface Au atoms ($N_{\mathrm{Ce-Au_s}}$ and $N_{\mathrm{O-Au_s}}$) plays a more dominant role than the local environment around surface Au atoms ($N_{\mathrm{Au_{Ce}-Au}}$ and $N_{\mathrm{Au_O-Au}}$). The coordination number of Au that is the closest to O ($N_{\mathrm{Au_O-Au}}$) has the least impact on $E_b$, $\Delta E$, and $Z_{\mathrm{eff,Ce}}$, which can be attributed to O only having short-range interaction with Au. It is worth mentioning that mixing all data from different surfaces worsens the random forest training compared to categorizing data by surfaces.

Lastly, we investigated the effects of $U_{\mathrm{eff}}$ of 3 eV and 5 eV. Configurations were not changed, but $E_b$ changed under different $U_{\mathrm{eff}}$. Changing $U_{\mathrm{eff}}$ from 3 eV to 5 eV makes $G_b$ more negative in value when a higher temperature is considered, while the effective charge of Ce $Z_{\mathrm{eff,Ce}}$ becomes more positive. However, there is no clear trend when it comes to how $U_{\mathrm{eff}}$ impacts $\Delta E$. SHAP values of four Au surface features impacting on $E_b$, $\Delta E$, and $Z_{\mathrm{eff,Ce}}$ changed significantly (from proportional relation to non-proportional relation) on Au(111), while slightly changed on Au(001) and Au(110) when using different $U_{\mathrm{eff}}$ values, indicating that Au(111) should be treated with care.

## 4. Concluding remarks

In this work, different configurations of the $CeO_2$ molecule in vacuum and on Au(001), Au(110), and Au(111) surfaces were studied by analyzing their binding energy, $E_\mathrm{b}$, stability, $\Delta E$, binding free energy, $G_\mathrm{b}$, molecule geometry, effective charge, $Z_\mathrm{eff}$, charge density difference, $\Delta\rho$, and electronic PDOS, $D_{\vec{R}L}(E)$. Moreover, the effects of four Au surface features, $N_{\mathrm{Ce-Au_s}}$, $N_{\mathrm{O-Au_s}}$, $N_{\mathrm{Au_{Ce}-Au}}$, and $N_{\mathrm{Au_O-Au}}$, on three target properties $E_\mathrm{b}$, $\Delta E$, and $Z_{\mathrm{eff,Ce}}$ were investigated in terms of SHAP values by training random forest models to regress features against each target property.

It has been verified that the $CeO_2$ molecule in vacuum favors a bent state rather than a linear one, without the Hubbard $U$ correction on the Ce $f$ orbital, or with Hubbard $U$ and thermal corrections together. However, the $CeO_2$ molecule prefers to be linear with the Hubbard $U$ correction. $CeO_2$ molecule binding on the Au surfaces was found to be more stable compared to its constituent atoms on the same Au surfaces by 3.5 eV to 6.5 eV. More specifically, $\Delta E$ was found to become less negative and less stable when the planar packing fraction and the symmetry of Au surfaces increase, but there is no clear trend between $\Delta E$ and $U_\mathrm{eff}$ across the three Au surfaces. The configurational probabilities of $CeO_2$ molecule on Au(001), Au(110), and Au(111) show that configurations with $E_\mathrm{b}$ stronger than or closer to -1 eV are significant when the temperature rises.

The binding of $CeO_2$ molecules on Au surfaces was conveniently categorized into three cases: case 1 is the chemisorption with $E_\mathrm{b}$ stronger than or close to -1 eV and symmetric bonds, case 2 is the chemisorption with $E_\mathrm{b}$ stronger than or close to -1 eV and asymmetric bonds, and case 3 is the physisorption with $E_\mathrm{b}$ weaker than -1 eV and symmetric bonds. The atomic configurations of these cases are reported in detail in Section 3. In cases 1 and 2, between the Ce and Au atoms, charge was found to accumulate closer to the Au atoms and deplete near Ce. In addition, charge depletion between the O and Au atoms was observed. In case 3, charge accumulation near the Au atom and depletion near the Ce between Ce and Au atoms was found, slightly to no charge depletion between the O and Au atoms was seen, while charge accumulated on the O atom. From electronic projected density of states of $CeO_2$ on Au(001), Au(110), and Au(111), it was seen that the spreading of Au *d* does not change significantly, while Ce *d*, Ce *f*, O1 *p*, and O2 *p* are as wide as the spreading of Au *d* when case 1 chemisorption happens due to the symmetric Ce-O bonds; Ce *d*, Ce *f*, O1 *p*, or O2 *p* are wide as the spreading of Au *d* in case 2 chemisorption for its asymmetric Ce-O bonds; Ce *d*, Ce *f*, O1 *p*, and O2 *p* are localized similar to the free $CeO_2$ molecule when case 3 physisorption occurs.

SHAP value analysis of random forest models helped us quantify how the Au surface features $N_{\mathrm{Ce-Au_s}}$, $N_{\mathrm{O-Au_s}}$, $N_{\mathrm{Au_{Ce}-Au}}$, and $N_{\mathrm{Au_O-Au}}$ affect $E_{\mathrm{b}}$, $\Delta E$, and $Z_{\mathrm{eff,Ce}}$. Comparison of random forest models regressed on data from all surfaces with those from individual surfaces shows that the former only provides general predictions, while the latter provides better predictions for each surface. This finding is crucial when surface anisotropy is considered in thin film growth.

While the current work focused on the binding and stability of $CeO_2$ molecules on high-symmetry Au surfaces, future investigations might include the mechanisms of $CeO_2$ molecule migration on Au surfaces and the effects of strain on both the binding and migration energies.

**Acknowledgments**

This work is supported by the U.S. Department of Energy, Office of Science, Basic Energy Sciences (BES) under award DE-SC0020077 at Purdue University.

**Conflict of Interests**

The authors declare that they have no known competing financial interests or personal relationships that could have appeared to influence the work reported in this paper.

**Data Availability**

The data that support the findings of this study can be made available from the corresponding authors upon request.

**References**

1 J. E. Krzanowski, Phase formation and phase separation in multiphase thin film hard coatings, *Surf. Coat. Technol.*, 2004, **188–189**, 376–383.

2 J. Yu, J. Xiong, B. Cheng and S. Liu, Fabrication and characterization of $Ag–TiO_2$ multiphase nanocomposite thin films with enhanced photocatalytic activity, *Appl. Catal. B*, 2005, **60**, 211–221.

3 J. Song and H. Wang, A material design guideline for self-assembled vertically aligned nanocomposite thin films, *Journal of Physics: Materials*, 2025, **8**, 012002.

4 J. Lu, D. Zhang, R. L. Paldi, Z. He, P. Lu, J. Deitz, A. Ahmad, H. Dou, X. Wang, J. Liu, Z. Hu, B. Yang, X. Zhang, A. A. El-Azab and H. Wang, Abnormal in-plane epitaxy and formation mechanism of vertically aligned Au nanopillars in self-assembled $CeO_2$–Au metamaterial systems, *Mater. Horiz.*, 2023, **10**, 3101–3113.

5 D. Zhang, S. Misra, L. Li, X. Wang, J. Jian, P. Lu, X. Gao, X. Sun, Z. Qi, M. Kalaswad, X. Zhang and H. Wang, Tunable Optical Properties in Self-Assembled Oxide-Metal Hybrid Thin Films via Au-Phase Geometry Control: From Nanopillars to Nanodisks, *Adv. Opt. Mater.*, DOI:10.1002/adom.201901359.

6 J. Gao, X. Wu, Q. Li, S. Du, F. Huang, L. Liang, H. Zhang, F. Zhuge, H. Cao and Y. Song, Template-Free Growth of Well-Ordered Silver Nano Forest/Ceramic Metamaterial Films with Tunable Optical Responses, *Advanced Materials*, DOI:10.1002/adma.201605324.

7 J. Graciani, A. B. Vidal, J. A. Rodriguez and J. Fdez. Sanz, Unraveling the Nature of the Oxide–Metal Interaction in Ceria-Based Noble Metal Inverse Catalysts, *The Journal of Physical Chemistry C*, 2014, **118**, 26931–26938.

8 W. Huang, Q. Liu, Z. Zhou, Y. Li, Y. Ling, Y. Wang, Y. Tu, B. Wang, X. Zhou, D. Deng, B. Yang, Y. Yang, Z. Liu, X. Bao and F. Yang, Tuning the activities of cuprous oxide nanostructures via the oxide-metal interaction, *Nat. Commun.*, 2020, **11**, 2312.

9 W. Kudernatsch, G. Peng, H. Zeuthen, Y. Bai, L. R. Merte, L. Lammich, F. Besenbacher, M. Mavrikakis and S. Wendt, Direct Visualization of Catalytically Active Sites at the FeO–Pt(111) Interface, *ACS Nano*, 2015, **9**, 7804–7814.

10 Y. Li, Z. Li, J. Hu and W. Huang, Electronic Oxide-Metal Strong Interactions (EOMSI) Localized at $CeO_x$-Ag Interface, *J. Phys. Chem. Lett.*, 2024, **15**, 8682–8688.

11 J. A. Rodriguez, S. Ma, P. Liu, J. Hrbek, J. Evans and M. Pérez, Activity of $CeO_x$ and $TiO_x$ Nanoparticles Grown on Au(111) in the Water-Gas Shift Reaction, *Science (1979).*, 2007, **318**, 1757–1760.

12 J. Graciani, K. Mudiyanselage, F. Xu, A. E. Baber, J. Evans, S. D. Senanayake, D. J. Stacchiola, P. Liu, J. Hrbek, J. F. Sanz and J. A. Rodriguez, Highly active copper-ceria and copper-ceria-titania catalysts for methanol synthesis from $CO_2$, *Science (1979).*, 2014, **345**, 546–550.

13 H. Ha, S. Yoon, K. An and H. Y. Kim, Catalytic CO Oxidation over Au Nanoparticles Supported on $CeO_2$ Nanocrystals: Effect of the Au–$CeO_2$ Interface, *ACS Catal.*, 2018, **8**, 11491–11501.

14 M. Jing, W. Song, L. Chen, S. Ma, J. Deng, H. Zheng, Y. Li, J. Liu and Z. Zhao, Density Functional Theory Study of the Formaldehyde Catalytic Oxidation Mechanism on a Au-Doped $CeO_2$ (111) Surface, *The Journal of Physical Chemistry C*, 2018, **122**, 438–448.

15 G. Righi, R. Magri and A. Selloni, $H_2$ Dissociation on Noble Metal Single Atom Catalysts Adsorbed on and Doped into $CeO_2$ (111), *The Journal of Physical Chemistry C*, 2019, **123**, 9875–9883.

16 A. S. K. Hashmi and G. J. Hutchings, Gold Catalysis, *Angew. Chem. Int. Ed.*, 2006, **45**, 7896–7936.

17 T. Montini, M. Melchionna, M. Monai and P. Fornasiero, Fundamentals and Catalytic Applications of $CeO_2$ -Based Materials, *Chem. Rev.*, 2016, **116**, 5987–6041.

18 C. Heinemann, H. H. Cornehl, D. Schröder, M. Dolg and H. Schwarz, The $CeO_2^+$ Cation: Gas-Phase Reactivity and Electronic Structure, *Inorg. Chem.*, 1996, **35**, 2463–2475.

19 T. K. Todorova, I. Infante, L. Gagliardi and J. M. Dyke, The chemiionization reactions Ce + O and Ce + $O_2$ : Assignment of the observed chemielectron bands, *Int. J. Quantum Chem.*, 2009, **109**, 2068–2079.

20 R. L. De Kock, M. A. Peterson, L. K. Timmer, E. J. Baerends and P. Vernooijs, A theoretical study of the linear versus bent geometry for several $MX_2$ molecules: $MgF_2$, $CaH_2$, $CaF_2$, $CeO_2$ and $YbCl_2$, *Polyhedron*, 1990, **9**, 1919–1934.

21 S. D. Gabelnick, G. T. Reedy and M. G. Chasanov, Infrared spectra and structure of some matrix-isolated lanthanide and actinide oxides, *J. Chem. Phys.*, 1974, **60**, 1167–1171.

22 A. Ahmad, J. Peng, K. SharafEldin, J. Lu, H. Wang and A. El-Azab, A DFT-based kinetic Monte Carlo simulation of multiphase oxide-metal thin film growth, *J. Appl. Phys.*, DOI:10.1063/5.0189427.

23 A. Ahmad, Y.-C. Chen, J. Peng and A. El-Azab, Strain effects on the binding and diffusion energies of Au adatoms and CeO2 admolcules on Au, CeO2, MgO and SrTiO3 surfaces, *Surf. Sci.*, 2026, **772**, 123008.

24 X. Liu, L. Sun and W.-Q. Deng, Theoretical Investigation of CO $_2$ Adsorption and Dissociation on Low Index Surfaces of Transition Metals, *The Journal of Physical Chemistry C*, 2018, **122**, 8306–8314.

25 B. Hagman, A. Posada-Borbón, A. Schaefer, M. Shipilin, C. Zhang, L. R. Merte, A. Hellman, E. Lundgren, H. Grönbeck and J. Gustafson, Steps Control the Dissociation of CO $_2$ on Cu(100), *J. Am. Chem. Soc.*, 2018, **140**, 12974–12979.

26 V. Piacente, G. Bardi, L. Malaspina and A. Desideri, Dissociation energy of $CeO_2$ and $Ce_2O_2$ molecules, *J. Chem. Phys.*, 1973, **59**, 31–36.

27 D. A. Andersson, S. I. Simak, B. Johansson, I. A. Abrikosov and N. V. Skorodumova, Modeling of $CeO_2$, $Ce_2O_3$, and $CeO_{2-x}$, in the LDA+U formalism, *Phys. Rev. B*, 2007, **75**, 035109.

28 N. V. Skorodumova, R. Ahuja, S. I. Simak, I. A. Abrikosov, B. Johansson and B. I. Lundqvist, Electronic, bonding, and optical properties of $CeO_2$ and $Ce_2O_3$ from first principles, *Phys. Rev. B*, 2001, **64**, 115108.

29 G. Kresse and J. Furthmüller, Efficiency of ab-initio total energy calculations for metals and semiconductors using a plane-wave basis set, *Comput. Mater. Sci.*, 1996, **6**, 15–50.

30 G. Kresse and J. Furthmüller, Efficient iterative schemes for *ab initio* total-energy calculations using a plane-wave basis set, *Phys. Rev. B*, 1996, **54**, 11169–11186.

31 P. E. Blöchl, Projector augmented-wave method, *Phys. Rev. B*, 1994, **50**, 17953–17979.

32 G. Kresse and D. Joubert, From ultrasoft pseudopotentials to the projector augmented-wave method, *Phys. Rev. B*, 1999, **59**, 1758–1775.

33 J. P. Perdew, K. Burke and M. Ernzerhof, Generalized gradient approximation made simple, *Phys. Rev. Lett.*, 1996, **77**, 3865–3868.

34 S. L. Dudarev, G. A. Botton, S. Y. Savrasov, C. J. Humphreys and A. P. Sutton, Electron-energy-loss spectra and the structural stability of nickel oxide: An LSDA+U study, *Phys. Rev. B*, 1998, **57**, 1505–1509.

35 Y. Li, T. Qin, Y. Wei, J. Xiong, P. Zhang, K. Lai, H. Chi, X. Liu, L. Chen, X. Yu, Z. Zhao, L. Li and J. Liu, A single site ruthenium catalyst for robust soot oxidation without platinum or palladium, *Nat. Commun.*, 2023, **14**, 7149.

36 M. P. Teter, M. C. Payne and D. C. Allan, Solution of Schrödinger's equation for large systems, *Phys. Rev. B*, 1989, **40**, 12255–12263.

37 D. M. Bylander, L. Kleinman and S. Lee, Self-consistent calculations of the energy bands and bonding properties of $B_{12}C_3$, *Phys. Rev. B*, 1990, **42**, 1394–1403.

38 P. Hohenberg and W. Kohn, Inhomogeneous electron gas, *Physical Review*, 1964, **136**, B864.

39 K. Kunc and R. M. Martin, *Ab Initio* Force Constants of GaAs: A New Approach to Calculation of Phonons and Dielectric Properties, *Phys. Rev. Lett.*, 1982, **48**, 406–409.

40 P. K. Lam, M. M. Dacorogna and M. L. Cohen, Self-consistent calculation of electron-phonon couplings, *Phys. Rev. B*, 1986, **34**, 5065–5069.

41 B. Monserrat, Electron–phonon coupling from finite differences, *Journal of Physics: Condensed Matter*, 2018, **30**, 083001.

42 V. Wang, N. Xu, J.-C. Liu, G. Tang and W.-T. Geng, VASPKIT: A user-friendly interface facilitating high-throughput computing and analysis using VASP code, *Comput. Phys. Commun.*, 2021, **267**, 108033.

43 D. A. . McQuarrie and J. D. . Simon, *Molecular thermodynamics*, University Science Books, 1999.

44 J. W. Ochterski, Thermochemistry in gaussian, *Gaussian Inc*.

45 A. Hjorth Larsen, J. Jørgen Mortensen, J. Blomqvist, I. E. Castelli, R. Christensen, M. Dułak, J. Friis, M. N. Groves, B. Hammer, C. Hargus, E. D. Hermes, P. C. Jennings, P. Bjerre Jensen, J. Kermode, J. R. Kitchin, E. Leonhard Kolsbjerg, J. Kubal, K. Kaasbjerg, S. Lysgaard, J. Bergmann Maronsson, T. Maxson, T. Olsen, L. Pastewka, A. Peterson, C. Rostgaard, J. Schiøtz, O. Schütt, M. Strange, K. S. Thygesen, T. Vegge, L. Vilhelmsen, M. Walter, Z. Zeng and K. W. Jacobsen, The atomic simulation environment—a Python library for working with atoms, *Journal of Physics: Condensed Matter*, 2017, **29**, 273002.

46 K. Momma and F. Izumi, *VESTA 3* for three-dimensional visualization of crystal, volumetric and morphology data, *J. Appl. Crystallogr.*, 2011, **44**, 1272–1276.

47 J. D. Hunter, Matplotlib: A 2D Graphics Environment, *Comput. Sci. Eng.*, 2007, **9**, 90–95.

48 A. Togo, [Phonopy-users] YCChen: inquiry about using Phonopy for binding free energy on slab models, https://sourceforge.net/p/phonopy/mailman/phonopy-users/thread/MW3PR22MB2187CBB94143EAF637316111A061A%40MW3PR22MB2187.namprd22.prod.outlook.com/#msg59189462, (accessed 4 October 2026).

49 A. Togo and I. Tanaka, First principles phonon calculations in materials science, *Scr. Mater.*, 2015, **108**, 1–5.

50 G. JONES, J. JAKOBSEN, S. SHIM, J. KLEIS, M. ANDERSSON, J. ROSSMEISL, F. ABILDPEDERSEN, T. BLIGAARD, S. HELVEG and B. HINNEMANN, First principles calculations and experimental insight into methane steam reforming over transition metal catalysts, *J. Catal.*, 2008, **259**, 147–160.

51 A. A. Peterson, F. Abild-Pedersen, F. Studt, J. Rossmeisl and J. K. Nørskov, How copper catalyzes the electroreduction of carbon dioxide into hydrocarbon fuels, *Energy Environ. Sci.*, 2010, **3**, 1311.

52 Feliciano. Giustino, *Materials modelling using density functional theory : properties and predictions*, Oxford University Press, 2014.

53 K. Sun, M. Kohyama, S. Tanaka and S. Takeda, Theoretical Study of Atomic Oxygen on Gold Surface by Hückel Theory and DFT Calculations, *J. Phys. Chem. A*, 2012, **116**, 9568–9573.

54 Y. Seminovski, P. Tereshchuk, A. Kiejna and J. L. F. Da Silva, The role of the cationic Pt sites in the adsorption properties of water and ethanol on the $Pt_4$/Pt(111) and $Pt_4/CeO_2$(111) substrates: A density functional theory investigation, *J. Chem. Phys.*, DOI:10.1063/1.4963162.

55 R. F. W. Bader, P. J. MacDougall and C. D. H. Lau, Bonded and nonbonded charge concentrations and their relation to molecular geometry and reactivity, *J. Am. Chem. Soc.*, 1984, **106**, 1594–1605.

56 G. Henkelman, A. Arnaldsson and H. Jónsson, A fast and robust algorithm for Bader decomposition of charge density, *Comput. Mater. Sci.*, 2006, **36**, 354–360.

57 R. F. W. Bader, Atoms in molecules, *Acc. Chem. Res.*, 1985, **18**, 9–15.

58 R. F. W. Bader, The Quantum Mechanical Basis of Conceptual Chemistry, *Monatshefte für Chemie - Chemical Monthly*, 2005, **136**, 819–854.

59 C. Rostgaard, The Projector Augmented-wave Method.

60 M. Schüler, O. E. Peil, G. J. Kraberger, R. Pordzik, M. Marsman, G. Kresse, T. O. Wehling and M. Aichhorn, Charge self-consistent many-body corrections using optimized projected localized orbitals, *Journal of Physics: Condensed Matter*, 2018, **30**, 475901.

61 C. Ben Mahmoud, A. Anelli, G. Csányi and M. Ceriotti, Learning the electronic density of states in condensed matter, *Phys. Rev. B*, 2020, **102**, 235130.

62 L. Breiman, Random Forests, *Mach. Learn.*, 2001, **45**, 5–32.

63 F. Pedregosa, G. Varoquaux, A. Gramfort, V. Michel, B. Thirion, O. Grisel, M. Blondel, P. Prettenhofer, R. Weiss, V. Dubourg, J. Vanderplas, A. Passos, D. Cournapeau, M. Brucher, M. Perrot and É. Duchesnay, Scikit-learn: Machine Learning in Python, *Journal of Machine Learning Research*, 2011, **12**, 2825–2830.

64 S. M. Lundberg, G. Erion, H. Chen, A. DeGrave, J. M. Prutkin, B. Nair, R. Katz, J. Himmelfarb, N. Bansal and S.-I. Lee, From local explanations to global understanding with explainable AI for trees, *Nat. Mach. Intell.*, 2020, **2**, 56–67.

65 S. M. Lundberg and S.-I. Lee, in *Advances in Neural Information Processing Systems*, eds. I. Guyon, U. Von Luxburg, S. Bengio, H. Wallach, R. Fergus, S. Vishwanathan and R. Garnett, Curran Associates, Inc., 2017, vol. 30.

66 B. Johansson, W. Luo, S. Li and R. Ahuja, Cerium; Crystal Structure and Position in The Periodic Table, *Sci. Rep.*, 2014, **4**, 6398.

**Supplementary Information for**

# On the binding and stability of $CeO_2$ molecules on low-index Au surfaces

Ying-Cheng Chen[a]*, Ahmad Ahmad[a], Anter El-Azab[a]*

[a] School of Materials Engineering, Purdue University, West Lafayette, IN 47907, USA

*Corresponding author: Ying-Cheng Chen, chen5305@purdue.edu; Anter El-Azab, aelazab@purdue.edu

This file contains supplementary information in the form of additional results from our investigation. The file is organized into three sections. Section S1 contains further results concerning the configurations and binding energies of dissociated and molecular oxygens on Au surfaces, which are mostly reproduced from the literature,[1] of which one result (

Figure S2 (a)) is uniquely found in this work. In Section S2, we present results on the electronic projected density of states (PDOS) of $CeO_2$ molecules on Ce *d* and Ce *f* orbitals in vacuum with 4 combinations of 2 configurations (bent and linear) and 2 $U_{\mathrm{eff}}$ values (3 eV and 5 eV), respectively. Section S3 summarizes information related to $CeO_2$ molecules on Au surfaces in terms of adsorption sites before and after optimization, binding energy $E_{\mathrm{b}}$, stability $\Delta E$, bond length, bond angle, categorized binding case, effective charge $Z_{\mathrm{eff,Ce}}$, $Z_{\mathrm{eff,O1}}$, and $Z_{\mathrm{eff,O2}}$, visualized charge density difference $\Delta\rho$, and electronic PDOS of orbitals Ce *d*, Ce *f*, O1 *p*, O2 *p*, and Au *d* of the nearest Au atom from one of the O atoms in $CeO_2$.

## S1. Oxygen molecule stability on Au surfaces

According to the literature, 2O chemisorbing on Au surfaces is the strongest by binding on hollow sites and forming a linear O-Au-O structure.[1] In this work, the most stable configurations of 2O on Au(001) and Au(111) match the literature,[1] as shown in

Figure S1(a) and

Figure S3(a), respectively. However, a -0.080 eV more stable configuration of 2O on Au(110) is found in this work (

Figure S2(a)), compared to that which matches the literature (

Figure S2(b)).[1]

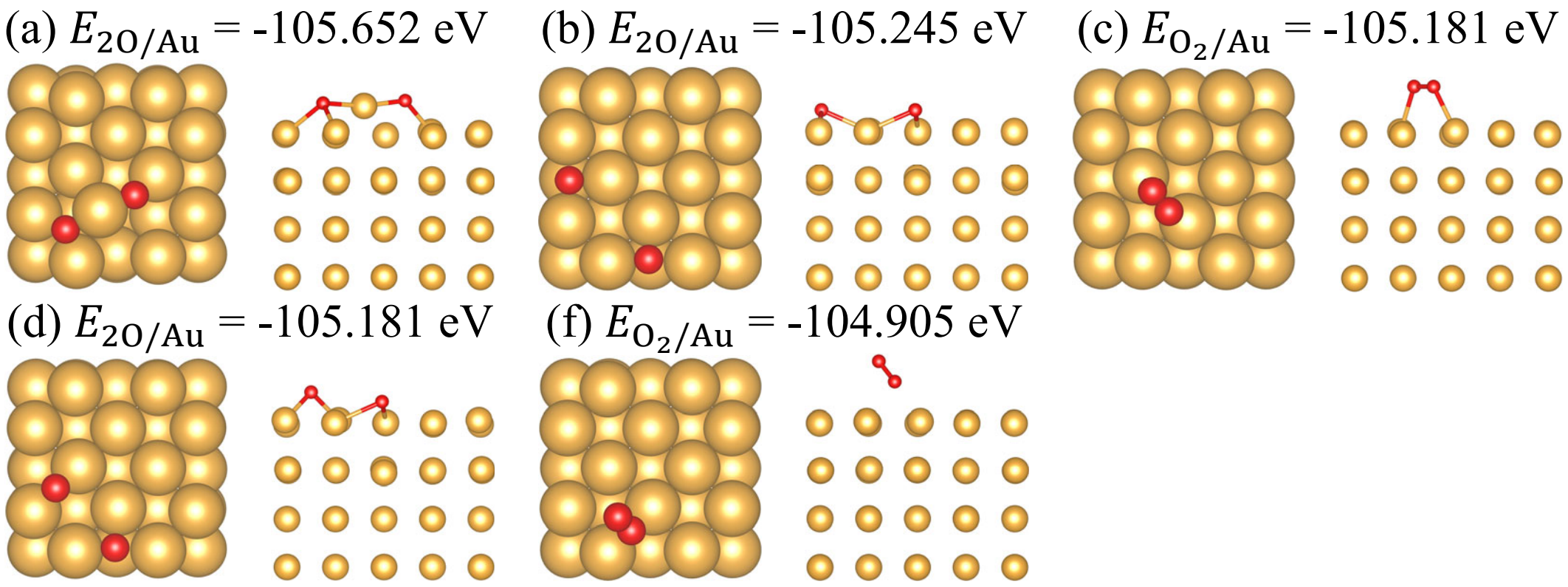

Figure S1. DFT total energy, top view, and side view of 2O and $O_2$ on Au(001). The more negative the DFT total energy, the stronger the adsorbate binding.

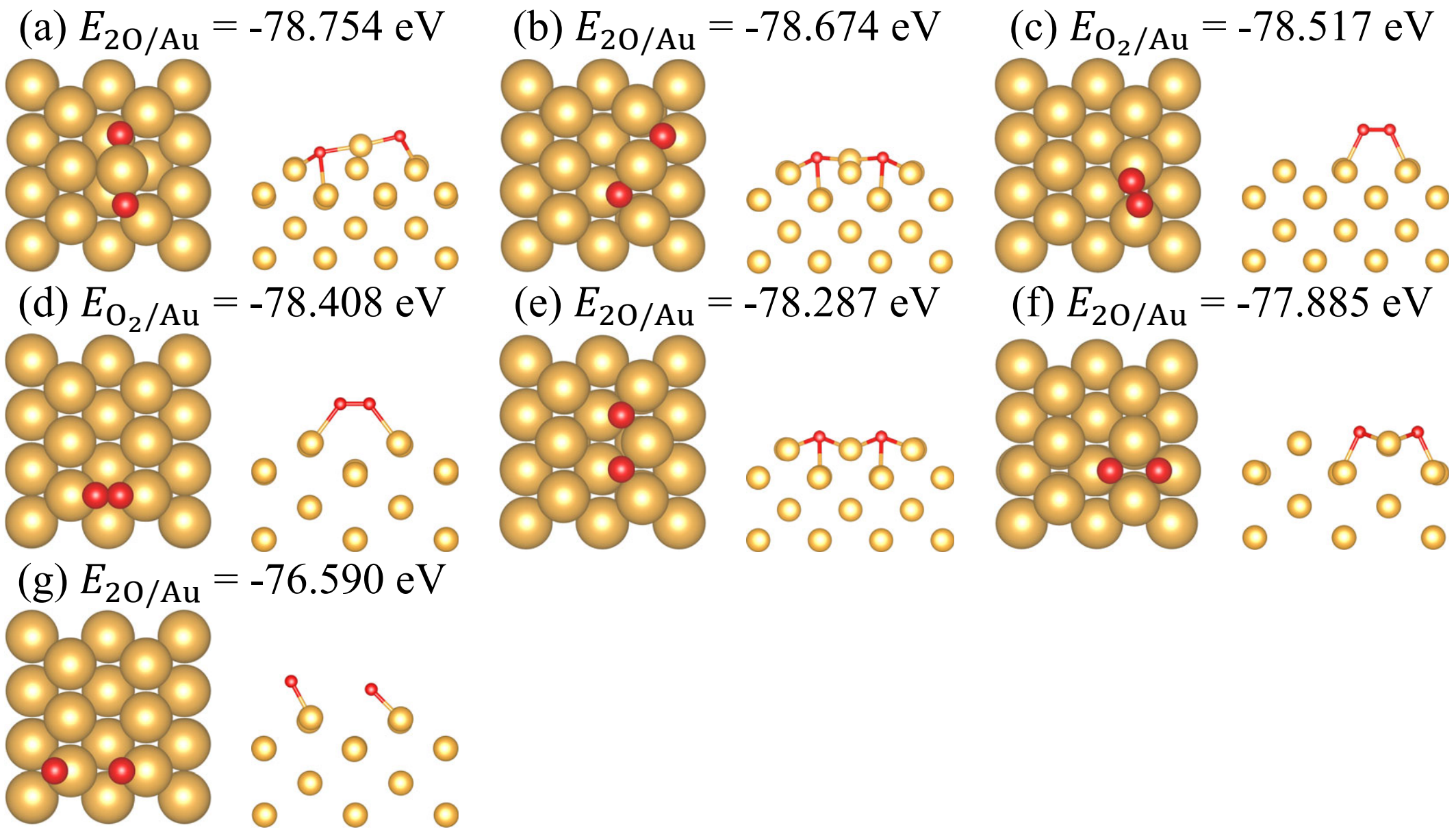

Figure S2. DFT total energy, top view, and side view of 2O and $O_2$ on Au(110).

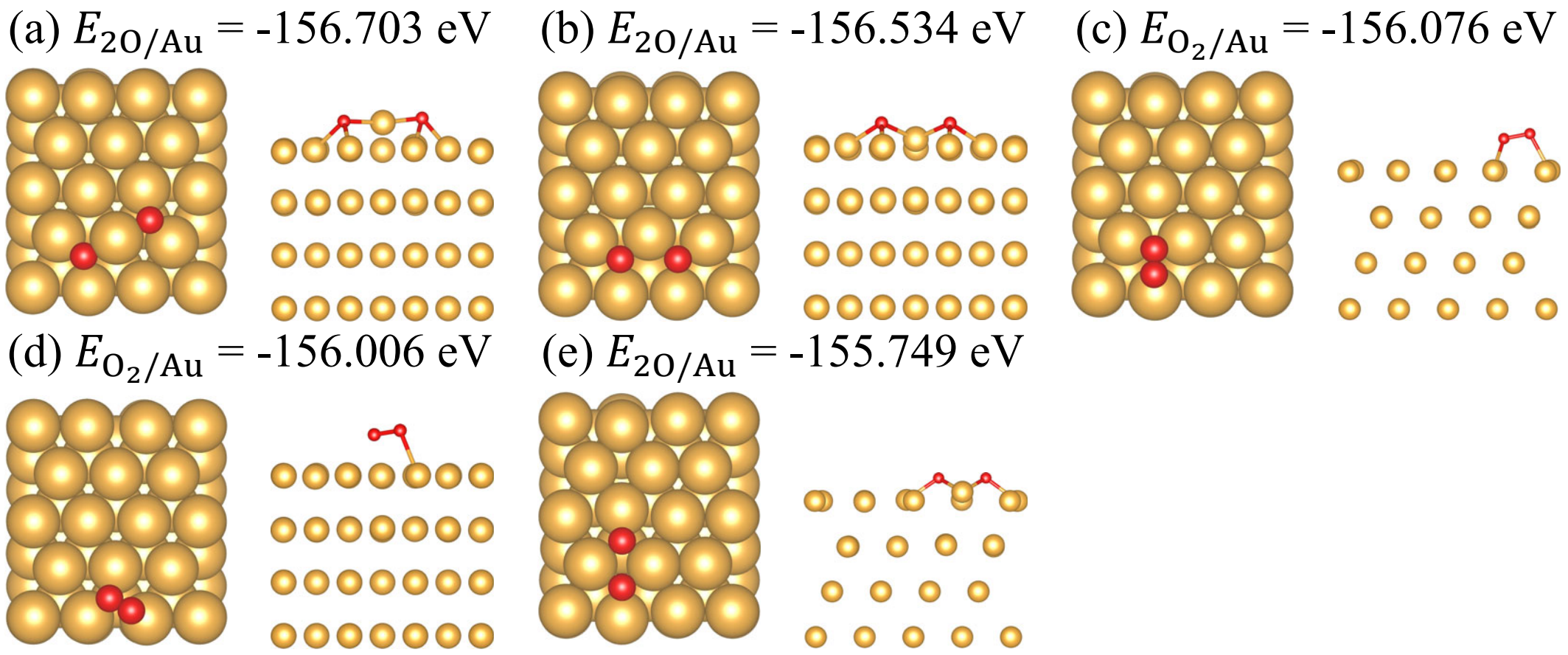

Figure S3. DFT total energy, top view, and side view of 2O and $O_2$ on Au(111).

**S2. $CeO_2$ molecules in vacuum**

According to the literature, the $CeO_2$ molecule prefers the bent state rather than the linear state.[2–5] Moreover, the bulk and slab $CeO_2$ are usually calculated using density functional theory (DFT) with the Hubbard $U$ correction (DFT$+U$) on the localized Ce $4f$ electrons to account for the electron-electron repulsion within the localized Ce $f$ orbitals.[6] However, using the DFT$+U$ to calculate the $CeO_2$ molecule can yield contradictory results, suggesting that the $CeO_2$ molecule prefers linear over bent unless thermal corrections are included, while using the DFT calculations without the Hubbard $U$ correction can give results aligned with the literature. Without the Hubbard $U$ correction, the DFT calculations of the $CeO_2$ molecule binding on the Au surfaces become difficult to converge. Therefore, DFT$+U$ calculations are conducted for all calculations in this study, except to prove that the bent $CeO_2$ molecule is more stable than the linear one. The electronic PDOS analysis in Figure S4 is intended to show how Ce $d$ and Ce $f$ orbitals of the $CeO_2$ molecule change without and with the Hubbard $U$ correction.

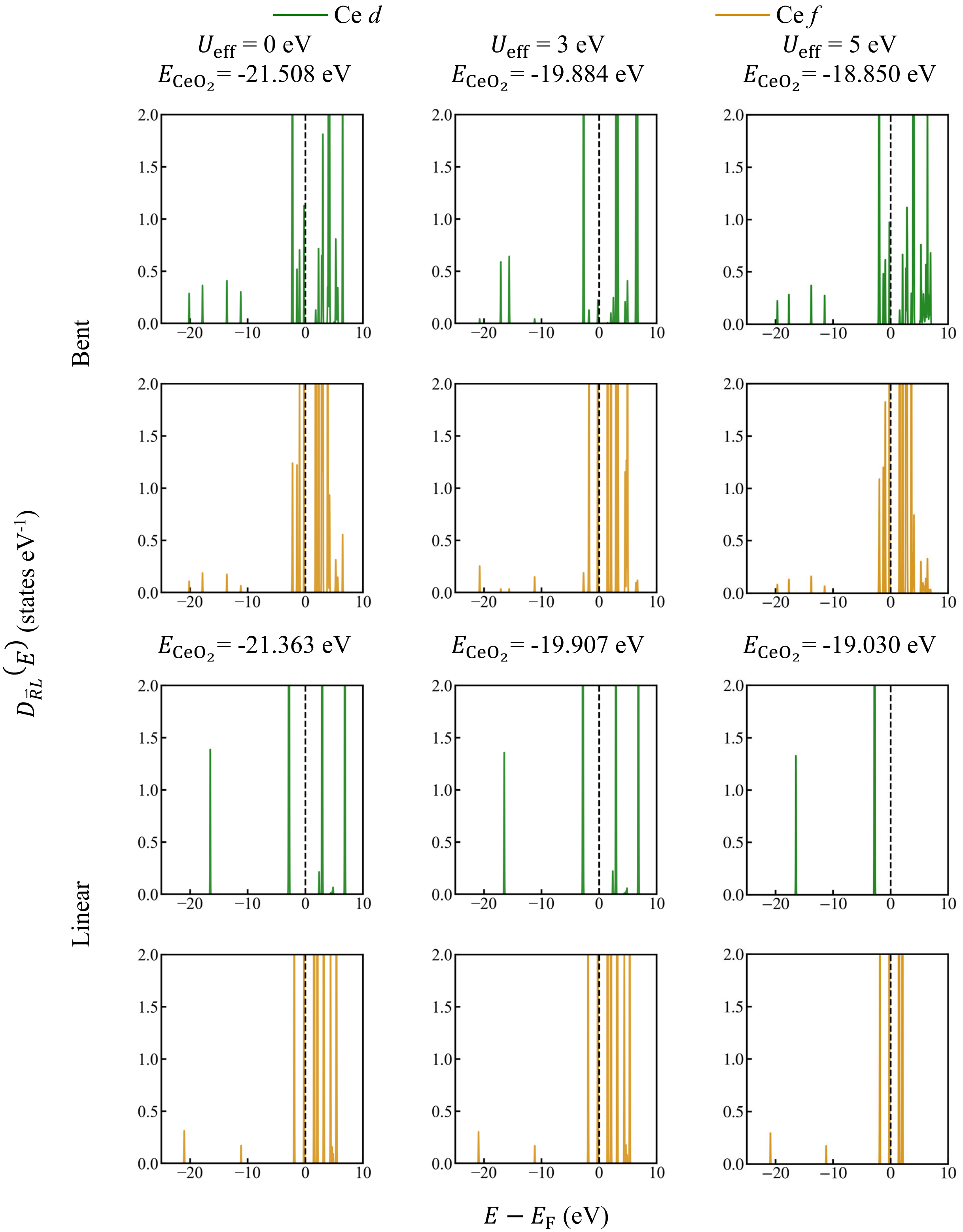

Figure S4. Electronic PDOS $D_{\vec{R}L}(E)$ of Ce $d$ (solid green) and Ce $f$ (solid orange) orbitals and DFT total energy of the $CeO_2$ molecule in vacuum, $E_{CeO_2}$, without (left column) and with a $U_{eff}$ of 3 eV (middle column) and 5 eV (right column).

## S3. $CeO_2$ molecules on Au

Section S3.1, S3.2, and S3.3 contain a number of results for various configurations of a $CeO_2$ molecule on Au(100), Au(110), and Au(111) surfaces, respectively, with a $U_{\mathrm{eff}}$ of 3 eV and 5 eV. These results include: (i) The adsorption sites before and after optimization, binding energy $E_{\mathrm{b}}$, stability $\Delta E$, bond length Ce-O1 and Ce-O2, bond angle O1-Ce-O2, and binding cases. (ii) How $E_{\mathrm{b}}$ correlates with bond length and bond angle. (iii) Adsorption sites after optimization, effective charge $Z_{\mathrm{eff,Ce}}$, $Z_{\mathrm{eff,O1}}$, and $Z_{\mathrm{eff,O2}}$. (iv) Visualized charge density difference $\Delta\rho$. (v) The electronic PDOS $D_{\vec{R}L}(E)$ for orbitals Ce $d$, Ce $f$, O1 $p$, O2 $p$, and Au $d$ of the nearest Au atom from one of the O atoms.

### S3.1 $CeO_2$ molecules on Au(001)

Table S1. The adsorption sites after and before optimization, the binding energy $E_b$, the stability $\Delta E$, the bond length Ce-O1 and Ce-O2, and the bond angle O1-Ce-O2 of the $CeO_2$ molecule on the Au(001) slab model with a $U_{eff}$ of 3 eV.

| After opt. | Before opt. | $E_b$ (eV) | $\Delta E$ (eV) | Ce-O1 (Å) | Ce-O2 (Å) | O1-Ce-O2 (°) | Case |
|---|---|---|---|---|---|---|---|
| HTT-B1 | HTT-B | -1.227 | -4.431 | 1.944 | 1.944 | 106.355 | 1 |
| HTV-B1 | THH-B | -1.162 | -4.366 | 1.999 | 1.851 | 121.904 | 2 |
| HTV-B2 | TTV-B1 | -1.156 | -4.360 | 1.844 | 1.996 | 145.392 | 2 |
| HTV-B3 | TTV-B2 | -1.155 | -4.359 | 1.851 | 1.999 | 119.577 | 2 |
| HTT-L | HTT-L | -1.151 | -4.355 | 1.933 | 1.933 | 176.642 | 1 |
| HTT-B2 | HHH-B | -1.088 | -4.292 | 1.939 | 1.939 | 121.706 | 1 |
| HHH-L | HHH-L | -0.758 | -3.962 | 1.894 | 1.894 | 176.053 | 3 |
| BTT-L | BTT-L | -0.715 | -3.919 | 1.888 | 1.888 | 171.438 | 3 |
| TTT-L | TTT-L | -0.661 | -3.865 | 1.909 | 1.909 | 176.291 | 3 |
| BHH-L | BHH-L | -0.581 | -3.785 | 1.885 | 1.885 | 179.528 | 3 |
| THH-L | THH-L | -0.369 | -3.573 | 1.859 | 1.859 | 171.107 | 3 |

Table S2. The adsorption sites after and before optimization, the binding energy $E_b$, the stability $\Delta E$, the bond length Ce-O1 and Ce-O2, and the bond angle O1-Ce-O2 of the $CeO_2$ molecule on the Au(001) slab model with a $U_{eff}$ of 5 eV.

| After opt. | Before opt. | $E_b$ (eV) | $\Delta E$ (eV) | Ce-O1 (Å) | Ce-O2 (Å) | O1-Ce-O2 (°) | Case |
|---|---|---|---|---|---|---|---|
| HTT-B1 | HTT-B | -1.238 | -4.950 | 1.960 | 1.960 | 105.667 | 1 |
| HTT-L | HTT-L | -1.198 | -4.910 | 1.946 | 1.946 | 176.391 | 1 |
| HTV-B1 | TTV-B2 | -1.148 | -4.859 | 1.867 | 2.018 | 117.674 | 2 |
| HTV-B2 | THH-B | -1.142 | -4.853 | 2.018 | 1.868 | 119.929 | 2 |
| HTV-B3 | TTV-B1 | -1.140 | -4.852 | 1.859 | 2.011 | 147.018 | 2 |
| HTT-B2 | HHH-B | -1.081 | -4.792 | 1.956 | 1.956 | 119.764 | 1 |
| HHH-L | HHH-L | -0.781 | -4.492 | 1.907 | 1.907 | 176.788 | 3 |
| BTT-L | BTT-L | -0.727 | -4.438 | 1.897 | 1.897 | 174.342 | 3 |
| TTT-L | TTT-L | -0.677 | -4.388 | 1.921 | 1.921 | 177.634 | 3 |
| BHH-L | BHH-L | -0.589 | -4.300 | 1.896 | 1.896 | 179.411 | 3 |
| THH-L | THH-L | -0.349 | -4.061 | 1.868 | 1.868 | 174.361 | 3 |

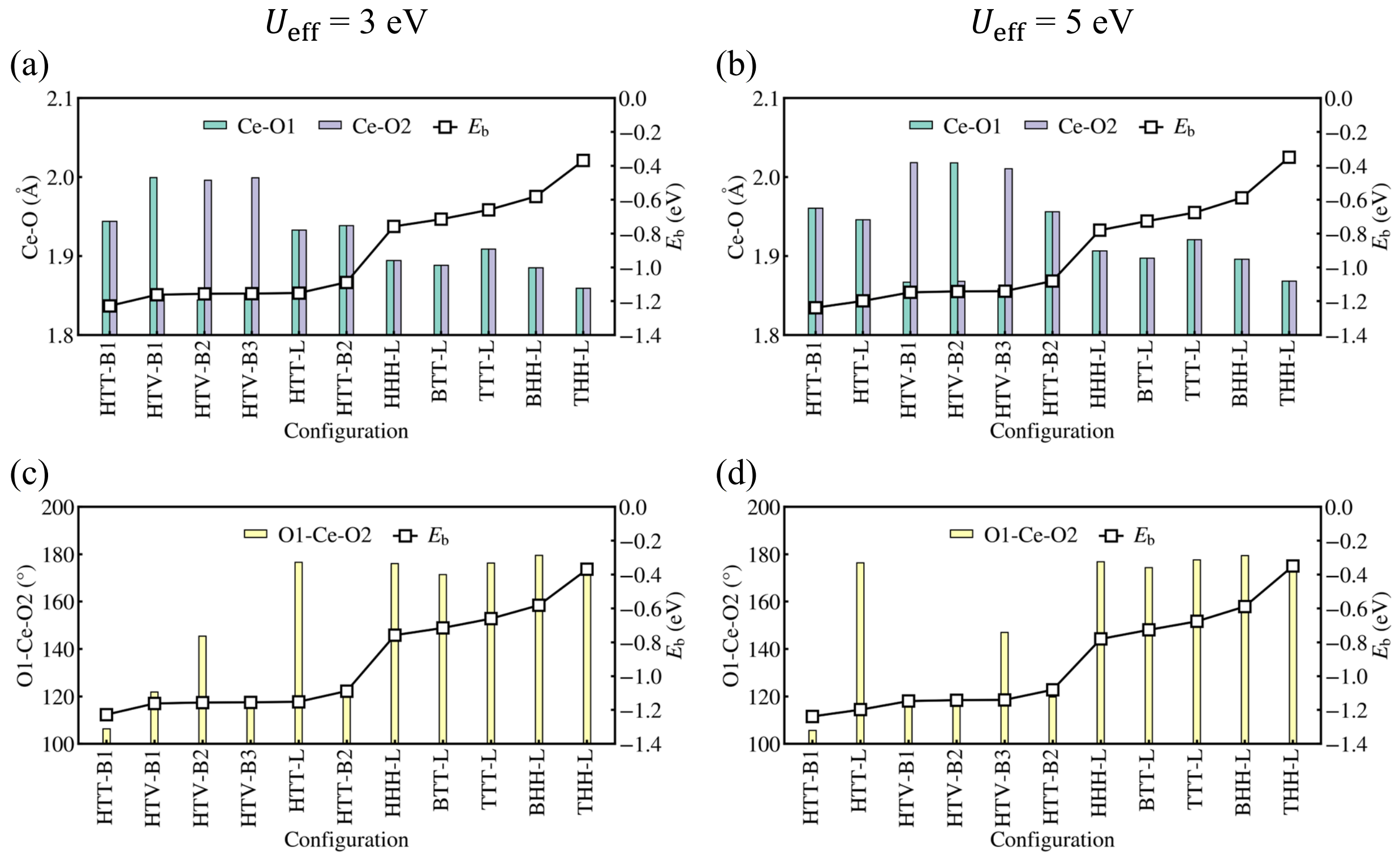

Figure S5. The correlation between the binding energy $E_b$ of the $CeO_2$ molecule on the Au(001) slab model with respect to (a, b) Ce-O1 and Ce-O2 bond lengths and (c, d) O1-Ce-O2 bond angle, where (a, c) have a $U_{eff}$ of 3 eV while (b, d) have a $U_{eff}$ of 5 eV.

Table S3. The adsorption sites after optimization, the effective charge $Z_{\mathrm{eff,Ce}}$, $Z_{\mathrm{eff,O1}}$, and $Z_{\mathrm{eff,O2}}$ of Ce, O1, and O2 atoms in the $CeO_2$ molecule on the Au(001) slab model with a $U_{\mathrm{eff}}$ of 3 eV are provided, along with their respective cases of adsorption.

| After opt. | $Z_{\mathrm{eff,Ce}}$ (e) | $Z_{\mathrm{eff,O1}}$ (e) | $Z_{\mathrm{eff,O2}}$ (e) | Case |
|---|---|---|---|---|
| HTT-B1 | 1.973 | -0.933 | -0.933 | 1 |
| HTV-B1 | 2.034 | -0.955 | -0.944 | 2 |
| HTV-B2 | 2.060 | -0.955 | -0.959 | 2 |
| HTV-B3 | 2.032 | -0.944 | -0.954 | 2 |
| HTT-L | 2.005 | -0.936 | -0.936 | 1 |
| HTT-B2 | 1.998 | -0.947 | -0.947 | 1 |
| HHH-L | 2.059 | -0.980 | -0.980 | 3 |
| BTT-L | 2.073 | -0.981 | -0.981 | 3 |
| TTT-L | 2.068 | -0.998 | -0.998 | 3 |
| BHH-L | 2.096 | -1.020 | -1.020 | 3 |
| THH-L | 2.112 | -1.014 | -1.014 | 3 |

Table S4. The adsorption sites after optimization, the effective charge $Z_{\mathrm{eff,Ce}}$, $Z_{\mathrm{eff,O1}}$, and $Z_{\mathrm{eff,O2}}$ of Ce, O1, and O2 atoms in the $CeO_2$ molecule on the Au(001) slab model with a $U_{\mathrm{eff}}$ of 5 eV are provided, along with their respective cases of adsorption.

| After opt. | $Z_{\mathrm{eff,Ce}}$ (e) | $Z_{\mathrm{eff,O1}}$ (e) | $Z_{\mathrm{eff,O2}}$ (e) | Case |
|---|---|---|---|---|
| HTT-B1 | 1.997 | -0.939 | -0.939 | 1 |
| HTT-L | 2.041 | -0.950 | -0.950 | 1 |
| HTV-B1 | 2.047 | -0.947 | -0.957 | 2 |
| HTV-B2 | 2.054 | -0.960 | -0.953 | 2 |
| HTV-B3 | 2.089 | -0.981 | -0.960 | 2 |
| HTT-B2 | 2.025 | -0.950 | -0.949 | 1 |
| HHH-L | 2.049 | -0.974 | -0.974 | 3 |
| BTT-L | 2.101 | -0.997 | -0.997 | 3 |
| TTT-L | 2.032 | -0.979 | -0.979 | 3 |
| BHH-L | 2.108 | -1.030 | -1.030 | 3 |
| THH-L | 2.167 | -1.050 | -1.050 | 3 |

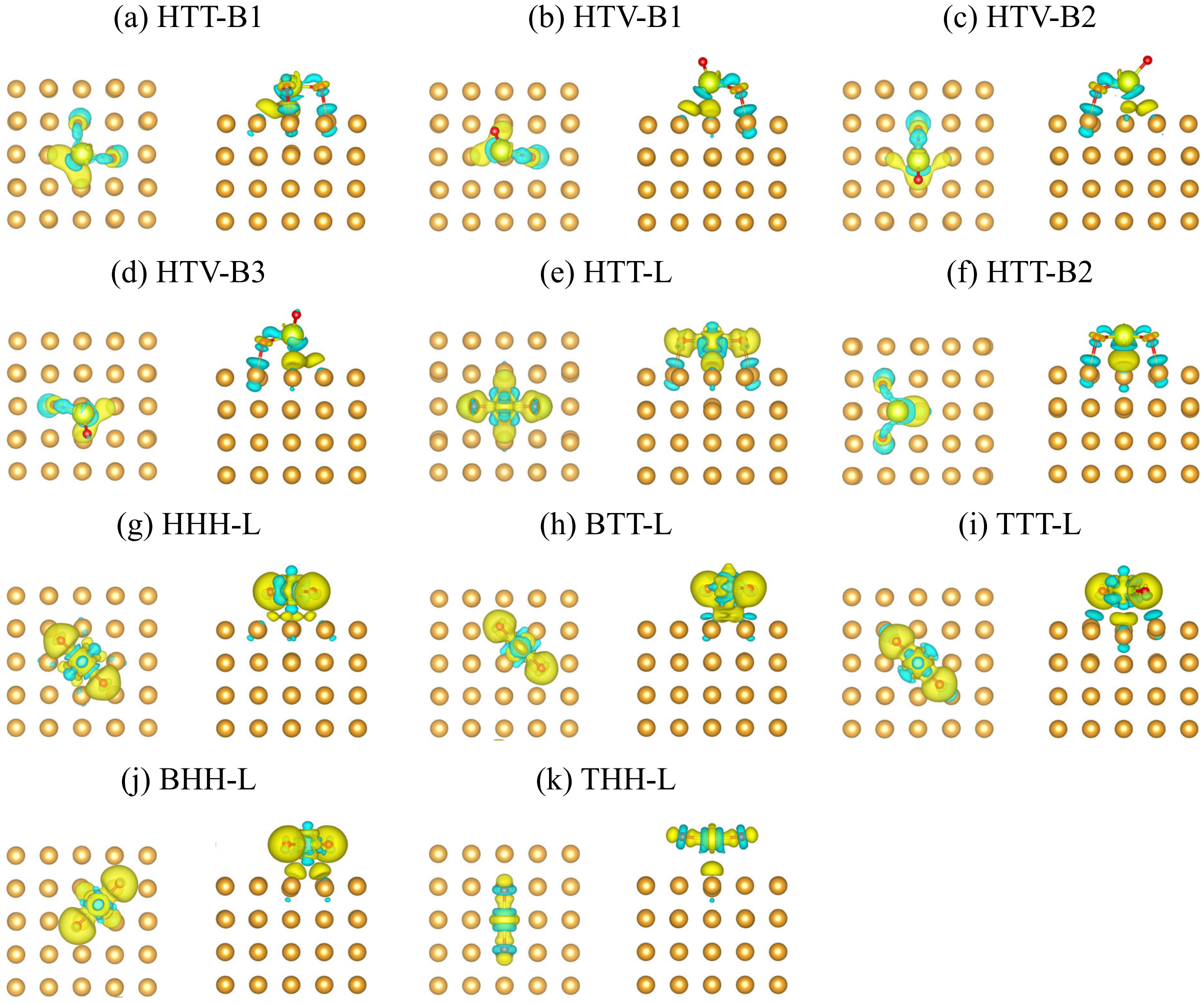

Figure S6. Different configurations and corresponding charge density difference $\Delta\rho$ of the $CeO_2$ molecule on the Au(001) slab model with a $U_{\text{eff}}$ of 3 eV. Yellow and blue isosurfaces mean 0.005 e $Bohr^{-3}$ charge accumulation and -0.005 e $Bohr^{-3}$ charge depletion of $\Delta\rho$, respectively. The green, red, and orange balls represent Ce, O, and Au atoms. The maximum plotted length of the Ce-O bond is 2.864 Å.

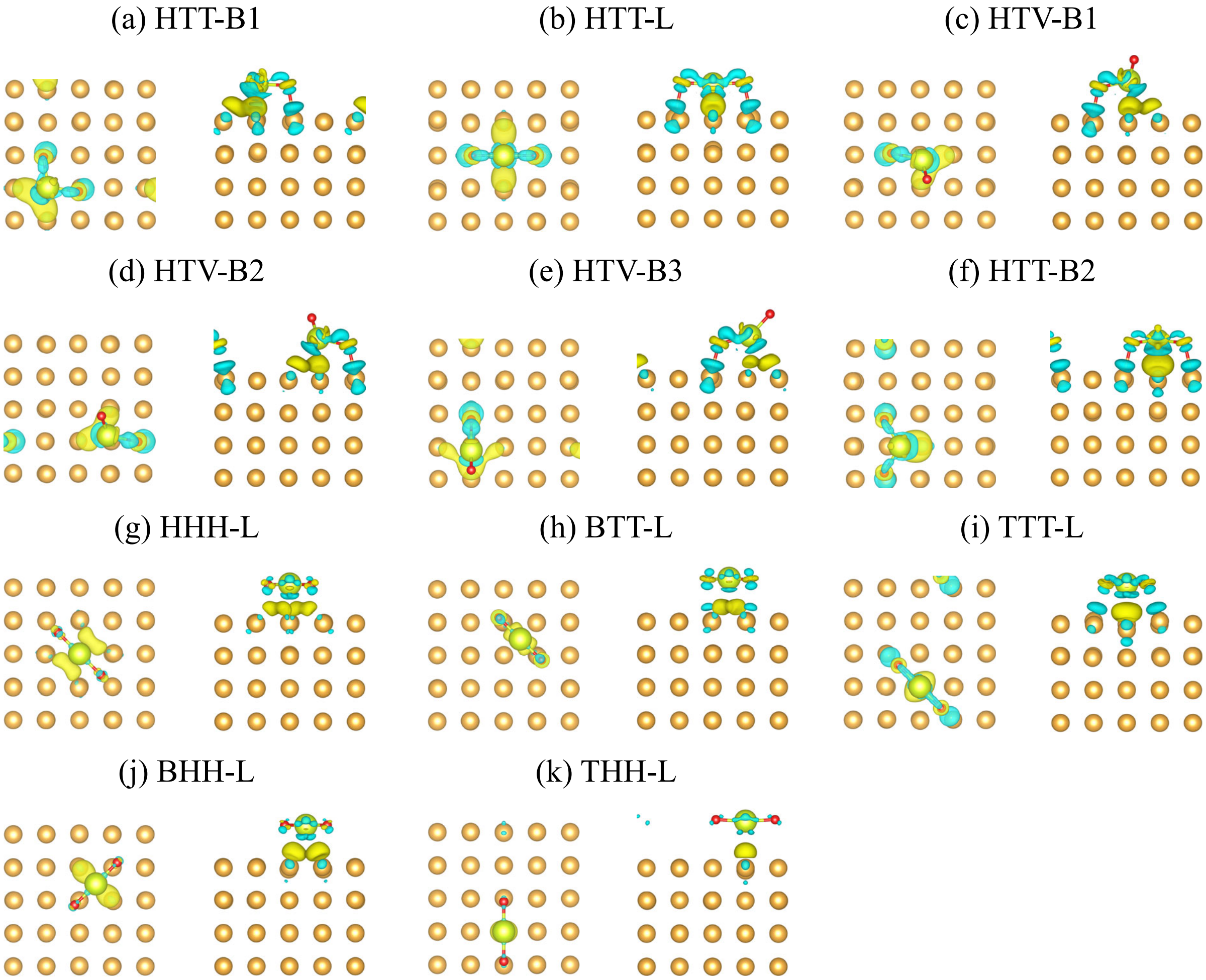

Figure S7. Different configurations and corresponding charge density difference $\Delta\rho$ of the $CeO_2$ molecule on the Au(001) slab model with a $U_{\text{eff}}$ of 5 eV. Yellow and blue isosurfaces mean 0.005 e Bohr$^{-3}$ charge accumulation and -0.005 e Bohr$^{-3}$ charge depletion of $\Delta\rho$, respectively. The green, red, and orange balls represent Ce, O, and Au atoms. The maximum plotted length of the Ce-O bond is 2.864 Å.

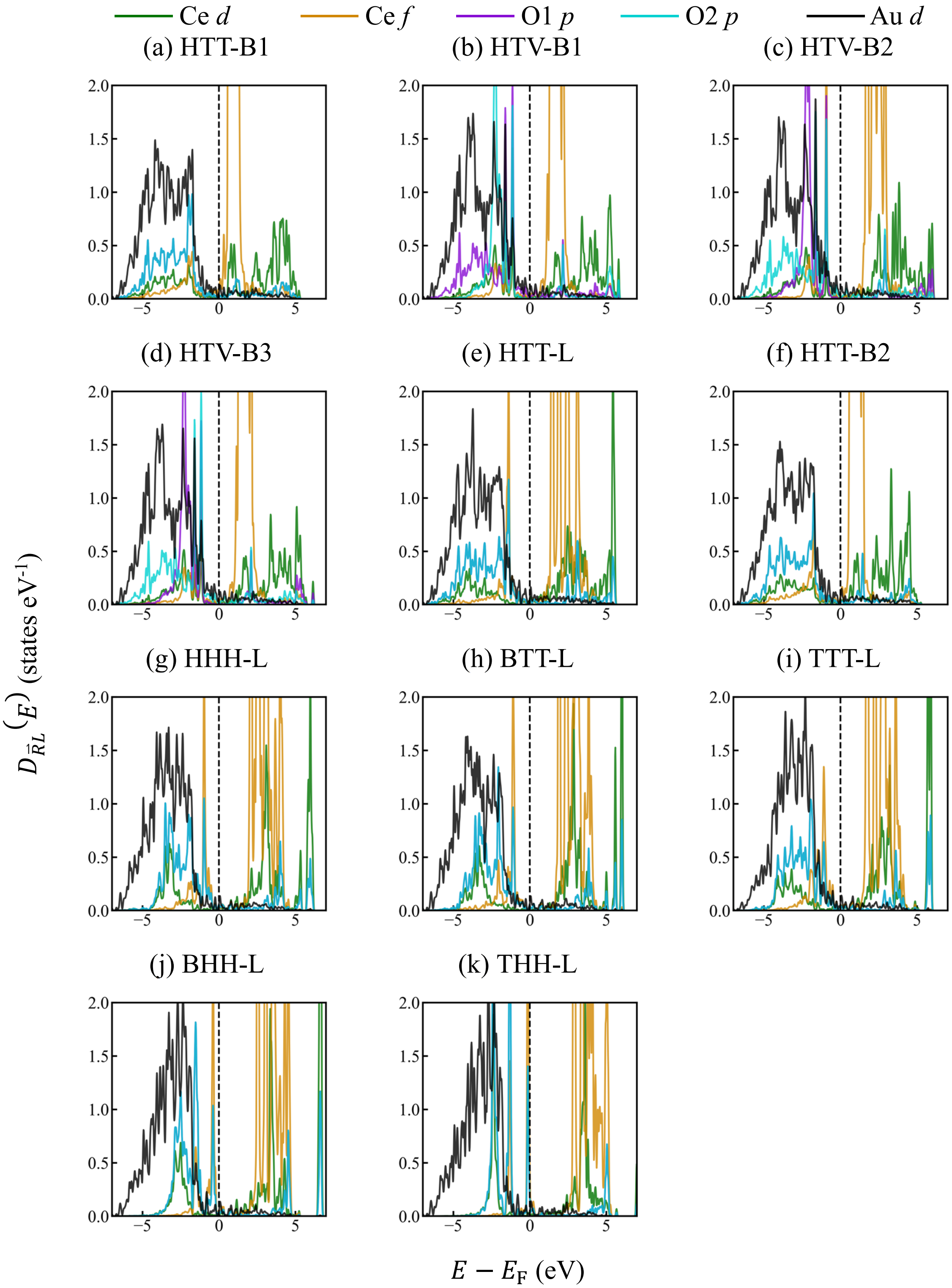

Figure S8. The electronic projected density of states $D_{\vec{R}L}(E)$ for Ce, O1, O2, and the nearest Au atom from one of the O atoms in different $CeO_2$ configurations on the Au(001) slab model with a $U_{\mathrm{eff}}$ of 3 eV. The Ce $d$, Ce $f$, O1 $p$, O2 $p$, and Au $d$ orbitals are denoted by solid green, solid red, dashed blue, dotted orange, and solid black lines, respectively.

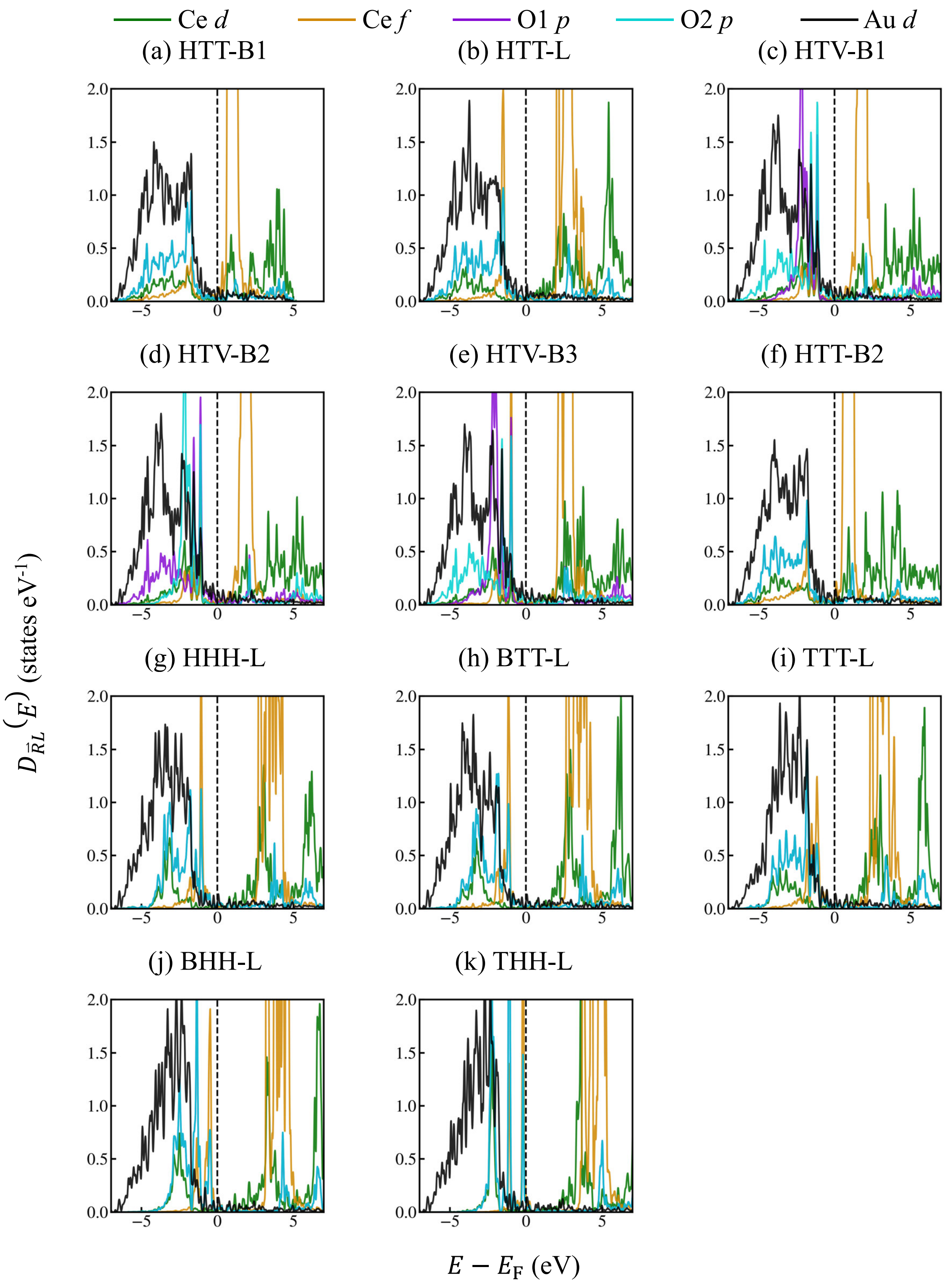

Figure S9. The electronic projected density of states $D_{\vec{R}L}(E)$ for Ce, O1, O2, and the nearest Au atom from one of the O atoms in different $CeO_2$ configurations on the Au(001) slab model with a $U_{eff}$ of 5 eV. The Ce *d*, Ce *f*, O1 *p*, O2 *p*, and Au *d* orbitals are denoted by solid green, solid red, dashed blue, dotted orange, and solid black lines, respectively.

## S3.2 CeO$_2$ molecules on Au(110)

Table S5. The adsorption sites after and before optimization, the binding energy $E_b$, the stability $\Delta E$, the bond length Ce-O1 and Ce-O2, and the bond angle O1-Ce-O2 of the CeO$_2$ molecule on the Au(110) slab model with a $U_{eff}$ of 3 eV.

| After opt. | Before opt. | $E_b$ (eV) | $\Delta E$ (eV) | Ce-O1 (Å) | Ce-O2 (Å) | O1-Ce-O2 (°) | Case |
|---|---|---|---|---|---|---|---|
| HTV-B | TTV-B1 | -1.322 | -4.232 | 1.846 | 2.001 | 122.981 | 2 |
| HBB-L | HBB-L | -1.215 | -4.125 | 1.930 | 1.930 | 174.149 | 1 |
| TTV-B | TTV-B2 | -1.041 | -3.951 | 1.848 | 1.979 | 130.151 | 2 |
| HHH-L | HHH-L | -0.862 | -3.772 | 1.906 | 1.906 | 175.189 | 3 |
| BTT-L | BTT-L | -0.794 | -3.704 | 1.897 | 1.897 | 166.666 | 3 |
| TTT-L | TTT-L | -0.620 | -3.530 | 1.909 | 1.909 | 170.230 | 3 |
| BHH-L | BHH-L | -0.328 | -3.238 | 1.858 | 1.858 | 167.710 | 3 |
| THH-L | TBB-L | -0.287 | -3.197 | 1.856 | 1.856 | 169.211 | 3 |

Table S6. The adsorption sites after and before optimization, the binding energy $E_b$, the stability $\Delta E$, the bond length Ce-O1 and Ce-O2, and the bond angle O1-Ce-O2 of the CeO$_2$ molecule on the Au(110) slab model with a $U_{eff}$ of 5 eV.

| After opt. | Before opt. | $E_b$ (eV) | $\Delta E$ (eV) | Ce-O1 (Å) | Ce-O2 (Å) | O1-Ce-O2 (°) | Case |
|---|---|---|---|---|---|---|---|
| HTV-B | TTV-B1 | -1.292 | -3.469 | 1.862 | 2.019 | 121.106 | 2 |
| HBB-L | HBB-L | -1.262 | -3.439 | 1.941 | 1.941 | 174.182 | 1 |
| TTV-B | TTV-B2 | -0.991 | -3.168 | 1.865 | 1.998 | 132.191 | 2 |
| HHH-L | HHH-L | -0.892 | -3.069 | 1.916 | 1.916 | 175.519 | 3 |
| BTT-L | BTT-L | -0.801 | -2.978 | 1.904 | 1.904 | 170.108 | 3 |
| TTT-L | TTT-L | -0.615 | -2.792 | 1.918 | 1.918 | 172.516 | 3 |
| BHH-L | BHH-L | -0.288 | -2.465 | 1.867 | 1.867 | 171.007 | 3 |
| THH-L | TBB-L | -0.271 | -2.448 | 1.865 | 1.865 | 169.817 | 3 |

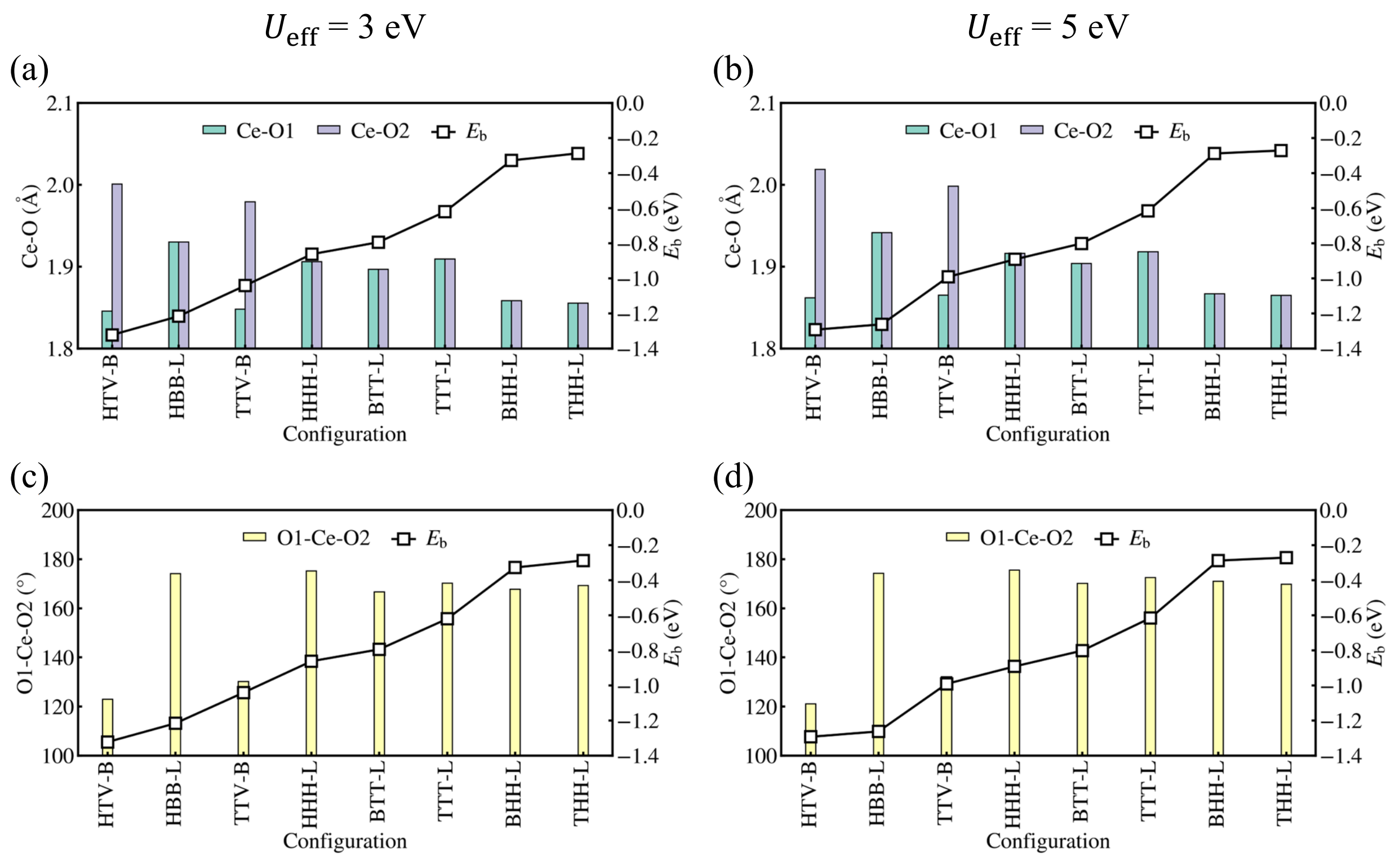

Figure S10. The correlation between the binding energy $E_b$ of the $CeO_2$ molecule on the Au(110) slab model with respect to (a, b) Ce-O1 and Ce-O2 bond lengths and (c, d) O1-Ce-O2 bond angle, where (a, c) have a $U_{\mathrm{eff}}$ of 3 eV and (b, d) have a $U_{\mathrm{eff}}$ of 5 eV.

Table S7. The adsorption sites after optimization, the effective charge $Z_{\mathrm{eff,Ce}}$, $Z_{\mathrm{eff,O1}}$, and $Z_{\mathrm{eff,O2}}$ of Ce, O1, and O2 atoms in the $CeO_2$ molecule on the Au(110) slab model with a $U_{\mathrm{eff}}$ of 3 eV are provided, along with their respective cases of adsorption.

| After opt. | $Z_{\mathrm{eff,Ce}}$ (e) | $Z_{\mathrm{eff,O1}}$ (e) | $Z_{\mathrm{eff,O2}}$ (e) | Case |
|---|---|---|---|---|
| HTV-B | 2.053 | -0.937 | -0.954 | 2 |
| HBB-L | 1.992 | -0.923 | -0.923 | 1 |
| TTV-B | 2.058 | -0.951 | -0.972 | 2 |
| HHH-L | 2.007 | -0.959 | -0.959 | 3 |
| BTT-L | 2.035 | -0.951 | -0.951 | 3 |
| TTT-L | 2.016 | -0.968 | -0.968 | 3 |
| BHH-L | 2.077 | -0.987 | -0.987 | 3 |
| THH-L | 2.083 | -0.999 | -0.999 | 3 |

Table S8. The adsorption sites after optimization, the effective charge $Z_{\mathrm{eff,Ce}}$, $Z_{\mathrm{eff,O1}}$, and $Z_{\mathrm{eff,O2}}$ of Ce, O1, and O2 atoms in the $CeO_2$ molecule on the Au(110) slab model with a $U_{\mathrm{eff}}$ of 5 eV are provided, along with their respective cases of adsorption.

| After opt. | $Z_{\mathrm{eff,Ce}}$ (e) | $Z_{\mathrm{eff,O1}}$ (e) | $Z_{\mathrm{eff,O2}}$ (e) | Case |
|---|---|---|---|---|
| HTV-B | 2.073 | -0.941 | -0.962 | 2 |
| HBB-L | 2.023 | -0.934 | -0.934 | 1 |
| TTV-B | 2.074 | -0.955 | -0.981 | 2 |
| HHH-L | 2.038 | -0.971 | -0.971 | 3 |
| BTT-L | 2.021 | -0.945 | -0.945 | 3 |
| TTT-L | 2.073 | -0.991 | -0.990 | 3 |
| BHH-L | 2.128 | -1.029 | -1.029 | 3 |
| THH-L | 2.125 | -1.023 | -1.023 | 3 |

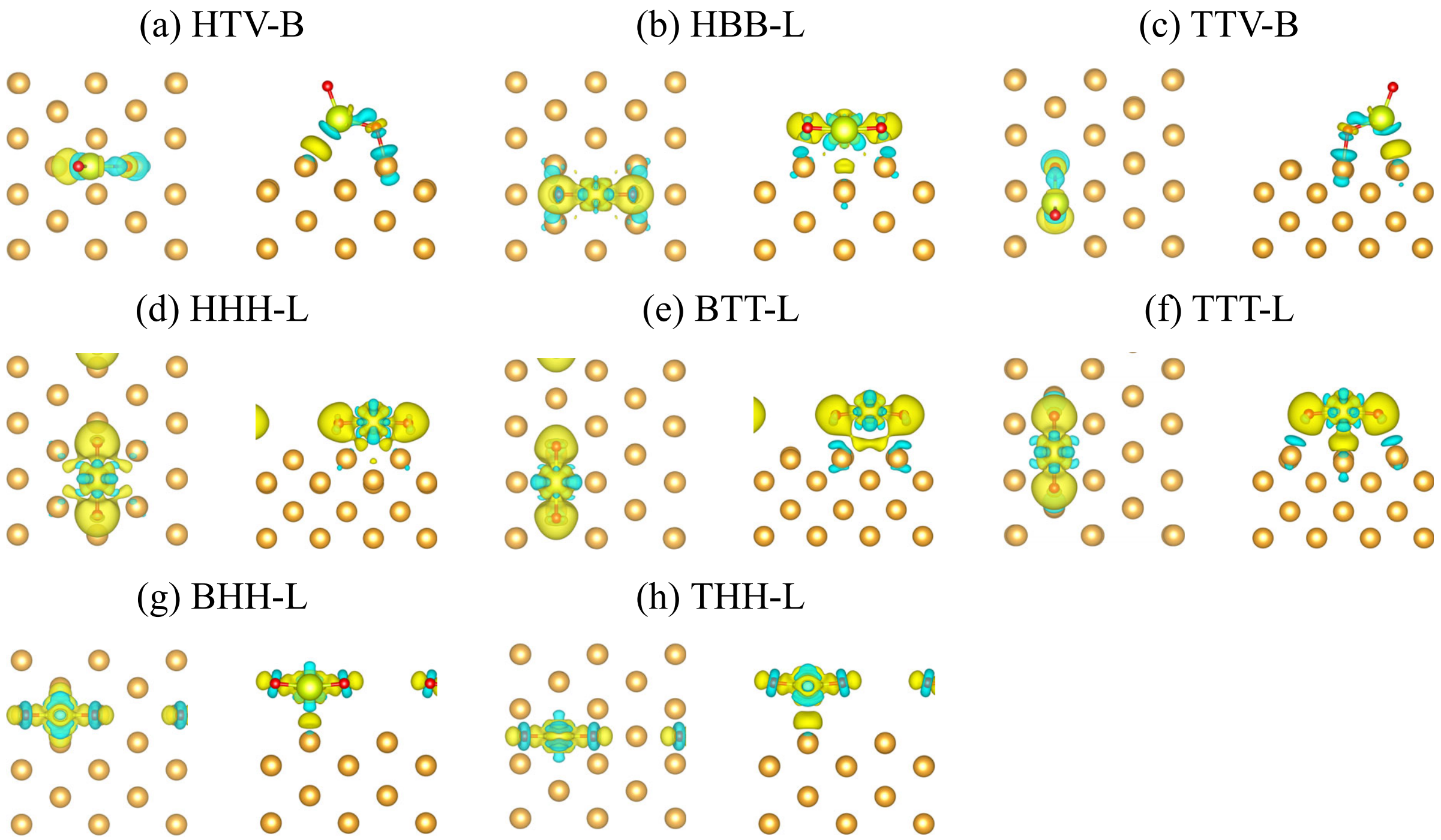

Figure S11. Different configurations and corresponding charge density difference $\Delta\rho$ of the $CeO_2$ molecule on the Au(110) slab model with a $U_{\mathrm{eff}}$ of 3 eV. Yellow and blue isosurfaces mean 0.005 e $Bohr^{-3}$ and -0.005 e $Bohr^{-3}$ of $\Delta\rho$, respectively. The green, red, and orange balls represent Ce, O, and Au atoms. The maximum plotted length of the Ce-O bond is 2.864 Å.

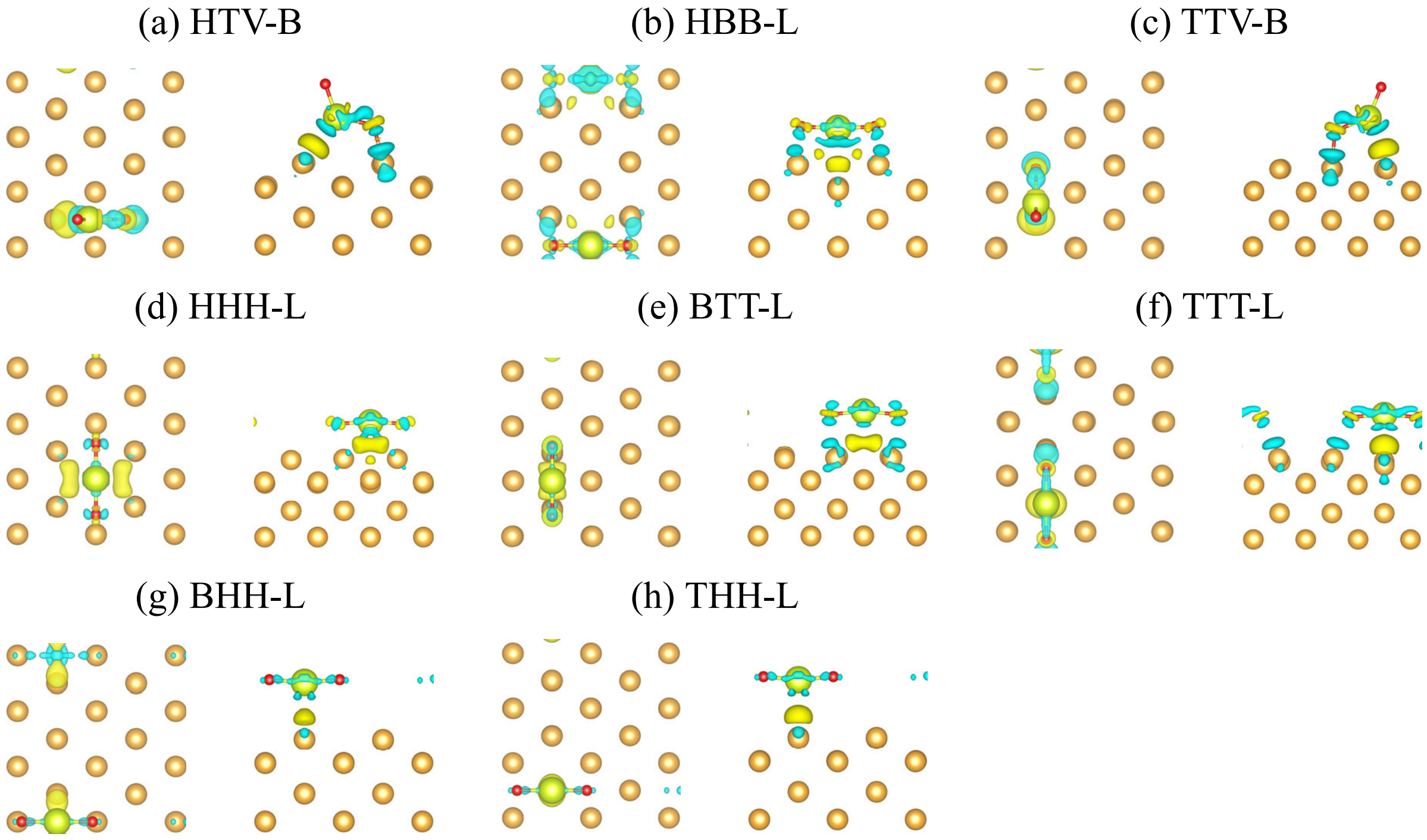

Figure S12. Different configurations and corresponding charge density difference $\Delta\rho$ of the $CeO_2$ molecule on the Au(110) slab model with a $U_{\mathrm{eff}}$ of 5 eV. Yellow and blue isosurfaces mean 0.005 e $Bohr^{-3}$ and -0.005 e $Bohr^{-3}$ of $\Delta\rho$, respectively. The green, red, and orange balls represent Ce, O, and Au atoms. The maximum plotted length of the Ce-O bond is 2.864 Å.

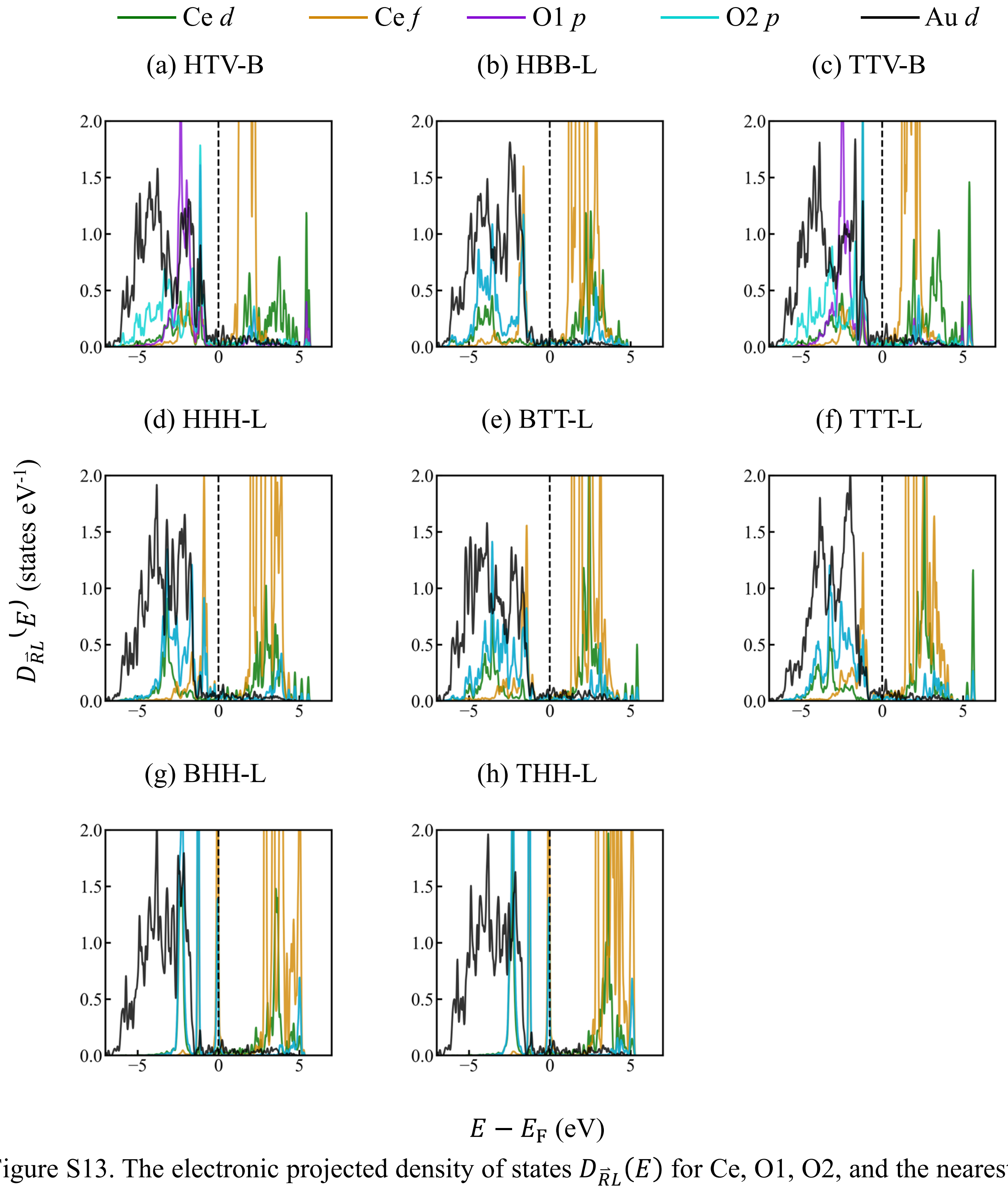

Figure S13. The electronic projected density of states $D_{\vec{R}L}(E)$ for Ce, O1, O2, and the nearest Au atom from one of the O atoms in different $CeO_2$ configurations on the Au(110) slab model with a $U_{\mathrm{eff}}$ of 3 eV. The Ce *d*, Ce *f*, O1 *p*, O2 *p*, and Au *d* orbitals are denoted by solid green, solid red, dashed blue, dotted orange, and solid black lines, respectively.

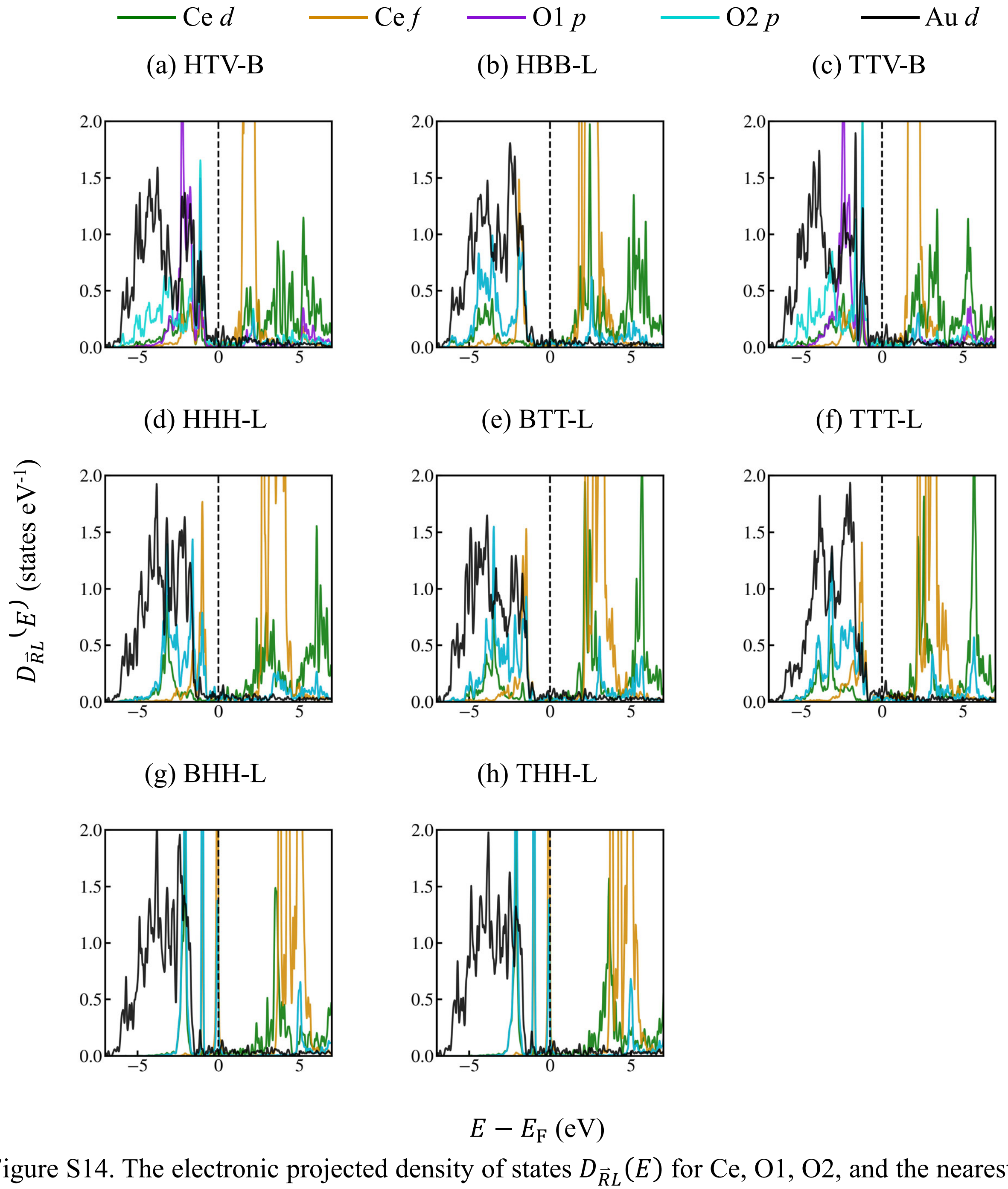

Figure S14. The electronic projected density of states $D_{\vec{R}L}(E)$ for Ce, O1, O2, and the nearest Au atom from one of the O atoms in different $CeO_2$ configurations on the Au(110) slab model with a $U_{\mathrm{eff}}$ of 5 eV. The Ce *d*, Ce *f*, O1 *p*, O2 *p*, and Au *d* orbitals are denoted by solid green, solid red, dashed blue, dotted orange, and solid black lines, respectively

### S3.3 $CeO_2$ molecules on Au(111)

Table S9. The adsorption sites after and before optimization, the binding energy $E_b$, the stability $\Delta E$, the bond length Ce-O1 and Ce-O2, and the bond angle O1-Ce-O2 of the $CeO_2$ molecule on the Au(111) slab model with a $U_{eff}$ of 3 eV.

| After opt. | Before opt. | $E_b$ (eV) | $\Delta E$ (eV) | Ce-O1 (Å) | Ce-O2 (Å) | O1-Ce-O2 (°) | Case |
|---|---|---|---|---|---|---|---|
| BTV-B1 | TTT-L | -1.044 | -4.651 | 1.849 | 1.995 | 128.610 | 2 |
| BTV-B2 | BBB-L | -1.028 | -4.635 | 1.992 | 1.849 | 123.044 | 2 |
| BTV-B3 | THH-L | -1.025 | -4.632 | 1.995 | 1.849 | 131.527 | 2 |
| BTV-B4 | TTT-B | -1.018 | -4.625 | 1.842 | 1.986 | 148.982 | 2 |
| BTV-B5 | THH-B | -1.017 | -4.624 | 1.849 | 1.992 | 124.919 | 2 |
| TTV-B | TTV-B | -0.991 | -4.598 | 1.849 | 1.985 | 130.765 | 2 |
| HTT-B | HTT-B | -0.944 | -4.551 | 1.925 | 1.926 | 111.431 | 3 |
| BTT-L | BTT-L | -0.873 | -4.480 | 1.902 | 1.933 | 178.405 | 3 |
| HHH-B | HHH-L | -0.609 | -4.216 | 1.912 | 1.912 | 119.938 | 3 |

Table S10. The adsorption sites after and before optimization, the binding energy $E_b$, the stability $\Delta E$, the bond length Ce-O1 and Ce-O2, and the bond angle O1-Ce-O2 of the $CeO_2$ molecule on the Au(111) slab model with a $U_{eff}$ of 5 eV.

| After opt. | Before opt. | $E_b$ (eV) | $\Delta E$ (eV) | Ce-O1 (Å) | Ce-O2 (Å) | O1-Ce-O2 (°) | Case |
|---|---|---|---|---|---|---|---|
| BTV-B1 | TTT-L | -1.011 | -3.874 | 1.866 | 2.016 | 126.412 | 2 |
| BTV-B2 | TTT-B | -1.006 | -3.869 | 1.854 | 1.997 | 154.769 | 2 |
| BTV-B3 | BBB-L | -0.999 | -3.862 | 2.012 | 1.866 | 120.566 | 2 |
| BTV-B4 | THH-L | -0.990 | -3.853 | 2.015 | 1.865 | 130.479 | 2 |
| BTV-B5 | THH-B | -0.986 | -3.849 | 1.867 | 2.014 | 124.585 | 2 |
| BTV-B6 | HHH-B | -0.981 | -3.844 | 1.866 | 2.013 | 129.022 | 2 |
| TTV-B | TTV-B | -0.945 | -3.809 | 1.867 | 2.003 | 130.197 | 2 |
| HTT-B | HTT-B | -0.935 | -3.799 | 1.942 | 1.943 | 110.921 | 3 |
| BTT-L | BTT-L | -0.901 | -3.764 | 1.911 | 1.944 | 179.138 | 3 |

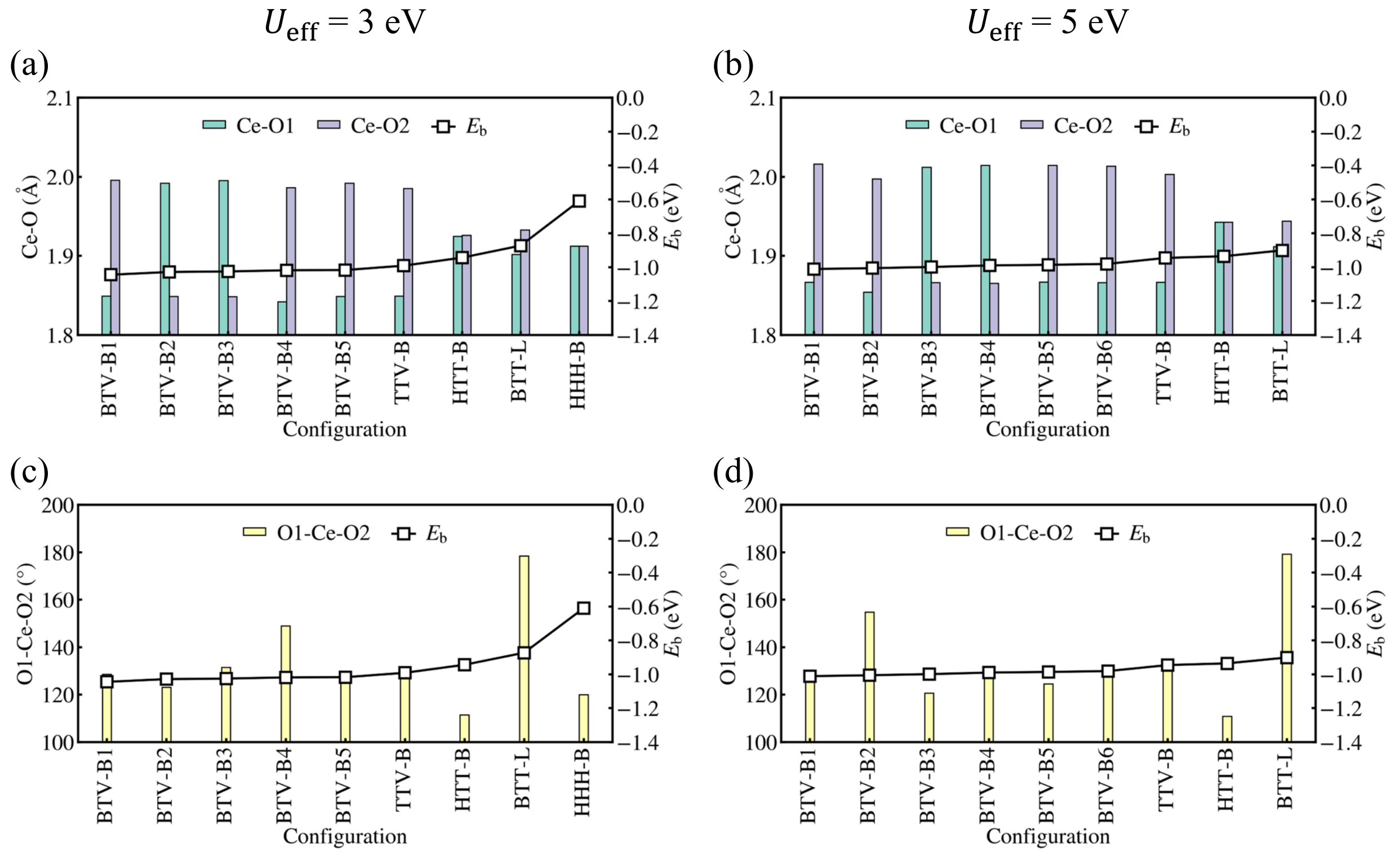

Figure S15. The correlation between the binding energy $E_b$ of the $CeO_2$ molecule on the Au(111) slab model with respect to (a, b) Ce-O1 and Ce-O2 bond lengths and (c, d) O1-Ce-O2 bond angle, where (a, c) have a $U_{eff}$ of 3 eV and (b, d) have a $U_{eff}$ of 5 eV.

Table S11. The adsorption sites after optimization, the effective charge $Z_{\mathrm{eff,Ce}}$, $Z_{\mathrm{eff,O1}}$, and $Z_{\mathrm{eff,O2}}$ of Ce, O1, and O2 atoms in the $CeO_2$ molecule on the Au(111) slab model with a $U_{\mathrm{eff}}$ of 3 eV are provided, along with their respective cases of adsorption.

| After opt. | $Z_{\mathrm{eff,Ce}}$ (e) | $Z_{\mathrm{eff,O1}}$ (e) | $Z_{\mathrm{eff,O2}}$ (e) | Case |
|---|---|---|---|---|
| BTV-B1 | 2.041 | -0.945 | -0.958 | 2 |
| BTV-B2 | 2.025 | -0.952 | -0.940 | 2 |
| BTV-B3 | 2.041 | -0.960 | -0.942 | 2 |
| BTV-B4 | 2.068 | -0.961 | -0.959 | 2 |
| BTV-B5 | 2.035 | -0.941 | -0.954 | 2 |
| TTV-B | 2.059 | -0.943 | -0.962 | 2 |
| HTT-B | 1.979 | -0.942 | -0.938 | 3 |
| BTT-L | 2.040 | -0.965 | -0.962 | 3 |
| HHH-B | 1.976 | -0.953 | -0.952 | 3 |

Table S12. The adsorption sites after optimization, the effective charge $Z_{\mathrm{eff,Ce}}$, $Z_{\mathrm{eff,O1}}$, and $Z_{\mathrm{eff,O2}}$ of Ce, O1, and O2 atoms in the $CeO_2$ molecule on the Au(111) slab model with a $U_{\mathrm{eff}}$ of 5 eV are provided, along with their respective cases of adsorption.

| After opt. | $Z_{\mathrm{eff,Ce}}$ (e) | $Z_{\mathrm{eff,O1}}$ (e) | $Z_{\mathrm{eff,O2}}$ (e) | Case |
|---|---|---|---|---|
| BTV-B1 | 2.057 | -0.947 | -0.966 | 2 |
| BTV-B2 | 2.082 | -0.964 | -0.964 | 2 |
| BTV-B3 | 2.046 | -0.958 | -0.949 | 2 |
| BTV-B4 | 2.054 | -0.969 | -0.959 | 2 |
| BTV-B5 | 2.056 | -0.951 | -0.961 | 2 |
| BTV-B6 | 2.063 | -0.944 | -0.954 | 2 |
| TTV-B | 2.077 | -0.949 | -0.965 | 2 |
| HTT-B | 2.004 | -0.949 | -0.948 | 3 |
| BTT-L | 2.049 | -0.969 | -0.960 | 3 |

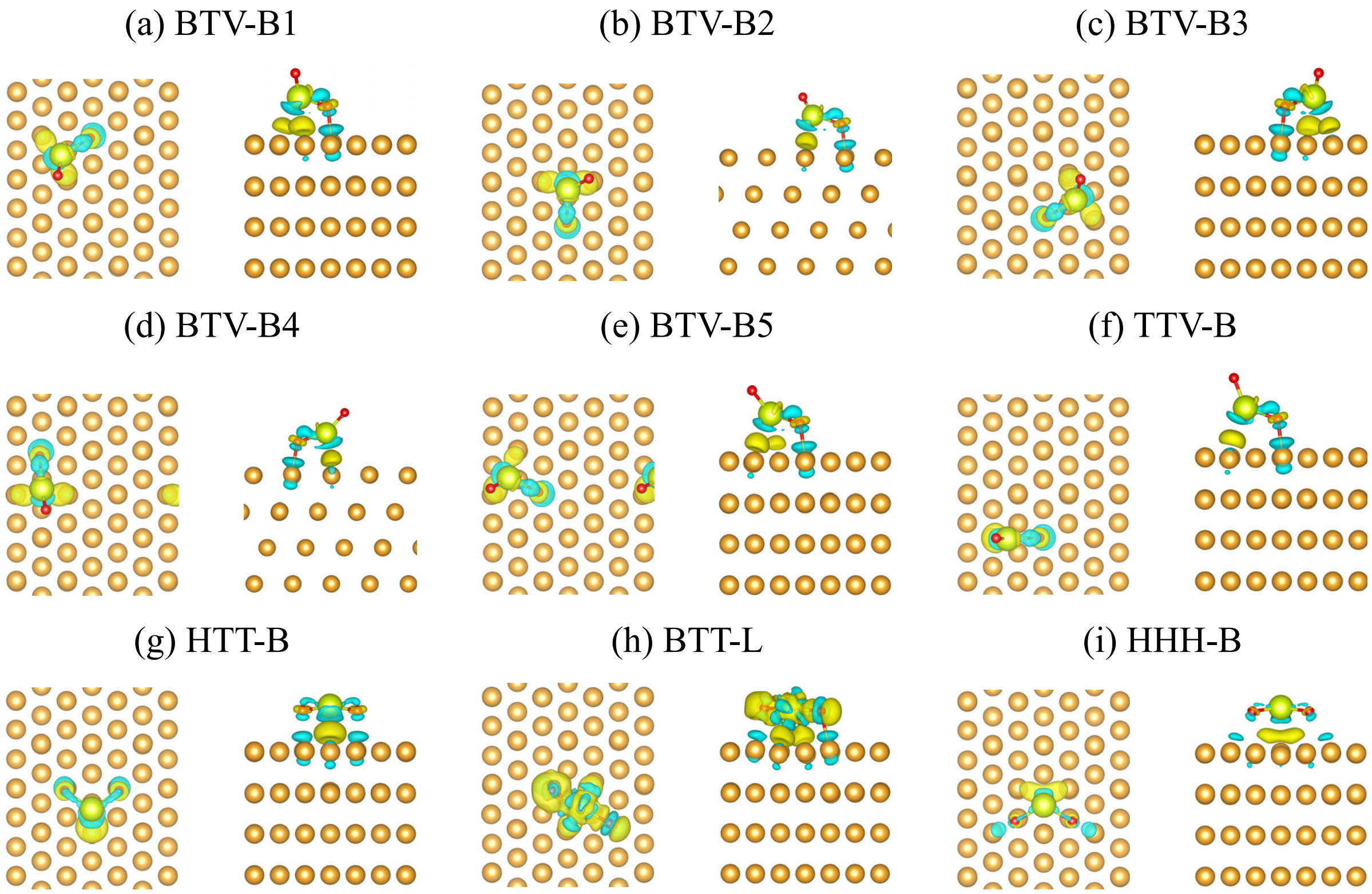

Figure S16. Different configurations and corresponding charge density difference $\Delta\rho$ of the $CeO_2$ molecule on the Au(111) slab model with a $U_{\text{eff}}$ of 3 eV. Yellow and blue isosurfaces mean 0.005 e $Bohr^{-3}$ and -0.005 e $Bohr^{-3}$ of $\Delta\rho$, respectively. The green, red, and orange balls represent Ce, O, and Au atoms. The maximum plotted length of the Ce-O bond is 2.864 Å.

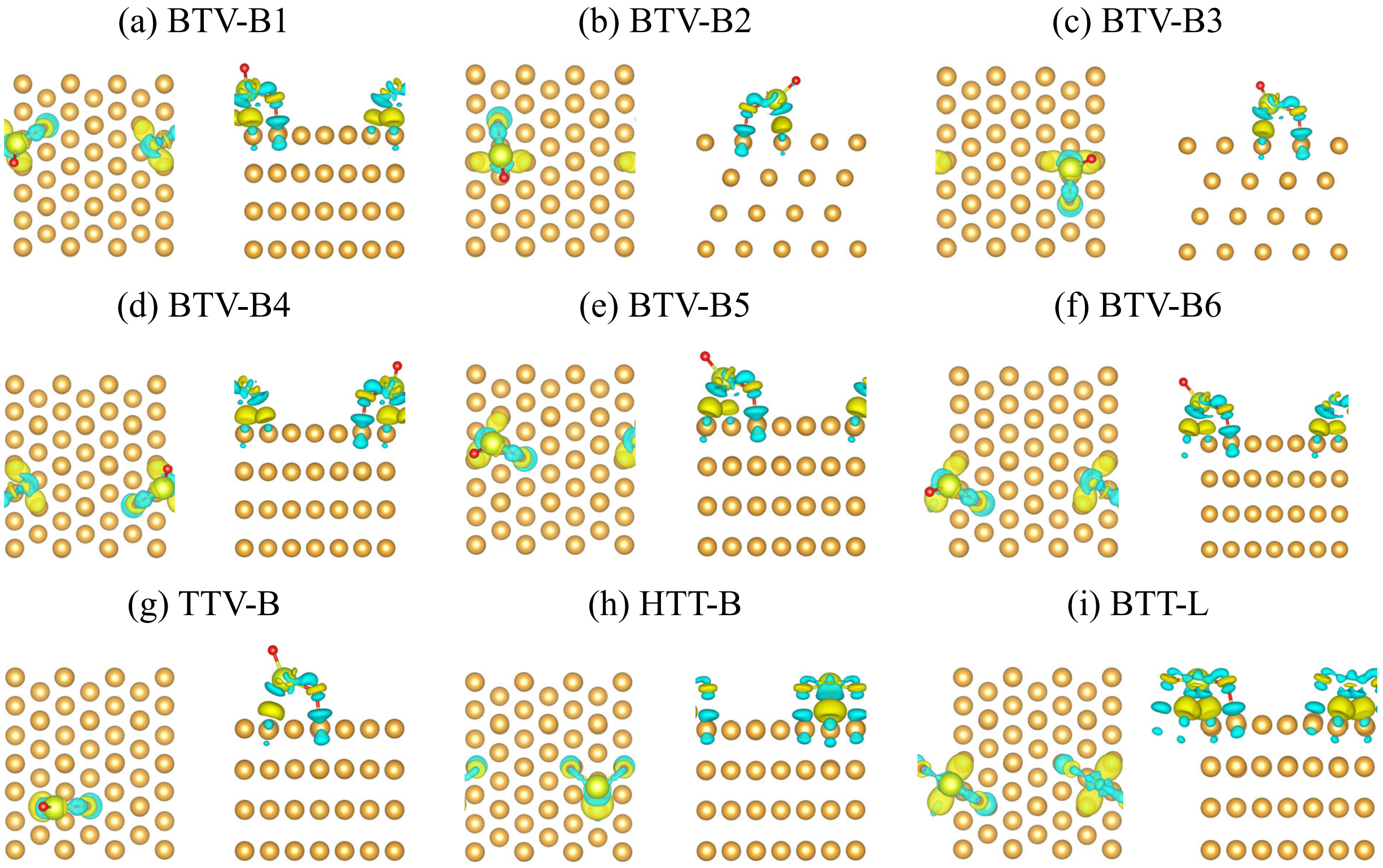

Figure S17. Different configurations and corresponding charge density difference $\Delta\rho$ of the $CeO_2$ molecule on the Au(111) slab model with a $U_{\mathrm{eff}}$ of 5 eV. Yellow and blue isosurfaces mean 0.005 e $Bohr^{-3}$ and -0.005 e $Bohr^{-3}$ of $\Delta\rho$, respectively. The green, red, and orange balls represent Ce, O, and Au atoms. The maximum plotted length of the Ce-O bond is 2.864 Å.

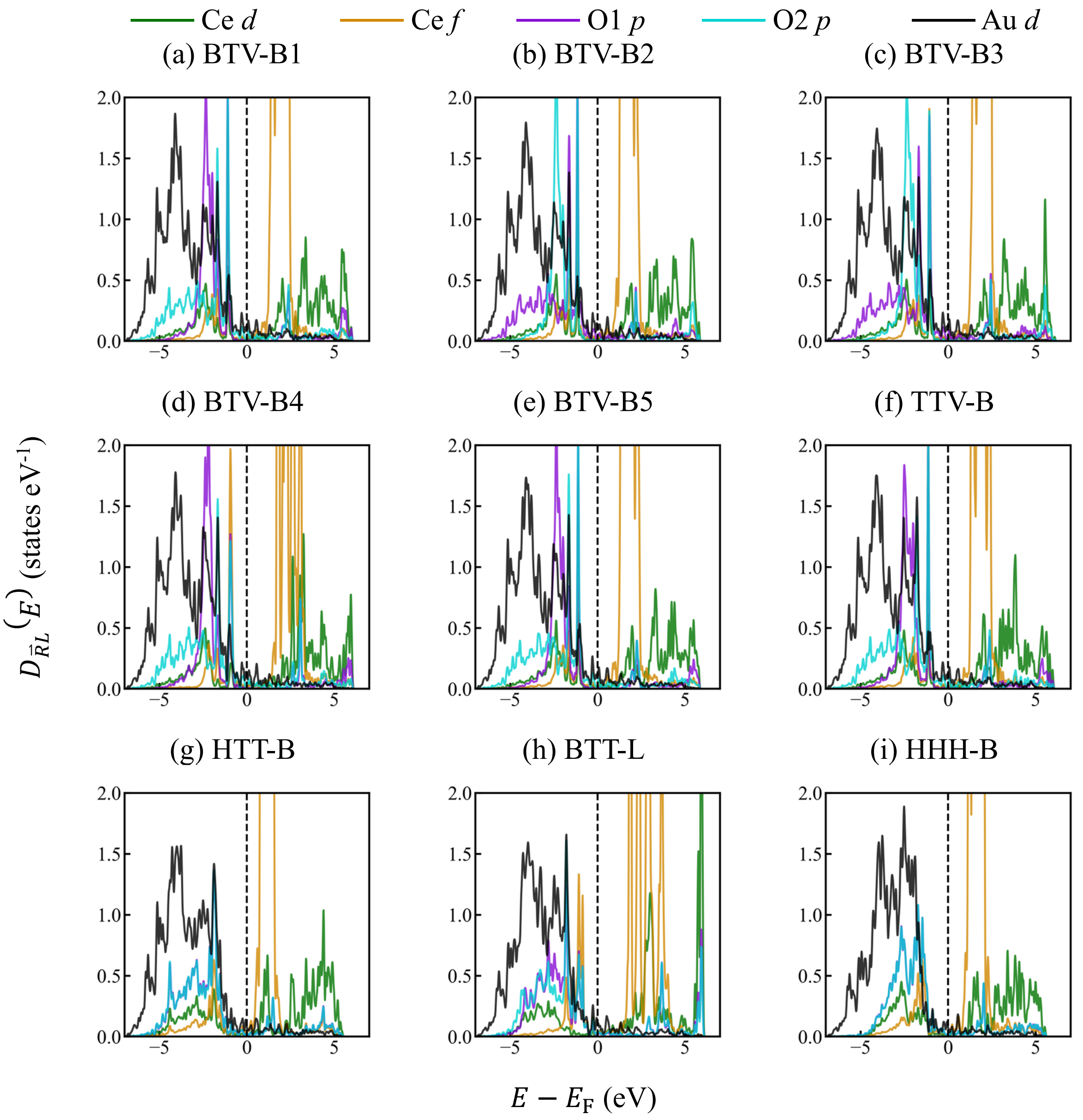

Figure S18. The electronic projected density of states $D_{\vec{R}L}(E)$ for Ce, O1, O2, and the nearest Au atom from one of the O atoms in different $CeO_2$ configurations on the Au(111) slab model with a $U_{\mathrm{eff}}$ of 3 eV. The Ce $d$, Ce $f$, O1 $p$, O2 $p$, and Au $d$ orbitals are denoted by solid green, solid red, dashed blue, dotted orange, and solid black lines, respectively.

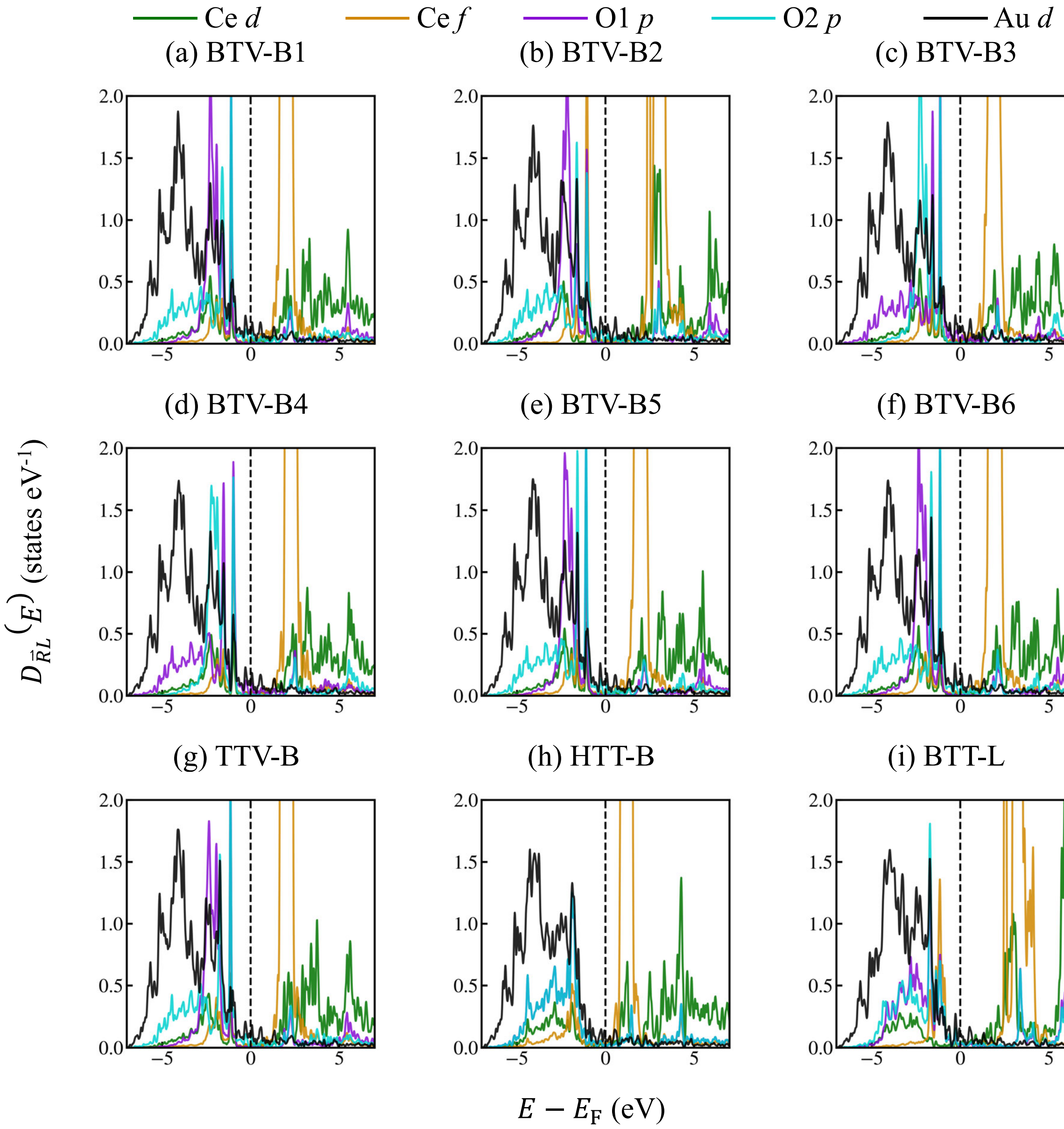

Figure S19. The electronic projected density of states $D_{\vec{R}L}(E)$ for Ce, O1, O2, and the nearest Au atom from one of the O atoms in different $CeO_2$ configurations on the Au(111) slab model with a $U_{\mathrm{eff}}$ of 5 eV. The Ce *d*, Ce *f*, O1 *p*, O2 *p*, and Au *d* orbitals are denoted by solid green, solid red, dashed blue, dotted orange, and solid black lines, respectively.

**References**

1 K. Sun, M. Kohyama, S. Tanaka and S. Takeda, Theoretical Study of Atomic Oxygen on Gold Surface by Hückel Theory and DFT Calculations, *J. Phys. Chem. A*, 2012, **116**, 9568–9573.

2 C. Heinemann, H. H. Cornehl, D. Schröder, M. Dolg and H. Schwarz, The $CeO_2^+$ Cation: Gas-Phase Reactivity and Electronic Structure, *Inorg. Chem.*, 1996, **35**, 2463–2475.

3 T. K. Todorova, I. Infante, L. Gagliardi and J. M. Dyke, The chemiionization reactions Ce + O and Ce + $O_2$ : Assignment of the observed chemielectron bands, *Int. J. Quantum Chem.*, 2009, **109**, 2068–2079.

4 R. L. De Kock, M. A. Peterson, L. K. Timmer, E. J. Baerends and P. Vernooijs, A theoretical study of the linear versus bent geometry for several $MX_2$ molecules: $MgF_2$, $CaH_2$, $CaF_2$, $CeO_2$ and $YbCl_2$, *Polyhedron*, 1990, **9**, 1919–1934.

5 S. D. Gabelnick, G. T. Reedy and M. G. Chasanov, Infrared spectra and structure of some matrix-isolated lanthanide and actinide oxides, *J. Chem. Phys.*, 1974, **60**, 1167–1171.

6 D. A. Andersson, S. I. Simak, B. Johansson, I. A. Abrikosov and N. V. Skorodumova, Modeling of $CeO_2$, $Ce_2O_3$, and $CeO_{2-x}$, in the LDA+U formalism, *Phys. Rev. B*, 2007, **75**, 035109.